\documentstyle[sprocl]{article}

\newcommand{\be}{\begin{equation}}
\newcommand{\ee}{\end{equation}}
\newcommand{\bearray}{\begin{eqnarray}}
\newcommand{\eearray}{\end{eqnarray}}
\newcommand{\nl}{\nonumber \\}

\newenvironment{exercise}{\small\begin{description}
                         \item[{Exercise:}]}{\end{description}}

\newcommand{\eq}[1]{{Eq.~(\protect\ref{#1})}}

\newcommand{\order}{{\cal O}}

\newcommand{\half}{\mbox{$\frac{1}{2}$}}
\newcommand{\third}{\mbox{$\frac{1}{3}$}}
\renewcommand{\Re}{{\rm\,Re\,}}
\newcommand{\Tr}{{\rm\,Tr\,}}

\newcommand{\ket}[1]{\left|#1\right\rangle}
\newcommand{\bra}[1]{\left\langle#1\right|}
\newcommand{\braket}[2]{\left\langle#1\right|\left.#2\right\rangle}

\newcommand{\e}{{\mathrm e}}
\newcommand{\I}{{i}}

\newcommand{\ti}{{t_{\mathrm i}}}
\newcommand{\tf}{{t_{\mathrm f}}}
\newcommand{\xxi}{{x_{\mathrm i}}}
\newcommand{\xxf}{{x_{\mathrm f}}}
\newcommand{\lat}{{\mathrm lat}}
\newcommand{\imp}{{\mathrm imp}}
\newcommand{\St}{\tilde{S}}
\newcommand{\Vt}{\tilde{V}}

\newcommand{\expval}[1]{\langle\langle #1 \rangle\rangle}

\newcommand{\Ht}{\tilde{H}}
\newcommand{\xt}{\tilde{x}}
\newcommand{\Dxt}{{\cal D}x(t)}
\newcommand{\Nconf}{{N_{\mathrm cf}}}
\newcommand{\Ncorr}{{N_{\mathrm cor}}}
\newcommand{\Gammab}{{\overline{\Gamma}}}
\newcommand{\xalpha}{{x^{(\alpha)}}}
\newcommand{\Galpha}{{G^{(\alpha)}}}

\newcommand{\psib}{\overline{\psi}}
\newcommand{\Porder}{{\cal P}}
\renewcommand{\P}{\Porder}
\newcommand{\dd}{d}

\newcommand{\Lag}{{\cal L}}

\newcommand{\F}[2]{F_{#1#2}}
\newcommand{\Pl}[2]{P_{#1#2}}
\newcommand{\Rt}[2]{R_{#1#2}}
\newcommand{\su}{{${\rm SU}_3$}}
\newcommand{\C}{{\cal C}}

\newcommand{\D}{{\mathrm D}}
\newcommand{\Dv}{{\bf D}}
\newcommand{\xv}{{\bf x}}

\def\U#1{U_{#1}}
\def\Udag#1{{U^\dagger_{#1}}}
\def\A#1{A_{#1}}

\begin{document}

\title{LATTICE QCD FOR NOVICES}
\author{G. PETER LEPAGE}
\address{Newman Laboratory of Nuclear Studies\\ Cornell University,
Ithaca, NY 14853\\E-mail: gpl@mail.lns.cornell.edu}
\date{May 1998}
\maketitle\abstracts{These lectures are for novices to lattice QCD.
They introduce a
set of simple ideas and numerical techniques that can be implemented
in a short period of time and that are capabable of generating
nontrivial, nonperturbative results from lattice QCD. The
simplest of these calculations can be completed on a standard
workstation or high-end personal computer.}

%\newpage
\section{Introduction}

These lectures are for novices who are interested in learning
how to do lattice QCD simulations. My intent is to describe in detail
everything that one needs to know in order to create and run a simple
lattice QCD simulation. My focus here is not on lengthy derivations or
detailed comparisons of algorithms. Rather I want to introduce 
a set of simple ideas and techniques that one can implement 
in a relatively short time and that are capable of generating
nontrivial results from lattice QCD.

We begin in Section~2 with simple one-dimensional quantum
mechanics. Most of the 
simulation techniques can be applied to ordinary quantum mechanics, and
the simulations require only seconds or minutes of computer time,
rather than the hours or days needed for QCD simulations. Consequently
such applications are ideal for learning the simulation technology. 
We broaden the discussion to include quantum field theories in
Section~3. Here we discuss the theoretical techniques needed to obtain
accurate results on the relatively coarse lattices well suited to
smaller computers. Finally, in Section~4, we adapt our techniques for
simulations of gluon dynamics in QCD.

\section{Numerical Path Integrals}

\subsection{Discretizing the Path Integral}
We begin with numerical techniques for evaluating {\em
path integrals}. Recall what path integrals tell us. In
one-dimensional quantum mechanics, for example, the evolution of a 
position eigenstate $\ket{\xxi}$ from time~$\ti$ to time~$\tf$ can be
computed using a path integral\,\cite{baym}:
\be \label{1d-pathint}
\bra{\xxf}\e^{-\Ht(\tf-\ti)}\ket{\xxi} = \int \Dxt\, \e^{-S[x]}.
\ee
Here the $\int \Dxt$ designates a sum over all possible particle paths
\be
\{x(t)\quad \mbox{for}\quad t=\ti \to \tf\}
\ee
 with
\be
x(\ti) = \xxi \quad\quad\quad x(\tf) = \xxf.
\ee
The hamiltonian is $\Ht$, and $S[x]$ is the classical action,
\be \label{onedqm}
S[x] \equiv \int_{\ti}^{\tf} dt\,  L(x,\dot{x})
\equiv \int_{\ti}^{\tf} dt\, \left[ \frac{m\,\dot{x}(t)^2}{2} +
V(x(t)) \right],
\ee
evaluated for each path~$x(t)$. There are no $\I$'s in these formulas
because we are using ``euclidean'' path integrals. These are derived in
the same way as standard path integrals but with $\I t\to
t$. Euclidean path integrals are much better for numerical work since
the integrands do not oscillate wildly in sign.

Knowledge of the propagator, \eq{1d-pathint}, as a function of
$\xxi,\ti,\xxf,\tf$ gives us complete information about the
quantum theory. For example, we can easily
determine the groundstate energy and wavefunction. Setting
\be \label{1d-periodicbc}
\xxi = \xxf \equiv x \quad\quad\quad \tf-\ti\equiv T,
\ee
the propagator can be rewritten
\be
\bra{x}\e^{-\Ht T}\ket{x} = \sum_n \braket{x}{E_n} \e^{-E_n T}
\braket{E_n}{x}
\ee
where $\ket{E_n}$ is the energy eigenstate with eigenvalue~$E_n$.
The sum is dominated by the lowest-energy states when $T$ is large,
because of the exponentials, 
and in the limit of very large~$T$ only the groundstate, $\ket{E_0}$,
contributes: 
\be
\bra{x}\e^{-\Ht T}\ket{x}
\,\,\stackrel{T\to\infty}{\longrightarrow}\,\,
 \e^{-E_0
T}\,\left|\braket{x}{E_0}\right|^2.
\ee
We  extract the groundstate energy~$E_0$ by integrating over~$x$,
\be
\int dx\, \bra{x}\e^{-\Ht T}\ket{x}
\,\,\stackrel{T\to\infty}{\longrightarrow}\,\,
\e^{-E_0 T},
\ee 
and then, going back to the previous equation, we determine the
groundstate wavefunction $\psi_{E_0}(x)\equiv\braket{x}{E_0}$.

Our goal therefore is to develop a numerical procedure for evaluating
the propagator using a path integral. There are two issues we must
address. First we must find a way to represent an arbitrary particle
path $\{x(t),\,\ti\le t\le \tf\}$ in the computer. A path is specified
by a function~$x(t)$ which, in principle, can be infinitely complex
(and therefore too much for any computer). We approximate this
function by specifying~$x(t)$ only at the nodes or sites on a
discretized $t$~axis:
\be
t_j = \ti + j\,a \quad\quad \mbox{for $j=0,1 \ldots N$}
\ee
where $a$ is the grid spacing,
\be
a \equiv \frac{\tf-\ti}{N}.
\ee
Then a path is described by a vector of numbers,
\be
x = \{x(t_0),x(t_1)\ldots x(t_N)\}.
\ee
It is common practice to refer to such a path as a ``configuration''.
The integral over all paths in this approximation becomes an ordinary
integral over all possible values for each of the $x(t_j)$'s: that is,
\be
\int \Dxt \to A \int_{-\infty}^{\infty} dx_1\,dx_2\ldots dx_{N-1}
\ee
where we have adopted the notation $x_j\equiv x(t_j)$. We don't
integrate over the endpoints since they are held fixed; for example,
for boundary conditions~(\ref{1d-periodicbc}),
\be
x_0 = x_N = x.
\ee
We won't need the normalization factor~$A$ for most of our work, but
for our one-dimensional problem it is\,\cite{baym}
\be
A \equiv \left(\frac{m}{2\pi a}\right)^{N/2}
\ee

The second issue we must address concerns the evaluation of the action
given only a discretized path~$\{x_j\}$. Focusing just on the
contribution from $t_j\le t\le t_{j+1}$, the obvious approximation is
\be
\int_{t_j}^{t_{j+1}} dt\,L \,\,\approx\,\, a
\left[ \frac{m}{2}\,\left(\frac{x_{j+1}-x_j}{a}\right)^2
+ \frac{1}{2}\left(V(x_{j+1}) + V(x_j)\right) \right]
\ee
With this approximation, our numerical representation of the path
integral is complete, and we have an approximate expression for the
quantum mechanical propagator: for example,
\be \label{1d-approxprop}
\bra{x}\e^{-\Ht T}\ket{x} \approx A \int_{-\infty}^\infty
dx_1\ldots dx_{N-1}\, \e^{-S_\lat[x]}
\ee
where
\be 
S_\lat[x] \equiv  \sum_{j=0}^{N-1} \left[ 
\frac{m}{2a}(x_{j+1}-x_j)^2 + a V(x_j) \right] ,
\ee
$x_0=x_N=x$, and $a=T/N$. We have reduced quantum mechanics to a
problem in numerical integration.

One might worry about approximating $\dot{x}$ with $(x_{j+1}-x_j)/a$
in our formula for the lattice action~$S_\lat[x]$. It is not obvious
that this is a good approximation given that $x_{j+1}-x_j$ can be
arbitrarily large in our path integral; that is, paths can be
arbitrarily rough. While not so important for our one-dimensional
problem,
this becomes a crucial issue for four-dimensional field
theories. It is dealt with using renormalization theory, which we 
discuss in later sections.

\begin{figure}
\begin{center}
% GNUPLOT: LaTeX picture
\setlength{\unitlength}{0.240900pt}
\ifx\plotpoint\undefined\newsavebox{\plotpoint}\fi
\sbox{\plotpoint}{\rule[-0.200pt]{0.400pt}{0.400pt}}%
\begin{picture}(900,600)(0,0)
\font\gnuplot=cmr10 at 10pt
\gnuplot
\sbox{\plotpoint}{\rule[-0.200pt]{0.400pt}{0.400pt}}%
\put(221.0,123.0){\rule[-0.200pt]{4.818pt}{0.400pt}}
\put(201,123){\makebox(0,0)[r]{$0$}}
\put(859.0,123.0){\rule[-0.200pt]{4.818pt}{0.400pt}}
\put(221.0,342.0){\rule[-0.200pt]{4.818pt}{0.400pt}}
\put(201,342){\makebox(0,0)[r]{$0.05$}}
\put(859.0,342.0){\rule[-0.200pt]{4.818pt}{0.400pt}}
\put(221.0,560.0){\rule[-0.200pt]{4.818pt}{0.400pt}}
\put(201,560){\makebox(0,0)[r]{$0.1$}}
\put(859.0,560.0){\rule[-0.200pt]{4.818pt}{0.400pt}}
\put(221.0,123.0){\rule[-0.200pt]{0.400pt}{4.818pt}}
\put(221,82){\makebox(0,0){$0$}}
\put(221.0,540.0){\rule[-0.200pt]{0.400pt}{4.818pt}}
\put(550.0,123.0){\rule[-0.200pt]{0.400pt}{4.818pt}}
\put(550,82){\makebox(0,0){$1$}}
\put(550.0,540.0){\rule[-0.200pt]{0.400pt}{4.818pt}}
\put(879.0,123.0){\rule[-0.200pt]{0.400pt}{4.818pt}}
\put(879,82){\makebox(0,0){$2$}}
\put(879.0,540.0){\rule[-0.200pt]{0.400pt}{4.818pt}}
\put(221.0,123.0){\rule[-0.200pt]{158.512pt}{0.400pt}}
\put(879.0,123.0){\rule[-0.200pt]{0.400pt}{105.273pt}}
\put(221.0,560.0){\rule[-0.200pt]{158.512pt}{0.400pt}}
\put(40,461){\makebox(0,0){$\langle x | \e^{-\tilde{H}T}| x \rangle$}}
\put(550,21){\makebox(0,0){$x$}}
\put(221.0,123.0){\rule[-0.200pt]{0.400pt}{105.273pt}}
\put(695,473){\makebox(0,0)[r]{path integral}}
\put(221,467){\circle*{18}}
\put(287,450){\circle*{18}}
\put(353,419){\circle*{18}}
\put(418,360){\circle*{18}}
\put(484,306){\circle*{18}}
\put(550,249){\circle*{18}}
\put(616,206){\circle*{18}}
\put(682,173){\circle*{18}}
\put(747,150){\circle*{18}}
\put(813,137){\circle*{18}}
\put(765,473){\circle*{18}}
\sbox{\plotpoint}{\rule[-0.500pt]{1.000pt}{1.000pt}}%
\put(695,432){\makebox(0,0)[r]{exact}}
\multiput(715,432)(20.756,0.000){5}{\usebox{\plotpoint}}
\put(815,432){\usebox{\plotpoint}}
\put(221,457){\usebox{\plotpoint}}
\put(221.00,457.00){\usebox{\plotpoint}}
\put(241.60,454.91){\usebox{\plotpoint}}
\put(262.09,451.69){\usebox{\plotpoint}}
\put(281.90,445.55){\usebox{\plotpoint}}
\put(301.13,437.91){\usebox{\plotpoint}}
\put(319.28,427.98){\usebox{\plotpoint}}
\put(337.26,417.67){\usebox{\plotpoint}}
\put(354.30,405.83){\usebox{\plotpoint}}
\put(371.25,393.96){\usebox{\plotpoint}}
\put(387.74,381.37){\usebox{\plotpoint}}
\put(403.56,367.94){\usebox{\plotpoint}}
\put(419.39,354.61){\usebox{\plotpoint}}
\put(435.47,341.53){\usebox{\plotpoint}}
\put(450.89,327.66){\usebox{\plotpoint}}
\put(466.72,314.24){\usebox{\plotpoint}}
\put(482.18,300.44){\usebox{\plotpoint}}
\put(497.97,287.03){\usebox{\plotpoint}}
\put(513.65,273.53){\usebox{\plotpoint}}
\put(530.08,260.92){\usebox{\plotpoint}}
\put(546.53,248.34){\usebox{\plotpoint}}
\put(563.26,236.13){\usebox{\plotpoint}}
\put(580.47,224.66){\usebox{\plotpoint}}
\put(597.80,213.26){\usebox{\plotpoint}}
\put(615.55,202.54){\usebox{\plotpoint}}
\put(633.70,192.60){\usebox{\plotpoint}}
\put(651.90,182.63){\usebox{\plotpoint}}
\put(670.96,174.58){\usebox{\plotpoint}}
\put(690.15,166.81){\usebox{\plotpoint}}
\put(709.70,159.94){\usebox{\plotpoint}}
\put(729.58,153.98){\usebox{\plotpoint}}
\put(749.45,148.01){\usebox{\plotpoint}}
\put(769.67,143.48){\usebox{\plotpoint}}
\put(789.98,139.43){\usebox{\plotpoint}}
\put(810.51,136.36){\usebox{\plotpoint}}
\put(831.09,134.00){\usebox{\plotpoint}}
\put(851.62,131.06){\usebox{\plotpoint}}
\put(872.30,129.96){\usebox{\plotpoint}}
\put(879,129){\usebox{\plotpoint}}
\end{picture}
\end{center}
\caption{Euclidean harmonic-oscillator propagator at large time ($T=4$)
computed exactly 
(dotted line), and computed using numerical integration to evaluate the
discretized path integral (points). The path integral was approximated
by an 8~dimensional integral which was evaluated numerically, using 
\protect\verb|vegas|, at the points indicated. The exact result is
approximated by the square of the ground state wavefunction 
multiplied by $\exp(-E_0 T)$.}
\label{fig-vegashosc}
\end{figure}
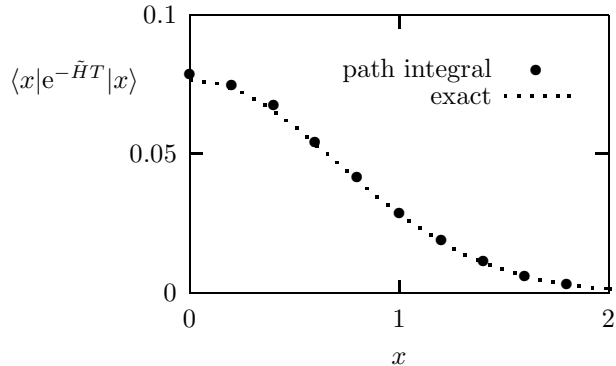

\begin{exercise} 
Set $a=1/2$ and $N=8$ in approximate formula~(\ref{1d-approxprop}) for
the propagator, and integrate the right-hand side numerically. The
seven dimensional integral that results is easily evaluated using
standard routines, such as \verb|vegas|\,\cite{vegas}. 
Do this first for the one-dimensional harmonic-oscillator potential
\be \label{onedhosc}
V(x) = \frac{x^2}{2} \quad\quad\mbox{with}\quad\quad m = 1.
\ee
Evaluate the propagator for several values of $x_0=x_N=x$,
and compare your results with those of standard quantum mechanics:
\be
\bra{x}\e^{-\Ht T}\ket{x} \approx \left| \braket{x}{E_0}\right|^2
\e^{-E_0 T}
\ee
where $E_0 = 1/2$ and
\be
\braket{x}{E_0} = \frac{e^{-x^2/2}}{\pi^{1/4}}.
\ee
Extract the energy and wavefunction from your numerical result.
Repeat this exercise for $V(x)=x^4/2$. 
(If you wish, you may restrict $x$ integrations to
the region $-5\to5$ rather than $-\infty\to\infty$; this has
negligible effect on the results of this exercise).

My results for the harmonic oscillator case are shown in
Fig.~\ref{fig-vegashosc}. 
\end{exercise}

\subsection{Monte Carlo Evaluation of Path Integrals}
Our analysis in the previous section focused on the groundstate. 
In quantum field theory, where the groundstate is the vacuum, we are
generally interested in excited states. To
analyze excited states using path integrals, we interrupt the
propagation of the groundstate by introducing new operators
at intermediate times. Consider, for example, the quantity
\be
\expval{x(t_2)x(t_1)} \equiv
\frac{\int\Dxt\,x(t_2)x(t_1)\,\e^{-S[x]}}{\int\Dxt\,\e^{-S[x]}}
\ee
where now we integrate over all $\xxi=\xxf=x$ as well as
the intermediate~$x(t)$'s. This quantity is a weighted average of
$x(t_2)x(t_1)$ over all paths, with weight $\exp(-S[x])$.
The numerator on the right-hand side equals, in quantum mechanics,
\be
\int dx\,
\bra{x}\e^{-\Ht(\tf-t_2)}\,\xt\,\e^{-\Ht(t_2-t_1)}\,\xt\,
\e^{-\Ht(t_1-\ti)}\ket{x}.
\ee
Setting $T=\tf-\ti$ and $t=t_2-t_1$ we can rewrite the full expression
as
\be \label{onedprop}
\expval{x(t_2)x(t_1)} = 
\frac{\sum \e^{-E_n T}\bra{E_n}\xt\,\e^{-(\Ht-E_n)t}\,\xt\ket{E_n}}%
{\sum \e^{-E_n T}}.
\ee
If $T\gg t$ and large, then the groundstate $\ket{E_0}$ dominates the
sums and
\be
G(t)\equiv\expval{x(t_2)x(t_1)} \to \bra{E_0}\xt\,\e^{-(\Ht-E_0)t}\,
\xt\ket{E_0}.
\ee
In our harmonic oscillator example, the state propagating between the
two~$\xt$'s cannot be $\ket{E_0}$ since $\xt$~switches the parity of
the state. Thus if we now make $t$~large (but still $\ll T$)
\be
G(t)\,\,\stackrel{{t\,\,{\rm large}}}{\longrightarrow}\,\,
\left|\bra{E_0}\xt\ket{E_1}\right|^2\, \e^{-(E_1-E_0)t}
\ee
where $\ket{E_1}$ is the first excited state.
Consequently we can extract the first excitation energy from the large-$t$
dependence of $G(t)$,
\be
\log(G(t)/G(t+a)) \to (E_1-E_0)a,
\ee
and then, going back to $G(t)$, we can determine the quantum
mechanical transition matrix element $\bra{E_0}\xt\ket{E_1}$.

In principle, path integral averages $\expval{\Gamma[x]}$ of
arbitrary functionals~$\Gamma[x]$ can be used to compute any physical
property of the excited states in the quantum theory. Also we note in
passing that 
\be
\expval{\Gamma[x]} = 
\frac{\sum\e^{-E_n T}\bra{E_n}\Gamma[\xt]\ket{E_n}}{\sum\e^{-E_n T}}
\ee
becomes a (quantum mechanical) thermal average if we replace
\be
T\to\beta\equiv 1/k_B\,T_{\mathrm temp}. 
\ee
Thus any computer code designed to compute path integral averages can be used
for thermal physics as well. Here we focus on the zero-temperature
limit of large~$T$.

We could evaluate the path integrals in $\expval{\Gamma[x]}$ using a
standard multidimensional integration code like~\verb|vegas|, at least
for one-dimensional systems. Here, instead, we employ a more generally
useful Monte Carlo procedure. Noting that
\be
\expval{\Gamma[x]} = 
\frac{\int\Dxt\,\Gamma[x]\,\e^{-S[x]}}{\int\Dxt\,\e^{-S[x]}},
\ee
is a weighted average over paths with weight~$\exp(-S[x])$, we
generate a large number, $\Nconf$, of random paths or
configurations, 
\be
\xalpha \equiv \{x_0^{(\alpha)}x_1^{(\alpha)}\ldots
x_{N-1}^{(\alpha)}\} \quad\quad\alpha=1,2\ldots \Nconf,
\ee
on our grid in such a way that the probability~$P[\xalpha]$ for
obtaining any particular path~$\xalpha$ is
\be \label{path-probab}
P[\xalpha] \propto \e^{-S[\xalpha]}.
\ee
Then an unweighted average of $\Gamma[x]$ over this set of  paths
approximates the weighted average over uniformly distributed paths:
\be
\expval{\Gamma[x]} \approx \Gammab \equiv
\frac{1}{\Nconf}\sum_{\alpha=1}^\Nconf \Gamma[\xalpha].
\ee
$\Gammab$ is our ``Monte Carlo estimator'' for~$\expval{\Gamma[x]}$ on
our lattice. Of course the estimate will never be exact since the
number of paths~$\Nconf$ will never be infinite. The Monte Carlo
uncertainty~$\sigma_\Gammab$ in our estimate is a potential source of
error; it is estimated in the usual fashion\cite{tasi-alg}:
\be \label{sigma-estimator}
\sigma_\Gammab^2 \approx \frac{1}{\Nconf}\left\{
\frac{1}{\Nconf}\sum_{\alpha=1}^\Nconf \Gamma^2[\xalpha] 
- \Gammab^2 \right\}.
\ee
This becomes
\be
\sigma_\Gammab^2 = \frac{\expval{\Gamma^2} -
\expval{\Gamma}^2}{\Nconf}
\ee
for large $\Nconf$. Since the numerator in this expression is
independent of~$\Nconf$ (in principle, it can be determined directly
from quantum mechanics), the statistical uncertainties vanish as
$1/\sqrt{\Nconf}$ when $\Nconf$ increases.

We need some sort of specialized random-vector generator to create our set
of random paths~$\xalpha$ with
probability~(\ref{path-probab}). Possibly the simplest procedure,
though not always the best, is the Metropolis
Algorithm\,\cite{metropolis}.
In this procedure, we start with an arbitrary path~$x^{(0)}$ and
modify it by visiting each of the sites on the
lattice, and randomizing the~$x_j$'s at those sites, one at a time, in
a particular fashion that is described below. In this way we generate 
a new random path from the old one: $x^{(0)}\to x^{(1)}$. This is
called ``updating'' the path.
Applying the algorithm to $x^{(1)}$ we generate path $x^{(2)}$, and so
on until we have $\Nconf$ random paths. This set of random paths
has the correct distribution if $\Nconf$~is sufficiently large.

\begin{figure}
\begin{verbatim}
def update(x):
    for j in range(0,N):
        old_x = x[j]                    # save original value
        old_Sj = S(j,x)                     
        x[j] = x[j] + uniform(-eps,eps) # update x[j]
        dS = S(j,x) - old_Sj            # change in action
        if dS>0 and exp(-dS)<uniform(0,1):
            x[j] = old_x                # restore old value

def S(j,x):                             # harm. osc. S
    jp = (j+1)%N                        # next site 
    jm = (j-1)%N                        # previous site 
    return a*x[j]**2/2 + x[j]*(x[j]-x[jp]-x[jm])/a
\end{verbatim}
\caption{Python code for one Metropolis update of
path~$\protect\{x_j,\,j=0\ldots N-1\protect\}$. The 
path is stored in array \protect\verb|x[j]|. 
Function \protect\verb|S(j,x)| returns
the value of the part of the action that depends on~$x_j$. Function
\protect\verb|uniform(a,b)| returns a random number between $a$
and~$b$. A sample \protect\verb|S(j,x)| is shown, for a harmonic
oscillator with~$x_N=x_0$.}
\label{py-update}
\end{figure}

The algorithm for randomizing $x_j$ at the $j^{\mathrm th}$~site is:
\begin{itemize}
\item generate a random number $\zeta$, with probability uniformly
distributed between $-\epsilon$ and $\epsilon$ for some
constant~$\epsilon$; 

\item replace $x_j\to x_j+\zeta$ and compute the change $\Delta S$ in
the action caused by this replacement (generally only a few terms in
the lattice action involve $x_j$, since lagrangians are local;
only these need be examined);

\item if $\Delta S<0$ (the action is reduced) retain the new value
for~$x_j$, and  proceed to the next site; 

\item if $\Delta S>0$ generate a random number $\eta$ unformly
distributed between~0 and~1; retain the new value for~$x_j$ 
if $\exp(-\Delta S)>\eta$, otherwise restore the old value;
proceed to the next site.  
\end{itemize}
An implementation of this algorithm, in the Python computer
language\,\cite{python}, is shown in Fig.~\ref{py-update}. The code
examples and Python are discussed in the Appendix.

There are two important details concerning the tuning and use of this
algorithm.  First, in general some or many of the $x_j$'s will be the
same in two successive random paths. The amount of such overlap is
determined by the parameter~$\epsilon$: when $\epsilon$~is very large,
changes in the $x_j$'s are usually large and most will be rejected;
when $\epsilon$ is very small, changes are small and most are
accepted, but the new~$x_j$'s will be almost equal to the old
ones. Neither extreme is desirable since each leads to very small
changes in $x$, thereby slowing down the numerical 
exploration of the space of all
important paths. Typically $\epsilon$~should be
tuned so that 40\%--60\% of the $x_j$'s are changed on each pass (or
``sweep'') through the lattice. Then $\epsilon$~is of order the
typical quantum fluctuations expected in the theory. Whatever
the~$\epsilon$, however, successive paths are going to be quite
similar (that is ``highly correlated'') and so contain rather similar
information about the theory. Thus when we accumulate
random paths~$\xalpha$ for our Monte Carlo estimates we should keep
only every $\Ncorr$-th path; the intervening sweeps
erase correlations, giving us configurations that are
statistically independent. The optimal value for $\Ncorr$ depends upon
the theory, and can be found by experimentation.  It also depends on
the lattice spacing~$a$, going roughly as
\be
\Ncorr \propto \frac{1}{a^2}.
\ee
Other algorithms exist for which $\Ncorr$ grows only as $1/a$ when
$a$~is reduced, but since our interest is in large~$a$'s we will not
discuss these further.

The second detail concerns the procedure for starting the
algorithm. The very first configuration used to seed the whole process
is usually fairly atypical. Consequently we should
discard some number of configurations at the beginning, before
starting to collect~$\xalpha$'s. Discarding $5\Ncorr$ to $10\Ncorr$
configurations is usually adequate. This is called ``thermalizing the
lattice.'' 

To summarize, a computer code for a complete Monte Carlo calculation of
$\expval{\Gamma[x]}$ for some function~$\Gamma[x]$ of a path~$x$
consists of the following steps:
\begin{itemize}
\item initialize the path, for example, by setting all $x_j$'s to
zero;

\item update the path $5\Ncorr$--$10\Ncorr$ times to thermalize it;

\item update the path $\Ncorr$ times, then compute $\Gamma[x]$ and save it; 
repeat $\Nconf$~times.

\item average the $\Nconf$ values of~$\Gamma[x]$ saved in the previous
step to obtain a Monte Carlo estimate~$\Gammab$
for~$\expval{\Gamma[x]}$.
\end{itemize}
A Python implementation of this procedure is shown in
Fig.~\ref{py-MC}.

\begin{figure}
\begin{verbatim}
def compute_G(x,n):
    g = 0
    for j in range(0,N):
        g = g + x[j]*x[(j+n)%N]
    return g/N

def MCaverage(x,G):
    for j in range(0,N):                # initialize x
        x[j] = 0
    for j in range(0,5*N_cor):          # thermalize x
        update(x)
    for alpha in range(0,N_cf):         # loop on random paths
        for j in range(0,N_cor):
            update(x)
        for n in range(0,N):              
            G[alpha][n] = compute_G(x,n)
    for n in range(0,N):                # compute MC averages
        avg_G = 0
        for alpha in range(0,N_cf):
            avg_G = avg_G + G[alpha][n]
        avg_G = avg_G/N_cf
        print "G(%d) = %g" % (n,avg_G)
\end{verbatim}
\caption{Sample Python code for a Monte Carlo evaluation of 
of $G(t)$ (\protect\eq{hosc-Gt}). Function \protect\verb|computeG(x,t)|
computes $G(t)$ for a given path~\protect\verb|x|. Function
\protect\verb|MCaverage(x,G)| computes the Monte Carlo average over
random paths~\protect\verb|x|. The results for path~$\xalpha$ are
stored in the array~\protect\verb|G[alpha][t]|, and the averages are computed
and printed. Function~\protect\verb|update(x)| does one Metropolis
sweep through the lattice (see Fig.~\protect\ref{py-update}).
}
\label{py-MC}
\end{figure}

\begin{exercise}
Write a computer program to implement the Metropolis Monte Carlo
algorithm for a one dimensional harmonic oscillator (\eq{onedhosc}),
and calculate 
\be \label{hosc-Gt}
G(t) = \frac{1}{N} \sum_j \expval{x(t_j+t)x(t_j)}
\ee
for all $t=0,a,2a\ldots(N-1)a$; that is calculate
\be
G_n = \frac{1}{N} \sum_j \expval{x_{(j+n){\mathrm mod} N}\,x_j}
\ee
for $n=0\ldots N-1$. The $(j+n){\mathrm mod}N$ in this last expression
reflects the periodic boundary conditions.
Try $N=20$ lattice sites with lattice spacing
$a=1/2$, and set $\epsilon=1.4$ and $\Ncorr=20$. Try $\Nconf$'s of 25,
100, 1000 and~10000. Use the results to
compute the excitation energy from
\be \label{deltaEn}
\Delta E_n \equiv
\log(G_n/G_{n+1}) \stackrel{n~{\mathrm large}}{\longrightarrow} (E_1-E_0) a
\ee
Try this for the harmonic oscillator potential and also for
anharmonic potentials. Vary the various parameters.

My results for the harmonic oscillator potential with $N=1000$
configurations are shown in Fig.~\ref{hoscMC}. These results required
less than a minute of personal computer time.
\begin{figure}
\begin{center}
% GNUPLOT: LaTeX picture
\setlength{\unitlength}{0.240900pt}
\ifx\plotpoint\undefined\newsavebox{\plotpoint}\fi
\sbox{\plotpoint}{\rule[-0.200pt]{0.400pt}{0.400pt}}%
\begin{picture}(900,600)(0,0)
\font\gnuplot=cmr10 at 10pt
\gnuplot
\sbox{\plotpoint}{\rule[-0.200pt]{0.400pt}{0.400pt}}%
\put(162.0,298.0){\rule[-0.200pt]{4.818pt}{0.400pt}}
\put(142,298){\makebox(0,0)[r]{1}}
\put(859.0,298.0){\rule[-0.200pt]{4.818pt}{0.400pt}}
\put(162.0,473.0){\rule[-0.200pt]{4.818pt}{0.400pt}}
\put(142,473){\makebox(0,0)[r]{2}}
\put(859.0,473.0){\rule[-0.200pt]{4.818pt}{0.400pt}}
\put(367.0,123.0){\rule[-0.200pt]{0.400pt}{4.818pt}}
\put(367,82){\makebox(0,0){1}}
\put(367.0,540.0){\rule[-0.200pt]{0.400pt}{4.818pt}}
\put(572.0,123.0){\rule[-0.200pt]{0.400pt}{4.818pt}}
\put(572,82){\makebox(0,0){2}}
\put(572.0,540.0){\rule[-0.200pt]{0.400pt}{4.818pt}}
\put(777.0,123.0){\rule[-0.200pt]{0.400pt}{4.818pt}}
\put(777,82){\makebox(0,0){3}}
\put(777.0,540.0){\rule[-0.200pt]{0.400pt}{4.818pt}}
\put(162.0,123.0){\rule[-0.200pt]{172.725pt}{0.400pt}}
\put(879.0,123.0){\rule[-0.200pt]{0.400pt}{105.273pt}}
\put(162.0,560.0){\rule[-0.200pt]{172.725pt}{0.400pt}}
\put(40,361){\makebox(0,0){$\Delta E(t)$}}
\put(520,21){\makebox(0,0){$t$}}
\put(162.0,123.0){\rule[-0.200pt]{0.400pt}{105.273pt}}
\put(264.0,294.0){\rule[-0.200pt]{0.400pt}{1.445pt}}
\put(254.0,294.0){\rule[-0.200pt]{4.818pt}{0.400pt}}
\put(254.0,300.0){\rule[-0.200pt]{4.818pt}{0.400pt}}
\put(367.0,293.0){\rule[-0.200pt]{0.400pt}{1.686pt}}
\put(357.0,293.0){\rule[-0.200pt]{4.818pt}{0.400pt}}
\put(357.0,300.0){\rule[-0.200pt]{4.818pt}{0.400pt}}
\put(469.0,290.0){\rule[-0.200pt]{0.400pt}{3.854pt}}
\put(459.0,290.0){\rule[-0.200pt]{4.818pt}{0.400pt}}
\put(459.0,306.0){\rule[-0.200pt]{4.818pt}{0.400pt}}
\put(572.0,302.0){\rule[-0.200pt]{0.400pt}{4.336pt}}
\put(562.0,302.0){\rule[-0.200pt]{4.818pt}{0.400pt}}
\put(562.0,320.0){\rule[-0.200pt]{4.818pt}{0.400pt}}
\put(674.0,250.0){\rule[-0.200pt]{0.400pt}{12.768pt}}
\put(664.0,250.0){\rule[-0.200pt]{4.818pt}{0.400pt}}
\put(664.0,303.0){\rule[-0.200pt]{4.818pt}{0.400pt}}
\put(777.0,304.0){\rule[-0.200pt]{0.400pt}{25.535pt}}
\put(767.0,304.0){\rule[-0.200pt]{4.818pt}{0.400pt}}
\put(264,297){\circle*{12}}
\put(367,296){\circle*{12}}
\put(469,298){\circle*{12}}
\put(572,311){\circle*{12}}
\put(674,276){\circle*{12}}
\put(777,357){\circle*{12}}
\put(767.0,410.0){\rule[-0.200pt]{4.818pt}{0.400pt}}
\put(162,298){\usebox{\plotpoint}}
\put(162.0,298.0){\rule[-0.200pt]{172.725pt}{0.400pt}}
\end{picture}
\end{center}
\caption{Monte Carlo values $\Delta E(t)\equiv \log(G(t)/G(t+a))/a$
plotted versus~$t$ for an harmonic oscillator. The exact asymptotic
result, $\Delta E(\infty) = 1$, is indicated by a line. Results are for a one
dimensional lattice with $N=20$ sites, lattice spacing $a=1/2$,
and $\Nconf=1000$ configurations, keeping configurations only every
$\Ncorr=20$ sweeps. The Metropolis step size $\epsilon$ was 1.4,
resulting in a Metropolis acceptance ratio of 0.5.}
\label{hoscMC}
\end{figure}
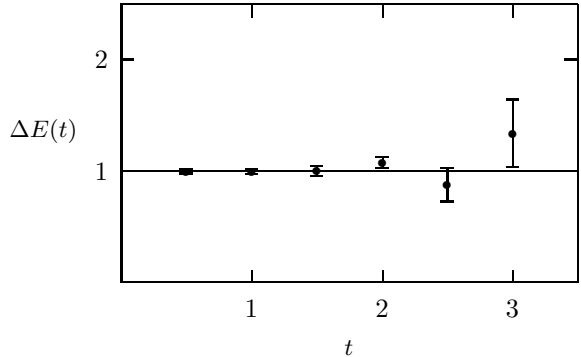
\end{exercise}

\begin{exercise} Redo the previous exercise but propagator
\be 
G(t) = \frac{1}{N} \sum_j \expval{x^3(t_j+t)x^3(t_j)}.
\ee
Here we use $x^3$ rather than $x$ to create
and destroy the excited stated; that is, we use $x^3$ rather than $x$
as the ``source'' and the ``sink''. 
Note that $\Delta E(t)$ converges to the same result, but only at much
larger $t$'s than before. Different sources and
sinks often lead to different asymptotic behavior. Choices that result in
fast convergence as $t$ increases are usually preferable because
statistical errors are smaller at smaller~$t$'s. Compare your best
estimate of the asymptotic value obtained using $x^3$ with that
obtained using~$x$.

Note also that $\Delta E(t)$ approaches its
its asymptotic value from above. Prove that this must be true in
general, provided source and sink are the same operator, using \eq{onedprop}.
This result is useful because it implies that each $\Delta E(t)$ gives
a rigorous upper bound on the asymptotic value, even
at small~$t$'s before convergence. 

My results with an $x^3$ source and sink are shown in
Fig.~\ref{hosc3MC}.
\begin{figure}
\begin{center}
% GNUPLOT: LaTeX picture
\setlength{\unitlength}{0.240900pt}
\ifx\plotpoint\undefined\newsavebox{\plotpoint}\fi
\sbox{\plotpoint}{\rule[-0.200pt]{0.400pt}{0.400pt}}%
\begin{picture}(900,600)(0,0)
\font\gnuplot=cmr10 at 10pt
\gnuplot
\sbox{\plotpoint}{\rule[-0.200pt]{0.400pt}{0.400pt}}%
\put(162.0,298.0){\rule[-0.200pt]{4.818pt}{0.400pt}}
\put(142,298){\makebox(0,0)[r]{1}}
\put(859.0,298.0){\rule[-0.200pt]{4.818pt}{0.400pt}}
\put(162.0,473.0){\rule[-0.200pt]{4.818pt}{0.400pt}}
\put(142,473){\makebox(0,0)[r]{2}}
\put(859.0,473.0){\rule[-0.200pt]{4.818pt}{0.400pt}}
\put(367.0,123.0){\rule[-0.200pt]{0.400pt}{4.818pt}}
\put(367,82){\makebox(0,0){1}}
\put(367.0,540.0){\rule[-0.200pt]{0.400pt}{4.818pt}}
\put(572.0,123.0){\rule[-0.200pt]{0.400pt}{4.818pt}}
\put(572,82){\makebox(0,0){2}}
\put(572.0,540.0){\rule[-0.200pt]{0.400pt}{4.818pt}}
\put(777.0,123.0){\rule[-0.200pt]{0.400pt}{4.818pt}}
\put(777,82){\makebox(0,0){3}}
\put(777.0,540.0){\rule[-0.200pt]{0.400pt}{4.818pt}}
\put(162.0,123.0){\rule[-0.200pt]{172.725pt}{0.400pt}}
\put(879.0,123.0){\rule[-0.200pt]{0.400pt}{105.273pt}}
\put(162.0,560.0){\rule[-0.200pt]{172.725pt}{0.400pt}}
\put(40,361){\makebox(0,0){$\Delta E(t)$}}
\put(520,21){\makebox(0,0){$t$}}
\put(162.0,123.0){\rule[-0.200pt]{0.400pt}{105.273pt}}
\put(264.0,385.0){\rule[-0.200pt]{0.400pt}{5.541pt}}
\put(254.0,385.0){\rule[-0.200pt]{4.818pt}{0.400pt}}
\put(254.0,408.0){\rule[-0.200pt]{4.818pt}{0.400pt}}
\put(367.0,320.0){\rule[-0.200pt]{0.400pt}{6.022pt}}
\put(357.0,320.0){\rule[-0.200pt]{4.818pt}{0.400pt}}
\put(357.0,345.0){\rule[-0.200pt]{4.818pt}{0.400pt}}
\put(469.0,286.0){\rule[-0.200pt]{0.400pt}{3.854pt}}
\put(459.0,286.0){\rule[-0.200pt]{4.818pt}{0.400pt}}
\put(459.0,302.0){\rule[-0.200pt]{4.818pt}{0.400pt}}
\put(572.0,301.0){\rule[-0.200pt]{0.400pt}{10.600pt}}
\put(562.0,301.0){\rule[-0.200pt]{4.818pt}{0.400pt}}
\put(562.0,345.0){\rule[-0.200pt]{4.818pt}{0.400pt}}
\put(674.0,281.0){\rule[-0.200pt]{0.400pt}{22.404pt}}
\put(664.0,281.0){\rule[-0.200pt]{4.818pt}{0.400pt}}
\put(664.0,374.0){\rule[-0.200pt]{4.818pt}{0.400pt}}
\put(777.0,193.0){\rule[-0.200pt]{0.400pt}{32.762pt}}
\put(767.0,193.0){\rule[-0.200pt]{4.818pt}{0.400pt}}
\put(264,396){\circle*{12}}
\put(367,333){\circle*{12}}
\put(469,294){\circle*{12}}
\put(572,323){\circle*{12}}
\put(674,328){\circle*{12}}
\put(777,261){\circle*{12}}
\put(767.0,329.0){\rule[-0.200pt]{4.818pt}{0.400pt}}
\put(162,298){\usebox{\plotpoint}}
\put(162.0,298.0){\rule[-0.200pt]{172.725pt}{0.400pt}}
\end{picture}
\end{center}
\caption{Monte Carlo values $\Delta E(t)\equiv \log(G(t)/G(t+a))/a$
plotted versus~$t$ for an harmonic oscillator using $x^3$ as the
source and sink. Parameters are the same as used to generate 
Fig.~\ref{hoscMC}. The energies take longer to reach their asymptotic value.}
\label{hosc3MC}
\end{figure}
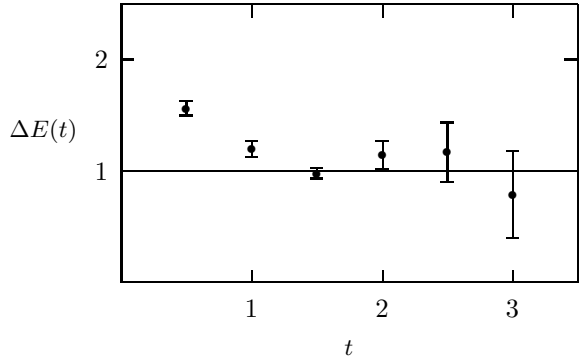
\end{exercise}

\subsection{Statistical Errors}
A Monte Carlo estimate $\Gammab$ of some expectation
value~$\expval\Gammab$ is never exact; there are always statistical
errors that vanish only in the limit where infinitely many
configurations are employed ($\Nconf\to\infty$).
An important part of any Monte Carlo analysis is the estimation of these
statistical errors. There is a simple but very powerful method,
called the ``statistical bootstrap,''  for making such
estimates.

In the previous exercises, for example, we assemble an ``ensemble'' of
measurements of the propagator~$\Galpha$, one for each
configuration~$\xalpha$. These are averaged to
obtain~$\overline{G}$, and, from it, an estimate 
for $\Delta E_n$ (\eq{deltaEn}).  
An obvious way to check the statistical errors on
this estimate for $\Delta E_n$ is to redo the whole calculation, say,
100~times, each time with different random numbers to generate
different random paths. With 100~copies of the entire calculation, we
could analyze the distribution of the 100~random $\Delta E_n$'s
obtained, and deduce the statistical uncertainty in our original
estimate. This, however, is exceedingly expensive in computer
time. The bootstrap procedure provides new, almost {\em
zero-cost} random ensembles of measurements by synthesizing them
from the original ensemble of $\Nconf$~measurements.

\begin{figure}
\begin{verbatim}
def bootstrap(G):
    N_cf = len(G)
    G_bootstrap = []                    # new ensemble
    for i in range(0,N_cf):
        alpha = int(uniform(0,N_cf))    # choose random config
        G_bootstrap.append(G[alpha])    # keep G[alpha]
    return G_bootstrap
\end{verbatim}
\caption{Sample Python code for producing a bootstrap copy of an
ensemble of measurements~\protect\verb|G|.
The original ensemble consists of individual
measurements~\protect\verb|G[alpha]|, one for each configuration.
The function~\protect\verb|bootstrap(G)| returns a single bootstrap copy of
ensemble~\protect\verb|G|, consisting of
\protect\verb|N\_cf|~measurements.
Function~\protect\verb|uniform(a,b)| returns a random number
between~\protect\verb|a| and~\protect\verb|b|.
}
\label{py-bootstrap}
\end{figure}

Given an ensemble~$\{\Galpha, \alpha=1\ldots \Nconf\}$ of
Monte Carlo measurements, we assemble a
``bootstrap copy'' of that ensemble by selecting $\Galpha$'s
at random from the original ensemble, taking $\Nconf$ in all while
allowing duplications and omissions. The resulting ensemble of
$G$'s might have two or three copies of
some~$\Galpha$'s, and no copies of others. This new ensemble can be
averaged and a new estimate obtained for~$\Delta E_n$.
This procedure can be repeated to generated as many
bootstrap copies of the original ensemble as we wish, and from each we
can generate a new estimate for~$\Delta E_n$. The distribution of these
$\Delta E_n$'s approximates the distribution of $\Delta E_n$'s that
would have been obtained from
the original Monte Carlo, and so can be used to estimate the statistical
error in our original estimate. A Python implementation of this
procedure is shown in Fig.~\ref{py-bootstrap}.

Another useful procedure related to statistical errors is ``binning.''
At the end of a large simulation we might have 100's or even 100,000's
of configurations~$\xalpha$, and for each a set of measurements
like~$G^{(\alpha)}$, our propagator. The measurements will
inevitably be averaged, but we want to save the separate
$G^{(\alpha)}$'s for making bootstrap error estimates and the like.  We can
save a lot of disk space, RAM, and CPU time by partially averaging or
binning the measurements: For example, instead of storing each of 
\be
G^{(1)}\quad G^{(2)}\quad G^{(3)}\quad G^{(4)} \quad G^{(5)}\dots
\ee
we might instead store
\bearray
\overline{G}^{(1)}& \equiv & \frac{G^{(1)}+G^{(2)}+G^{(3)}+G^{(4)}}{4}
\nl
\overline{G}^{(2)}& \equiv & \frac{G^{(5)}+G^{(6)}+G^{(7)}+G^{(8)}}{4}
\nl
\ldots &&
\eearray
The $\overline{G}^{(\beta)}$'s are far less numerous but have the same
average, standard deviation, and other statistical properties as the
original set. Typically the bin size is adjusted so that there are
only 50--100~$\overline{G}^{(\beta)}$'s. A Python implementation of
this procedure is shown in Fig.~\ref{py-bin}.

\begin{figure}
\begin{verbatim}
def bin(G,binsize):
    G_binned = []                       # binned ensemble
    for i in range(0,len(G),binsize):   # loop on bins
        G_avg = 0
        for j in range(0,binsize):      # loop on bin elements
            G_avg = G_avg + G[i+j]
        G_binned.append(G_avg/binsize)  # keep bin avg 
    return G_binned
\end{verbatim}
\caption{Sample Python code for producing a binned copy of an
ensemble of measurements~\protect\verb|G|.
The original ensemble consists of individual
measurements~\protect\verb|G[alpha]|, one for each configuration.
The function~\protect\verb|bin(G,binsize)| bins the ensemble into bins
of size~\protect\verb|binsize|, averages the~\protect\verb|G|'s within
each bin, and returns an ensemble consisting of the averages.
}
\label{py-bin}
\end{figure}

Binning has an important side effect: it reduces or can even remove
the effects of correlations between different configurations. If, when
generating configurations, $\Ncorr$ is too small, successive Monte
Carlo estimates are statistically correlated. This leads to error
estimates, using \eq{sigma-estimator}, that can be much smaller than
the true errors\,---\,a very bad situation. If, however, the Monte
Carlo estimates are binned with sufficiently large bins, the majority
of estimates in one bin will be uncorrelated from the majority in
adjacent bins. Consequently the bin averages will be uncorrelated, and
standard statistical formulas, like \eq{sigma-estimator}, are
reliable. To determine the bin size required to
remove correlations, first bin the measurements and compute the
errors. Then rebin with double the bin size and recompute the
errors. There are no correlations if the statistical errors are
roughly independent of bin size; if, on the other hand, the
statistical errors grow substantially with bin size (for example,
proportional to the square root of the bin size), then there are
strong correlations between bins. Continuing doubling the bin size
until the statistical errors stop growing. Note that measurements of
different physical quantities decorrelate at different rates;
different things may require different bin sizes.

\begin{exercise}
Rerun your Metropolis simulation of the harmonic oscillator with
$\Ncorr=1$. Do several different runs and compare your results. Do
they agree within statistical errors? Try binning the results from
each of the runs in bins of~20 and recompute the statistical
errors. Verify that different runs now agree within the errors
computed from the binned results.

In Fig.~\ref{hosccorr} I show results from $N=1000$ configurations
using $\Ncorr=1$. Compare this with results in Fig.~\ref{hoscMC},
which come from the same number of configurations but with
$\Ncorr=20$. The errorbars in the $\Ncorr=1$ plot are obviously
unreliable. 
\end{exercise}
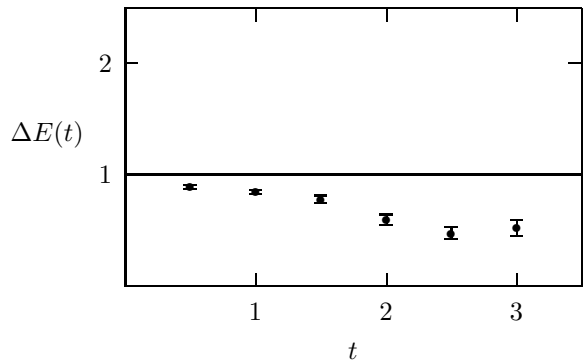
\begin{figure}
\begin{center}
% GNUPLOT: LaTeX picture
\setlength{\unitlength}{0.240900pt}
\ifx\plotpoint\undefined\newsavebox{\plotpoint}\fi
\sbox{\plotpoint}{\rule[-0.200pt]{0.400pt}{0.400pt}}%
\begin{picture}(900,600)(0,0)
\font\gnuplot=cmr10 at 10pt
\gnuplot
\sbox{\plotpoint}{\rule[-0.200pt]{0.400pt}{0.400pt}}%
\put(162.0,298.0){\rule[-0.200pt]{4.818pt}{0.400pt}}
\put(142,298){\makebox(0,0)[r]{1}}
\put(859.0,298.0){\rule[-0.200pt]{4.818pt}{0.400pt}}
\put(162.0,473.0){\rule[-0.200pt]{4.818pt}{0.400pt}}
\put(142,473){\makebox(0,0)[r]{2}}
\put(859.0,473.0){\rule[-0.200pt]{4.818pt}{0.400pt}}
\put(367.0,123.0){\rule[-0.200pt]{0.400pt}{4.818pt}}
\put(367,82){\makebox(0,0){1}}
\put(367.0,540.0){\rule[-0.200pt]{0.400pt}{4.818pt}}
\put(572.0,123.0){\rule[-0.200pt]{0.400pt}{4.818pt}}
\put(572,82){\makebox(0,0){2}}
\put(572.0,540.0){\rule[-0.200pt]{0.400pt}{4.818pt}}
\put(777.0,123.0){\rule[-0.200pt]{0.400pt}{4.818pt}}
\put(777,82){\makebox(0,0){3}}
\put(777.0,540.0){\rule[-0.200pt]{0.400pt}{4.818pt}}
\put(162.0,123.0){\rule[-0.200pt]{172.725pt}{0.400pt}}
\put(879.0,123.0){\rule[-0.200pt]{0.400pt}{105.273pt}}
\put(162.0,560.0){\rule[-0.200pt]{172.725pt}{0.400pt}}
\put(40,361){\makebox(0,0){$\Delta E(t)$}}
\put(520,21){\makebox(0,0){$t$}}
\put(162.0,123.0){\rule[-0.200pt]{0.400pt}{105.273pt}}
\put(264.0,275.0){\rule[-0.200pt]{0.400pt}{1.445pt}}
\put(254.0,275.0){\rule[-0.200pt]{4.818pt}{0.400pt}}
\put(254.0,281.0){\rule[-0.200pt]{4.818pt}{0.400pt}}
\put(367.0,267.0){\rule[-0.200pt]{0.400pt}{1.686pt}}
\put(357.0,267.0){\rule[-0.200pt]{4.818pt}{0.400pt}}
\put(357.0,274.0){\rule[-0.200pt]{4.818pt}{0.400pt}}
\put(469.0,253.0){\rule[-0.200pt]{0.400pt}{2.891pt}}
\put(459.0,253.0){\rule[-0.200pt]{4.818pt}{0.400pt}}
\put(459.0,265.0){\rule[-0.200pt]{4.818pt}{0.400pt}}
\put(572.0,219.0){\rule[-0.200pt]{0.400pt}{3.854pt}}
\put(562.0,219.0){\rule[-0.200pt]{4.818pt}{0.400pt}}
\put(562.0,235.0){\rule[-0.200pt]{4.818pt}{0.400pt}}
\put(674.0,196.0){\rule[-0.200pt]{0.400pt}{4.577pt}}
\put(664.0,196.0){\rule[-0.200pt]{4.818pt}{0.400pt}}
\put(664.0,215.0){\rule[-0.200pt]{4.818pt}{0.400pt}}
\put(777.0,201.0){\rule[-0.200pt]{0.400pt}{6.022pt}}
\put(767.0,201.0){\rule[-0.200pt]{4.818pt}{0.400pt}}
\put(264,278){\circle*{12}}
\put(367,271){\circle*{12}}
\put(469,259){\circle*{12}}
\put(572,227){\circle*{12}}
\put(674,205){\circle*{12}}
\put(777,214){\circle*{12}}
\put(767.0,226.0){\rule[-0.200pt]{4.818pt}{0.400pt}}
\put(162,298){\usebox{\plotpoint}}
\put(162.0,298.0){\rule[-0.200pt]{172.725pt}{0.400pt}}
\end{picture}
\end{center}
\caption{Monte Carlo values $\Delta E(t)\equiv \log(G(t)/G(t+a))/a$
plotted versus~$t$ for an harmonic oscillator, as in
Fig.~\ref{hoscMC} but with $\Ncorr=1$. The errorbars are unreliable.}
\label{hosccorr}
\end{figure}

\section{Field Theory on a Lattice}

\subsection{From Quantum Mechanics to Field Theory}
Field theories of the sort we are interested in have lagrangian
formulations and so can be quantized immediately using path
integrals. The procedure is precisely analogous to what we do in the
previous section when quantizing the harmonic oscillator. The analogues
of the coordinates $x(t)$ in quantum mechanics are just the fields
$\phi(x)$ or $A_\mu(x)$ where $x=(t,\vec{x})$ is a space-time
point. Indeed our quantum mechanical examples can be thought of as
field theory examples in 0~spatial and 1~temporal dimension:
$x(t)\to\phi(t)\to\phi(x)$. The analogue of the ground state in
quantum field theory is the vacuum state, $\ket{0}$, while the
analogues of the excited states, created when $\phi(x)$ or $\phi^3$
or\,\ldots\,acts on~$\ket{0}$, correspond to states with one or more
particles create in the vacuum.

In the lattice approximation both space and time are discrete:
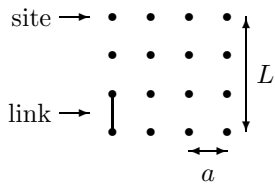
\begin{center}
\setlength{\unitlength}{.02in}
\begin{picture}(75,75)(0,0)
\multiput(25,25)(10,0){4}{\circle*{2}}
\multiput(25,35)(10,0){4}{\circle*{2}}
\multiput(25,45)(10,0){4}{\circle*{2}}
\multiput(25,55)(10,0){4}{\circle*{2}}

\put(60,40){\vector(0,1){15}}
\put(60,40){\vector(0,-1){15}}
\put(65,40){\makebox(0,0){$L$}}

\put(50,20){\vector(1,0){5}}
\put(50,20){\vector(-1,0){5}}
\put(50,15){\makebox(0,0)[t]{$a$}}

\put(11,55){\vector(1,0){8}}
\put(9,55){\makebox(0,0)[r]{site}}

\put(11,30){\vector(1,0){8}}
\put(9,30){\makebox(0,0)[r]{link}}
\put(25,25){\line(0,1){10}}

\end{picture}
\end{center}
The nodes or ``sites'' of the grid are separated by
lattice spacing~$a$, and the length of a side of the grid is~$L$; the
lines joining adjacent sites are called ``links.'' The quantum field
is specified by its values at the grid sites: a configuration is
describe by the set of numbers
$\{\phi(x_j),\,\,\forall  \,x_j\,\epsilon\,\,\mbox{grid}\}$.
The path integral generalizes in the obvious fashion:
\be
\expval{\Gamma[\phi]} \equiv
\frac{1}{{\cal Z}}\, \int 
\e^{-S[\phi]} \, \Gamma[\phi] 
\prod_{x_j\epsilon\,{\mathrm grid}}\!\!\!d\phi(x_j)
\ee
where
\be
{\cal Z} \equiv \int \e^{-S[\phi]} \,
\prod_{x_j} d\phi(x_j).
\ee
Here the action $S[\phi]$ is the continuum action with spatial and
temporal derivatives replaced by differences between field values at
the grid sites.
We study excitations of the field theory using operators like
\be
\Gamma(t) \equiv \frac{1}{\sqrt{N}} \sum_{\vec{x}_j} \phi(\vec{x}_j,t)
\ee
where the sum over the $N$ 
spatial~$\vec{x}_j$'s enforces zero three-momentum. 
Since the excitations correspond to particle
creation, their energies are the energies of particles: for example,
\be
\expval{\Gamma(t)\Gamma(0)} \,\stackrel{t\,{\mathrm
large}}{\longrightarrow}\,
|\bra{0}\Gamma(0)\ket{\phi:\,\vec{p}=0}|^2\,\e^{-m_\phi t},
\ee
where $\ket{\phi:\,\vec{p}=0}$ is a one $\phi$-particle state with
zero three momentum, and $m_\phi$ is the mass of the $\phi$~particle.

\begin{exercise}
Show that
\be
\bra{0}\Gamma(0)\ket{\phi:\,\vec{p}=0} = \frac{Z_2}{2m_\phi}
\ee
where $Z_2$ is the wavefunction renormalization parameter for the
$\phi$ field.
\end{exercise}

\subsection{Coarse Lattices}
We have, through the lattice approximation to the path integral,
turned the problem of solving a nonperturbative relativistic quantum field
theory, once again, into a problem of numerical integration. This is a major
development for theories, like QCD, where perturbation theory doesn't
suffice (at low energies).

Early enthusiasm for such an approach to QCD, back when QCD was first
invented, quickly gave
way to the grim realization that very large computers would be needed
to numerically integrate the path integral. In recent years,
however, two developments have made QCD simulations far more
accessible. One is that small computers have become much faster; the
other is that QCD simulations have become much faster\,---\,$10^3$
to~$10^6$ times faster. These developments imply that the simplest QCD
simulations can be done using no more than a single personal computer
or even a laptop.

What has changed to make QCD simulations faster? The cost of a QCD
calculation is given roughly by the formula 
\be
\mbox{cost} \approx \left(\frac{L}{a}\right)^4\,\frac{1}{a}\,
\frac{1}{m_\pi^2 a},
\ee
where the first factor is  the number of lattice sites, while the
second and third factors are due to ``critical-slowing-down'' of the
algorithms used for the simulation. From this formula, the single most
important determinant of the cost is the lattice spacing: the cost is
proportional to $1/a^6$. This means that one wants to keep $a$ as
large as possible. Until recently it was thought that
$a<0.05$--0.1\,fm would be essential for accurate QCD simulations. As
we shall see $a\approx0.3$--0.4\,fm works quite well. Given that the
cost varies as $1/a^6$, the coarser lattices should be $10^3$
to~$10^6$ times cheaper to simulate.

The size of the lattice spacing is limited by discretization errors.
The challenge is to make the lattice spacing as large as possible
while keeping the discretization errors of order, say, a few percent
or less. These errors have two sources: first, the lattice forces us
to use approximate derivatives, and, second, it imposes an ultraviolet
cutoff. We consider each in turn.

In the lattice approximation, we only know the fields at the lattice
sites. Thus all derivatives in field equations, the action, and the like
must be converted to differences. For example, the second derivative of field
$\phi$ at some point~$x_j$ on the lattice is given approximately by
\be
\frac{\partial^2\phi(x_j)}{\partial x^2} = \Delta_x^{(2)}\,\phi(x_j)
+\order(a^2)
\ee
where
\be
\Delta_x^{(2)}\,\phi(x) \equiv \frac{\phi(x+a)-2\phi(x)+\phi(x-a)}{a^2}.
\ee
We generally want more accurate
approximations for work on coarse lattices: for example, the
approximation 
\be \label{improvedlapl}
\frac{\partial^2\phi(x_j)}{\partial x^2} = \Delta_x^{(2)}\,\phi(x_j)
-\frac{a^2}{12}\,(\Delta_x^{(2)})^2\,\phi(x_j)
+\order(a^4)
\ee
is accurate to within a few percent even when acting on structures in
$\phi(x)$ that are only four or five lattice spacings across. With
such precision one might expect that lattice spacings as large
as a quarter the diameter of hadron, or about 0.4\,fm, would still be
quite useful. Our theories are quantum theories, however, and
therefore there is a second important consideration.

The shortest wavelength oscillation that can be modeled on a lattice
is one with wavelength~$\lambda_{\rm min} = 2a$; for example, the
function $\phi(x)=+1,-1,+1\ldots$ for $x=0,a,2a\ldots$ oscillates with
this wavelength. Thus gluons and quarks with momenta
$p\!=\!2\pi/\lambda$ larger than~$\pi/a$ are excluded from the lattice
theory by the lattice; that is, the lattice functions as an
ultraviolet cutoff. In simple classical field theories this is often
irrelevant: short-wavelength ultraviolet modes are either unexcited or
decouple from the long-wavelength infrared modes of interest. However,
in a noisy nonlinear theory, like an interacting quantum field theory,
ultraviolet modes strongly affect infrared modes.
Thus we cannot simply discard all particles
with momenta larger than~$\pi/a$; we must somehow mimic their effects
on infrared states. This is done by changing or ``renormalizing'' the
parameters in our discretized theory and by adding new
local interactions.

The new interactions complicate the improved discretizations
discussed above. For example, an interacting scalar theory on the
lattice would have a discretized kinetic lagrangian
\be
\sum_\mu \half \phi^\dagger \partial^2_\mu \phi \to
\sum_\mu
\half \left(\phi^\dagger \Delta^{(2)}_\mu \phi  +   a^2 c\,
\phi^\dagger (\Delta^{(2)}_\mu)^2\phi\right)
\ee
where parameter $c$ has two parts: $-1/12$ from numerical analysis
(\eq{improvedlapl}), and an additional renormalization due to the
cutoff. Typically the renormalization is completely context
dependent\,---\,for example, it is different for QED and QCD, or for
particles of different spin, and so on. It cannot be looked up in a
numerical analysis book; rather, it must be computed using quantum
field theory. In QCD these renormalizations can be computed using
(weak-coupling) 
perturbation theory, since the renormalizations are due to QCD physics
at large momenta, $p>\pi/a$, where the theory is perturbative:
\be
c = -\frac{1}{12} + c_1\alpha_s(\pi/a) + c_2\alpha_s^2(\pi/a)+\cdots.
\ee
This is
true, that is, provided the lattice spacing is small enough that
momentum $\pi/a$ is perturbative. Work in continuum QCD suggests that
lattice spacings of 0.5\,fm or smaller should suffice, but, until
recently, lattice simulations seemed to suggest that perturbation
theory only started to work for lattice spacings smaller than
0.05--0.1\,fm. 

\begin{exercise}
Our action for one-dimensional quantum mechanics, \eq{onedqm}, can be
rewritten 
\be
S[x] 
\equiv \int_{\ti}^{\tf} dt\, \left[ -\half m\,x(t)\ddot{x}(t) +
V(x(t)) \right],
\ee
by integrating by parts (assuming $x(\ti)=x(\tf)=x$). To discretize
we replace 
\be
\ddot{x}(t_j) \to \Delta^{(2)} x_j 
\equiv \frac{x_{j+1}-2x_j+x_{j-1}}{a^2},
\ee
where the $x_j$'s are periodic ($x_0=x_N$, $x_{-1}=x_{N-1}$, and so on);
this gives the same lattice action we used earlier. We can improve the
discretization, however, by using the corrected approximation,
\eq{improvedlapl}, for the second derivative:
\be
S_{\mathrm imp}[x] \equiv  \sum_{j=0}^{N-1} a \left[ 
-\half m\, x_j \left(\Delta^{(2)}-a^2(\Delta^{(2)})^2/12\right)
x_j + a V(x_j) \right]
\ee
Modify your Monte Carlo code for the harmonic oscillator to include
the correction term. Run high-statistics comparisons ($\Nconf = 10^4$
or~$10^5$) with and without the correction term.
\end{exercise}

\begin{exercise}
Discretized, euclidean classical equations of motion can be
derived from the actions in the previous exercise by setting
\be
\frac{\partial S[x]}{\partial x_j} = 0.
\ee
Using the improved action, for example, we obtain
\be
m \left(\Delta^{(2)}-a^2(\Delta^{(2)})^2/12\right)
x_j  =   \frac{d V(x_j)}{d x_j} .
\ee
Setting 
\be
V(x) = \half m\,\omega^2_0 x^2,
\ee
 find solutions of the form $x_j =
\exp(-\omega t_j)$, the euclidean-time version of an oscillatory
solution with frequency $\omega$. Show that the frequency
is given by
\be
\omega^2 =  \omega_0^2\left[ 1 - \frac{(a\omega_0)^2}{12}+
\order((a\omega)^4) \right]
\ee
for the unimproved action. The $(a\omega_0)^2$ correction is the error
caused by the finite lattice spacing. 

Repeat the exercise for the
improved action and show that it has two solutions. One,
\be
\omega^2 = \omega_0^2\left[ 1 + \frac{(a\omega_0)^4}{90} +\order((a\omega)^6)
\right],
\ee
is an improved version of the previous result; its errors are fourth
order in $a\omega_0$ rather than second order. The other solution,
however, is
\be
\omega^2 \approx \left(\frac{2.6}{a}\right)^2.
\ee
It corresponds to a new oscillation mode that does not appear in the
continuum; it is an artifact of the improved lattice theory. This new
mode is sometimes called a ``numerical ghost.'' In a quantum field
theory it would be a new, very massive particle ($m\propto 1/a$).

Our lattice theory was designed to be accurate for low-energies, and
so we should not be surprised when unphysical modes appear at high
energies. These
ghost modes, being high-energy, typically decouple from low-energy
physics and so can usually be ignored. However, they can have one
unfortunate effect on the numerical analysis. Returning to the
previous exercise, note
that the $\Delta E_n$'s for the improved action are below the
asymptotic result when $n$~is small (see Fig.~\ref{hoscimp}), 
in contradiction of the general result discussed in
Section~2. The general result ignored the possibility that spurious
states might be induced by the numerical analysis. Here these states
have negative norm (impossible for real quantum states), which is why
the energies rise from below. This is incovenient because it means
that the $\Delta E_n$'s cannot be used to rigorously bound the true
answer\,---\,they may be either above or below it\,---\,unlike the
case for the unimproved action, where they must always be above.
\begin{figure}
\begin{center}
% GNUPLOT: LaTeX picture
\setlength{\unitlength}{0.240900pt}
\ifx\plotpoint\undefined\newsavebox{\plotpoint}\fi
\sbox{\plotpoint}{\rule[-0.200pt]{0.400pt}{0.400pt}}%
\begin{picture}(900,600)(0,0)
\font\gnuplot=cmr10 at 10pt
\gnuplot
\sbox{\plotpoint}{\rule[-0.200pt]{0.400pt}{0.400pt}}%
\put(162.0,298.0){\rule[-0.200pt]{4.818pt}{0.400pt}}
\put(142,298){\makebox(0,0)[r]{1}}
\put(859.0,298.0){\rule[-0.200pt]{4.818pt}{0.400pt}}
\put(162.0,473.0){\rule[-0.200pt]{4.818pt}{0.400pt}}
\put(142,473){\makebox(0,0)[r]{2}}
\put(859.0,473.0){\rule[-0.200pt]{4.818pt}{0.400pt}}
\put(367.0,123.0){\rule[-0.200pt]{0.400pt}{4.818pt}}
\put(367,82){\makebox(0,0){1}}
\put(367.0,540.0){\rule[-0.200pt]{0.400pt}{4.818pt}}
\put(572.0,123.0){\rule[-0.200pt]{0.400pt}{4.818pt}}
\put(572,82){\makebox(0,0){2}}
\put(572.0,540.0){\rule[-0.200pt]{0.400pt}{4.818pt}}
\put(777.0,123.0){\rule[-0.200pt]{0.400pt}{4.818pt}}
\put(777,82){\makebox(0,0){3}}
\put(777.0,540.0){\rule[-0.200pt]{0.400pt}{4.818pt}}
\put(162.0,123.0){\rule[-0.200pt]{172.725pt}{0.400pt}}
\put(879.0,123.0){\rule[-0.200pt]{0.400pt}{105.273pt}}
\put(162.0,560.0){\rule[-0.200pt]{172.725pt}{0.400pt}}
\put(40,361){\makebox(0,0){$\Delta E(t)$}}
\put(520,21){\makebox(0,0){$t$}}
\put(162.0,123.0){\rule[-0.200pt]{0.400pt}{105.273pt}}
\put(264.0,269.0){\rule[-0.200pt]{0.400pt}{1.204pt}}
\put(254.0,269.0){\rule[-0.200pt]{4.818pt}{0.400pt}}
\put(254.0,274.0){\rule[-0.200pt]{4.818pt}{0.400pt}}
\put(367.0,288.0){\rule[-0.200pt]{0.400pt}{2.409pt}}
\put(357.0,288.0){\rule[-0.200pt]{4.818pt}{0.400pt}}
\put(357.0,298.0){\rule[-0.200pt]{4.818pt}{0.400pt}}
\put(469.0,286.0){\rule[-0.200pt]{0.400pt}{3.854pt}}
\put(459.0,286.0){\rule[-0.200pt]{4.818pt}{0.400pt}}
\put(459.0,302.0){\rule[-0.200pt]{4.818pt}{0.400pt}}
\put(572.0,284.0){\rule[-0.200pt]{0.400pt}{6.504pt}}
\put(562.0,284.0){\rule[-0.200pt]{4.818pt}{0.400pt}}
\put(562.0,311.0){\rule[-0.200pt]{4.818pt}{0.400pt}}
\put(674.0,261.0){\rule[-0.200pt]{0.400pt}{16.381pt}}
\put(664.0,261.0){\rule[-0.200pt]{4.818pt}{0.400pt}}
\put(664.0,329.0){\rule[-0.200pt]{4.818pt}{0.400pt}}
\put(777.0,224.0){\rule[-0.200pt]{0.400pt}{56.130pt}}
\put(767.0,224.0){\rule[-0.200pt]{4.818pt}{0.400pt}}
\put(264,271){\circle*{12}}
\put(367,293){\circle*{12}}
\put(469,294){\circle*{12}}
\put(572,298){\circle*{12}}
\put(674,295){\circle*{12}}
\put(777,340){\circle*{12}}
\put(767.0,457.0){\rule[-0.200pt]{4.818pt}{0.400pt}}
\put(162,298){\usebox{\plotpoint}}
\put(162.0,298.0){\rule[-0.200pt]{172.725pt}{0.400pt}}
\end{picture}
\end{center}
\caption{Monte Carlo values $\Delta E(t)\equiv \log(G(t)/G(t+a))/a$
plotted versus~$t$ for an harmonic oscillator, as in
Fig.~\ref{hoscMC} but with an improved action. The energies approach
their asymptotic value from below.}
\label{hoscimp}
\end{figure}
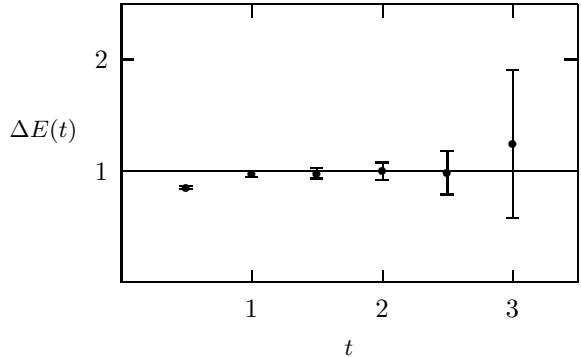

Ghost modes always arise when improved discretizations are used for
temporal derivatives. There is a trick, however, for correcting
temporal difference operators that avoids extra states. This is to
change integration variables in the path integral: for the harmonic
oscillator we replace
$x_j \to \xt_j$ where
\be
x_j = \xt_j + \delta\xt_j
\ee
and
\be
\delta\xt_j \equiv \xi_1 \,a^2\Delta^{(2)}\xt_j +
\xi_2\,a^2\omega_0^2\xt_j.
\ee
Substituting this into the action we obtain
\bearray
S[x] &=& S[\xt + \delta\xt] \nl
&=& S[\xt] + \sum_j \delta\xt_j\, \frac{\partial
S[\xt]}{\partial \xt_j} + \order(a^4)\nl
&\equiv&\St[\xt]
\eearray
Find  values for the $\xi_i$ such that the improved harmonic-oscillator
action, in terms of 
$\xt$, is 
\be
\St_{\mathrm imp}[\xt] =  \sum_{j=0}^{N-1} a \left[ 
-\half m\,\xt_j \Delta^{(2)}\xt_j
+ \Vt_{\mathrm imp}(x_j) \right]
\ee
where
\be
\Vt_{\mathrm imp}(\xt) \equiv \half m\omega^2_0 \xt_j^2
\left(1 + \frac{(a\omega_0)^2}{12}\right) 
\ee
and all $\order(a^4)$ terms are ignored. This action has no
$a^2$~errors, but also has no ghosts. The value of the path integral
is not changed by a simple change of variables (provided that the
jacobian is included\,---\,show that in this case it has no effect on
expectation values). Rerun your numerical tests from
the last exercise using this action.
\end{exercise}

\begin{exercise}
The field transformation trick in the previous exercise
is particularly simple for the harmonic
oscillator. Generalize the trick for the case of an anharmonic
oscillator with, for example,
\be
V(x) = \half m\,\omega_0^2 x^2 (1 + c m \omega_0 x^2).
\ee
where $c$ is a dimensionless parameter. Try a variable change with
\be
\delta\xt_j \equiv \xi_1\, a^2\Delta^{(2)}\xt_j +
\xi_2\,a^2\omega_0^2\xt_j + \xi_3\, a^2 m \omega^3_0\xt_j^3 .
\ee
The resulting action is as above but with a new $\Vt_\imp$:
\bearray
\Vt_\imp(\xt) &=&
\frac{m\omega_0^2}{2}\,\xt^2(1+cm\omega_0 \xt^2)
+ \frac{a^2m\omega_0^4}{24}\,\left(\xt+2cm\omega_0\xt^3\right)^2\nl
&-& a \delta v(\xt) + \frac{a^3}{2}\, \delta v(\xt)^2
\eearray
where the terms involving 
\be
\delta v(\xt)\equiv  c m\omega_0^3 \xt^2/4
\ee
are due to the jacobian (from the change of integration variables;
the jacobian matters here because the change is nonlinear). Test this
improved lattice action against the original unimproved action using
$m=\omega_0=1$ and $c=2$. (The asymptotic value for $\Delta E$ is
1.933 with these parameters.) 

One might expect errors of order $a^4$ with the improved
action. However renormalization effects
arise when interactions are anharmonic. In particular the coefficients
of the $x^2$ and $x^4$ 
interactions are renormalized away from their naive values. Such
effects enter at the same order as the corrections from the
jacobian. In our one-dimensional theory, unlike in QCD, these
corrections vanish like powers of $a$ when $a\to 0$. The leading
such correction is an $\order(a)$ shift in the coefficient of the
$x^2$ potential. Run high-precision simulations at $a=1/2$ and $a=1/4$
to compute the $\order(a)$ error due to renormalization. Try adjusting
the coefficient of $x^2$ to remove this error. (Alternatively one could
try using perturbation theory to compute the shift needed to eliminate
the error.)
\end{exercise}

\subsection{Perturbation Theory and Tadpole Improvement\protect\footnotemark}
\footnotetext{This  section is based upon work with Paul Mackenzie
that is described in~\cite{gpl93}.}
Improved discretizations and large lattice spacings are old ideas,
pioneered by Wilson, Symanzik and others\,\cite{wilson83}. 
However, perturbation theory is
essential; the lattice spacing~$a$ must be small enough so that
$p\!\approx\!\pi/a$~QCD is perturbative. This was the requirement that 
drove lattice QCD towards very costly simulations with tiny lattice
spacings. Traditional perturbation theory for
lattice QCD begins to fail at distances of order $1/20$
to $1/10$\,fm, and therefore lattice spacings must be at least this small
before improved actions are useful. This seems very odd since
phenomenological applications of continuum perturbative QCD suggest that
perturbation theory works well down to energies of order 1\,GeV, which
corresponds to a lattice spacing of~0.6\,fm. The breakthrough, in the early
1990's, was the discovery of a trivial modification of lattice QCD, called
``tadpole improvement,'' that allows perturbation theory to work even at
distances as large as $1/2$\,fm\,\cite{gpl93,gpl94,tsukuba}.

One can readily derive Feynman diagram rules for lattice QCD using the
same techniques as in the continuum, but applied to the lattice
lagrangian\,\cite{kawai81}. 
The particle propagators and interaction vertices are usually
complicated functions of the momenta that become identical to their
continuum analogues in the low-momentum limit. All loop momenta are cut
off at $p_\mu=\pm\pi/a$. 

Testing perturbation theory is also straightforward. One designs 
short-distance quantities that can be computed easily in a simulation (i.e.,
in a Monte Carlo evaluation of the lattice path integral). The Monte Carlo
gives the exact value which can then be compared with the perturbative
expansion for the same quantity. An example of such a quantity is the
expectation value of the Wilson loop operator,
\be
W({\cal C})\equiv\langle 0|\third\Re\Tr\Porder\e^{-\I g\oint_{\cal C}A\cdot
\dd x}|0\rangle,
\ee
where $A$ is the QCD vector potential, $\Porder$~denotes path ordering, and
${\cal C}$ is any small, closed path or loop on the lattice. $W({\cal
C})$~is perturbative for sufficiently small loops~$\cal C$. 
We can test the utility of perturbation theory over any range of distances 
by varying the loop size while comparing
numerical Monte Carlo results for~$W({\cal C})$ with perturbation theory.

\begin{figure}
\begin{center} 
% GNUPLOT: LaTeX picture
\setlength{\unitlength}{0.240900pt}
\ifx\plotpoint\undefined\newsavebox{\plotpoint}\fi
\sbox{\plotpoint}{\rule[-0.200pt]{0.400pt}{0.400pt}}%
\begin{picture}(1200,900)(0,0)
\font\gnuplot=cmr10 at 10pt
\gnuplot
\sbox{\plotpoint}{\rule[-0.200pt]{0.400pt}{0.400pt}}%
\put(220.0,113.0){\rule[-0.200pt]{220.664pt}{0.400pt}}
\put(220.0,401.0){\rule[-0.200pt]{4.818pt}{0.400pt}}
\put(198,401){\makebox(0,0)[r]{0.2}}
\put(1116.0,401.0){\rule[-0.200pt]{4.818pt}{0.400pt}}
\put(220.0,688.0){\rule[-0.200pt]{4.818pt}{0.400pt}}
\put(198,688){\makebox(0,0)[r]{0.4}}
\put(1116.0,688.0){\rule[-0.200pt]{4.818pt}{0.400pt}}
\put(449.0,113.0){\rule[-0.200pt]{0.400pt}{4.818pt}}
\put(449,68){\makebox(0,0){0.001}}
\put(449.0,812.0){\rule[-0.200pt]{0.400pt}{4.818pt}}
\put(678.0,113.0){\rule[-0.200pt]{0.400pt}{4.818pt}}
\put(678,68){\makebox(0,0){0.01}}
\put(678.0,812.0){\rule[-0.200pt]{0.400pt}{4.818pt}}
\put(907.0,113.0){\rule[-0.200pt]{0.400pt}{4.818pt}}
\put(907,68){\makebox(0,0){0.1}}
\put(907.0,812.0){\rule[-0.200pt]{0.400pt}{4.818pt}}
\put(220.0,113.0){\rule[-0.200pt]{220.664pt}{0.400pt}}
\put(1136.0,113.0){\rule[-0.200pt]{0.400pt}{173.207pt}}
\put(220.0,832.0){\rule[-0.200pt]{220.664pt}{0.400pt}}
\put(45,472){\makebox(0,0){$\chi_{22}$}}
%\put(678,-22){\makebox(0,0){loop size (fm)}}
\put(678,877){\makebox(0,0){$\chi_{22}$ Loop Ratio\,---\,$1^{\rm st}$ Order}}
\put(220.0,113.0){\rule[-0.200pt]{0.400pt}{173.207pt}}
\put(518,688){\makebox(0,0)[r]{exact}}
\put(562,688){\circle{24}}
\put(1029,650){\circle{24}}
\put(975,495){\circle{24}}
\put(946,448){\circle{24}}
\put(605,259){\circle{24}}
\put(254,208){\circle{24}}
\put(518,643){\makebox(0,0)[r]{new P.Th.}}
\put(540.0,643.0){\rule[-0.200pt]{15.899pt}{0.400pt}}
\put(254,209){\usebox{\plotpoint}}
\put(254,209.17){\rule{1.700pt}{0.400pt}}
\multiput(254.00,208.17)(4.472,2.000){2}{\rule{0.850pt}{0.400pt}}
\put(262,210.67){\rule{1.927pt}{0.400pt}}
\multiput(262.00,210.17)(4.000,1.000){2}{\rule{0.964pt}{0.400pt}}
\put(270,211.67){\rule{1.927pt}{0.400pt}}
\multiput(270.00,211.17)(4.000,1.000){2}{\rule{0.964pt}{0.400pt}}
\put(278,212.67){\rule{1.686pt}{0.400pt}}
\multiput(278.00,212.17)(3.500,1.000){2}{\rule{0.843pt}{0.400pt}}
\put(285,213.67){\rule{1.927pt}{0.400pt}}
\multiput(285.00,213.17)(4.000,1.000){2}{\rule{0.964pt}{0.400pt}}
\put(293,215.17){\rule{1.700pt}{0.400pt}}
\multiput(293.00,214.17)(4.472,2.000){2}{\rule{0.850pt}{0.400pt}}
\put(301,216.67){\rule{1.927pt}{0.400pt}}
\multiput(301.00,216.17)(4.000,1.000){2}{\rule{0.964pt}{0.400pt}}
\put(309,217.67){\rule{1.927pt}{0.400pt}}
\multiput(309.00,217.17)(4.000,1.000){2}{\rule{0.964pt}{0.400pt}}
\put(317,218.67){\rule{1.686pt}{0.400pt}}
\multiput(317.00,218.17)(3.500,1.000){2}{\rule{0.843pt}{0.400pt}}
\put(324,219.67){\rule{1.927pt}{0.400pt}}
\multiput(324.00,219.17)(4.000,1.000){2}{\rule{0.964pt}{0.400pt}}
\put(332,221.17){\rule{1.700pt}{0.400pt}}
\multiput(332.00,220.17)(4.472,2.000){2}{\rule{0.850pt}{0.400pt}}
\put(340,222.67){\rule{1.927pt}{0.400pt}}
\multiput(340.00,222.17)(4.000,1.000){2}{\rule{0.964pt}{0.400pt}}
\put(348,223.67){\rule{1.927pt}{0.400pt}}
\multiput(348.00,223.17)(4.000,1.000){2}{\rule{0.964pt}{0.400pt}}
\put(356,224.67){\rule{1.686pt}{0.400pt}}
\multiput(356.00,224.17)(3.500,1.000){2}{\rule{0.843pt}{0.400pt}}
\put(363,226.17){\rule{1.700pt}{0.400pt}}
\multiput(363.00,225.17)(4.472,2.000){2}{\rule{0.850pt}{0.400pt}}
\put(371,227.67){\rule{1.927pt}{0.400pt}}
\multiput(371.00,227.17)(4.000,1.000){2}{\rule{0.964pt}{0.400pt}}
\put(379,228.67){\rule{1.927pt}{0.400pt}}
\multiput(379.00,228.17)(4.000,1.000){2}{\rule{0.964pt}{0.400pt}}
\put(387,229.67){\rule{1.927pt}{0.400pt}}
\multiput(387.00,229.17)(4.000,1.000){2}{\rule{0.964pt}{0.400pt}}
\put(395,230.67){\rule{1.686pt}{0.400pt}}
\multiput(395.00,230.17)(3.500,1.000){2}{\rule{0.843pt}{0.400pt}}
\put(402,232.17){\rule{1.700pt}{0.400pt}}
\multiput(402.00,231.17)(4.472,2.000){2}{\rule{0.850pt}{0.400pt}}
\put(410,233.67){\rule{1.927pt}{0.400pt}}
\multiput(410.00,233.17)(4.000,1.000){2}{\rule{0.964pt}{0.400pt}}
\put(418,234.67){\rule{1.927pt}{0.400pt}}
\multiput(418.00,234.17)(4.000,1.000){2}{\rule{0.964pt}{0.400pt}}
\put(426,235.67){\rule{1.927pt}{0.400pt}}
\multiput(426.00,235.17)(4.000,1.000){2}{\rule{0.964pt}{0.400pt}}
\put(434,236.67){\rule{1.686pt}{0.400pt}}
\multiput(434.00,236.17)(3.500,1.000){2}{\rule{0.843pt}{0.400pt}}
\put(441,238.17){\rule{1.700pt}{0.400pt}}
\multiput(441.00,237.17)(4.472,2.000){2}{\rule{0.850pt}{0.400pt}}
\put(449,239.67){\rule{1.927pt}{0.400pt}}
\multiput(449.00,239.17)(4.000,1.000){2}{\rule{0.964pt}{0.400pt}}
\put(457,240.67){\rule{1.927pt}{0.400pt}}
\multiput(457.00,240.17)(4.000,1.000){2}{\rule{0.964pt}{0.400pt}}
\put(465,241.67){\rule{1.686pt}{0.400pt}}
\multiput(465.00,241.17)(3.500,1.000){2}{\rule{0.843pt}{0.400pt}}
\put(472,242.67){\rule{1.927pt}{0.400pt}}
\multiput(472.00,242.17)(4.000,1.000){2}{\rule{0.964pt}{0.400pt}}
\put(480,244.17){\rule{1.700pt}{0.400pt}}
\multiput(480.00,243.17)(4.472,2.000){2}{\rule{0.850pt}{0.400pt}}
\put(488,245.67){\rule{1.927pt}{0.400pt}}
\multiput(488.00,245.17)(4.000,1.000){2}{\rule{0.964pt}{0.400pt}}
\put(496,246.67){\rule{1.927pt}{0.400pt}}
\multiput(496.00,246.17)(4.000,1.000){2}{\rule{0.964pt}{0.400pt}}
\put(504,247.67){\rule{1.686pt}{0.400pt}}
\multiput(504.00,247.17)(3.500,1.000){2}{\rule{0.843pt}{0.400pt}}
\put(511,249.17){\rule{1.700pt}{0.400pt}}
\multiput(511.00,248.17)(4.472,2.000){2}{\rule{0.850pt}{0.400pt}}
\put(519,250.67){\rule{1.927pt}{0.400pt}}
\multiput(519.00,250.17)(4.000,1.000){2}{\rule{0.964pt}{0.400pt}}
\put(527,251.67){\rule{1.927pt}{0.400pt}}
\multiput(527.00,251.17)(4.000,1.000){2}{\rule{0.964pt}{0.400pt}}
\put(535,252.67){\rule{1.927pt}{0.400pt}}
\multiput(535.00,252.17)(4.000,1.000){2}{\rule{0.964pt}{0.400pt}}
\put(543,253.67){\rule{1.686pt}{0.400pt}}
\multiput(543.00,253.17)(3.500,1.000){2}{\rule{0.843pt}{0.400pt}}
\put(550,255.17){\rule{1.700pt}{0.400pt}}
\multiput(550.00,254.17)(4.472,2.000){2}{\rule{0.850pt}{0.400pt}}
\put(558,256.67){\rule{1.927pt}{0.400pt}}
\multiput(558.00,256.17)(4.000,1.000){2}{\rule{0.964pt}{0.400pt}}
\put(566,257.67){\rule{1.927pt}{0.400pt}}
\multiput(566.00,257.17)(4.000,1.000){2}{\rule{0.964pt}{0.400pt}}
\put(574,258.67){\rule{1.927pt}{0.400pt}}
\multiput(574.00,258.17)(4.000,1.000){2}{\rule{0.964pt}{0.400pt}}
\put(582,259.67){\rule{1.686pt}{0.400pt}}
\multiput(582.00,259.17)(3.500,1.000){2}{\rule{0.843pt}{0.400pt}}
\put(589,261.17){\rule{1.700pt}{0.400pt}}
\multiput(589.00,260.17)(4.472,2.000){2}{\rule{0.850pt}{0.400pt}}
\put(597,262.67){\rule{1.927pt}{0.400pt}}
\multiput(597.00,262.17)(4.000,1.000){2}{\rule{0.964pt}{0.400pt}}
\put(605,263.67){\rule{1.927pt}{0.400pt}}
\multiput(605.00,263.17)(4.000,1.000){2}{\rule{0.964pt}{0.400pt}}
\put(613,264.67){\rule{1.686pt}{0.400pt}}
\multiput(613.00,264.17)(3.500,1.000){2}{\rule{0.843pt}{0.400pt}}
\put(620,266.17){\rule{1.700pt}{0.400pt}}
\multiput(620.00,265.17)(4.472,2.000){2}{\rule{0.850pt}{0.400pt}}
\put(628,267.67){\rule{1.927pt}{0.400pt}}
\multiput(628.00,267.17)(4.000,1.000){2}{\rule{0.964pt}{0.400pt}}
\put(636,268.67){\rule{1.927pt}{0.400pt}}
\multiput(636.00,268.17)(4.000,1.000){2}{\rule{0.964pt}{0.400pt}}
\put(644,269.67){\rule{1.686pt}{0.400pt}}
\multiput(644.00,269.17)(3.500,1.000){2}{\rule{0.843pt}{0.400pt}}
\put(651,271.17){\rule{1.700pt}{0.400pt}}
\multiput(651.00,270.17)(4.472,2.000){2}{\rule{0.850pt}{0.400pt}}
\put(659,272.67){\rule{1.927pt}{0.400pt}}
\multiput(659.00,272.17)(4.000,1.000){2}{\rule{0.964pt}{0.400pt}}
\put(667,274.17){\rule{1.700pt}{0.400pt}}
\multiput(667.00,273.17)(4.472,2.000){2}{\rule{0.850pt}{0.400pt}}
\put(675,276.17){\rule{1.500pt}{0.400pt}}
\multiput(675.00,275.17)(3.887,2.000){2}{\rule{0.750pt}{0.400pt}}
\put(682,278.17){\rule{1.700pt}{0.400pt}}
\multiput(682.00,277.17)(4.472,2.000){2}{\rule{0.850pt}{0.400pt}}
\put(690,279.67){\rule{1.927pt}{0.400pt}}
\multiput(690.00,279.17)(4.000,1.000){2}{\rule{0.964pt}{0.400pt}}
\multiput(698.00,281.61)(1.579,0.447){3}{\rule{1.167pt}{0.108pt}}
\multiput(698.00,280.17)(5.579,3.000){2}{\rule{0.583pt}{0.400pt}}
\put(706,284.17){\rule{1.700pt}{0.400pt}}
\multiput(706.00,283.17)(4.472,2.000){2}{\rule{0.850pt}{0.400pt}}
\put(714,286.17){\rule{1.500pt}{0.400pt}}
\multiput(714.00,285.17)(3.887,2.000){2}{\rule{0.750pt}{0.400pt}}
\multiput(721.00,288.61)(1.579,0.447){3}{\rule{1.167pt}{0.108pt}}
\multiput(721.00,287.17)(5.579,3.000){2}{\rule{0.583pt}{0.400pt}}
\put(729,291.17){\rule{1.700pt}{0.400pt}}
\multiput(729.00,290.17)(4.472,2.000){2}{\rule{0.850pt}{0.400pt}}
\multiput(737.00,293.61)(1.579,0.447){3}{\rule{1.167pt}{0.108pt}}
\multiput(737.00,292.17)(5.579,3.000){2}{\rule{0.583pt}{0.400pt}}
\multiput(745.00,296.61)(1.355,0.447){3}{\rule{1.033pt}{0.108pt}}
\multiput(745.00,295.17)(4.855,3.000){2}{\rule{0.517pt}{0.400pt}}
\multiput(752.00,299.60)(1.066,0.468){5}{\rule{0.900pt}{0.113pt}}
\multiput(752.00,298.17)(6.132,4.000){2}{\rule{0.450pt}{0.400pt}}
\multiput(760.00,303.61)(1.579,0.447){3}{\rule{1.167pt}{0.108pt}}
\multiput(760.00,302.17)(5.579,3.000){2}{\rule{0.583pt}{0.400pt}}
\multiput(768.00,306.60)(1.066,0.468){5}{\rule{0.900pt}{0.113pt}}
\multiput(768.00,305.17)(6.132,4.000){2}{\rule{0.450pt}{0.400pt}}
\multiput(776.00,310.60)(0.920,0.468){5}{\rule{0.800pt}{0.113pt}}
\multiput(776.00,309.17)(5.340,4.000){2}{\rule{0.400pt}{0.400pt}}
\multiput(783.00,314.60)(1.066,0.468){5}{\rule{0.900pt}{0.113pt}}
\multiput(783.00,313.17)(6.132,4.000){2}{\rule{0.450pt}{0.400pt}}
\multiput(791.00,318.59)(0.821,0.477){7}{\rule{0.740pt}{0.115pt}}
\multiput(791.00,317.17)(6.464,5.000){2}{\rule{0.370pt}{0.400pt}}
\multiput(799.00,323.60)(1.066,0.468){5}{\rule{0.900pt}{0.113pt}}
\multiput(799.00,322.17)(6.132,4.000){2}{\rule{0.450pt}{0.400pt}}
\multiput(807.00,327.59)(0.710,0.477){7}{\rule{0.660pt}{0.115pt}}
\multiput(807.00,326.17)(5.630,5.000){2}{\rule{0.330pt}{0.400pt}}
\multiput(814.00,332.59)(0.671,0.482){9}{\rule{0.633pt}{0.116pt}}
\multiput(814.00,331.17)(6.685,6.000){2}{\rule{0.317pt}{0.400pt}}
\multiput(822.00,338.59)(0.821,0.477){7}{\rule{0.740pt}{0.115pt}}
\multiput(822.00,337.17)(6.464,5.000){2}{\rule{0.370pt}{0.400pt}}
\multiput(830.00,343.59)(0.671,0.482){9}{\rule{0.633pt}{0.116pt}}
\multiput(830.00,342.17)(6.685,6.000){2}{\rule{0.317pt}{0.400pt}}
\multiput(838.00,349.59)(0.492,0.485){11}{\rule{0.500pt}{0.117pt}}
\multiput(838.00,348.17)(5.962,7.000){2}{\rule{0.250pt}{0.400pt}}
\multiput(845.00,356.59)(0.671,0.482){9}{\rule{0.633pt}{0.116pt}}
\multiput(845.00,355.17)(6.685,6.000){2}{\rule{0.317pt}{0.400pt}}
\multiput(853.00,362.59)(0.569,0.485){11}{\rule{0.557pt}{0.117pt}}
\multiput(853.00,361.17)(6.844,7.000){2}{\rule{0.279pt}{0.400pt}}
\multiput(861.00,369.59)(0.494,0.488){13}{\rule{0.500pt}{0.117pt}}
\multiput(861.00,368.17)(6.962,8.000){2}{\rule{0.250pt}{0.400pt}}
\multiput(869.00,377.59)(0.492,0.485){11}{\rule{0.500pt}{0.117pt}}
\multiput(869.00,376.17)(5.962,7.000){2}{\rule{0.250pt}{0.400pt}}
\multiput(876.00,384.59)(0.494,0.488){13}{\rule{0.500pt}{0.117pt}}
\multiput(876.00,383.17)(6.962,8.000){2}{\rule{0.250pt}{0.400pt}}
\multiput(884.59,392.00)(0.488,0.560){13}{\rule{0.117pt}{0.550pt}}
\multiput(883.17,392.00)(8.000,7.858){2}{\rule{0.400pt}{0.275pt}}
\multiput(892.59,401.00)(0.488,0.560){13}{\rule{0.117pt}{0.550pt}}
\multiput(891.17,401.00)(8.000,7.858){2}{\rule{0.400pt}{0.275pt}}
\multiput(900.59,410.00)(0.485,0.645){11}{\rule{0.117pt}{0.614pt}}
\multiput(899.17,410.00)(7.000,7.725){2}{\rule{0.400pt}{0.307pt}}
\multiput(907.59,419.00)(0.488,0.626){13}{\rule{0.117pt}{0.600pt}}
\multiput(906.17,419.00)(8.000,8.755){2}{\rule{0.400pt}{0.300pt}}
\multiput(915.59,429.00)(0.488,0.626){13}{\rule{0.117pt}{0.600pt}}
\multiput(914.17,429.00)(8.000,8.755){2}{\rule{0.400pt}{0.300pt}}
\multiput(923.59,439.00)(0.488,0.692){13}{\rule{0.117pt}{0.650pt}}
\multiput(922.17,439.00)(8.000,9.651){2}{\rule{0.400pt}{0.325pt}}
\multiput(931.59,450.00)(0.485,0.798){11}{\rule{0.117pt}{0.729pt}}
\multiput(930.17,450.00)(7.000,9.488){2}{\rule{0.400pt}{0.364pt}}
\multiput(938.59,461.00)(0.488,0.692){13}{\rule{0.117pt}{0.650pt}}
\multiput(937.17,461.00)(8.000,9.651){2}{\rule{0.400pt}{0.325pt}}
\multiput(946.58,472.00)(0.491,0.756){17}{\rule{0.118pt}{0.700pt}}
\multiput(945.17,472.00)(10.000,13.547){2}{\rule{0.400pt}{0.350pt}}
\multiput(956.59,487.00)(0.489,0.961){15}{\rule{0.118pt}{0.856pt}}
\multiput(955.17,487.00)(9.000,15.224){2}{\rule{0.400pt}{0.428pt}}
\multiput(965.58,504.00)(0.491,1.017){17}{\rule{0.118pt}{0.900pt}}
\multiput(964.17,504.00)(10.000,18.132){2}{\rule{0.400pt}{0.450pt}}
\multiput(975.59,524.00)(0.485,1.408){11}{\rule{0.117pt}{1.186pt}}
\multiput(974.17,524.00)(7.000,16.539){2}{\rule{0.400pt}{0.593pt}}
\multiput(982.59,543.00)(0.488,1.352){13}{\rule{0.117pt}{1.150pt}}
\multiput(981.17,543.00)(8.000,18.613){2}{\rule{0.400pt}{0.575pt}}
\multiput(990.59,564.00)(0.488,1.484){13}{\rule{0.117pt}{1.250pt}}
\multiput(989.17,564.00)(8.000,20.406){2}{\rule{0.400pt}{0.625pt}}
\multiput(998.59,587.00)(0.488,1.550){13}{\rule{0.117pt}{1.300pt}}
\multiput(997.17,587.00)(8.000,21.302){2}{\rule{0.400pt}{0.650pt}}
\multiput(1006.59,611.00)(0.488,1.682){13}{\rule{0.117pt}{1.400pt}}
\multiput(1005.17,611.00)(8.000,23.094){2}{\rule{0.400pt}{0.700pt}}
\multiput(1014.59,637.00)(0.488,1.682){13}{\rule{0.117pt}{1.400pt}}
\multiput(1013.17,637.00)(8.000,23.094){2}{\rule{0.400pt}{0.700pt}}
\multiput(1022.59,663.00)(0.485,2.018){11}{\rule{0.117pt}{1.643pt}}
\multiput(1021.17,663.00)(7.000,23.590){2}{\rule{0.400pt}{0.821pt}}
\sbox{\plotpoint}{\rule[-0.500pt]{1.000pt}{1.000pt}}%
\put(518,598){\makebox(0,0)[r]{old P.Th.}}
\multiput(540,598)(20.756,0.000){4}{\usebox{\plotpoint}}
\put(606,598){\usebox{\plotpoint}}
\put(254,182){\usebox{\plotpoint}}
\put(254.00,182.00){\usebox{\plotpoint}}
\multiput(262,183)(20.756,0.000){0}{\usebox{\plotpoint}}
\put(274.66,183.58){\usebox{\plotpoint}}
\multiput(278,184)(20.756,0.000){0}{\usebox{\plotpoint}}
\multiput(285,184)(20.595,2.574){0}{\usebox{\plotpoint}}
\put(295.32,185.00){\usebox{\plotpoint}}
\multiput(301,185)(20.756,0.000){0}{\usebox{\plotpoint}}
\put(316.03,185.88){\usebox{\plotpoint}}
\multiput(317,186)(20.756,0.000){0}{\usebox{\plotpoint}}
\multiput(324,186)(20.595,2.574){0}{\usebox{\plotpoint}}
\put(336.71,187.00){\usebox{\plotpoint}}
\multiput(340,187)(20.595,2.574){0}{\usebox{\plotpoint}}
\multiput(348,188)(20.756,0.000){0}{\usebox{\plotpoint}}
\put(357.39,188.20){\usebox{\plotpoint}}
\multiput(363,189)(20.756,0.000){0}{\usebox{\plotpoint}}
\put(378.09,189.00){\usebox{\plotpoint}}
\multiput(379,189)(20.595,2.574){0}{\usebox{\plotpoint}}
\multiput(387,190)(20.756,0.000){0}{\usebox{\plotpoint}}
\put(398.74,190.53){\usebox{\plotpoint}}
\multiput(402,191)(20.756,0.000){0}{\usebox{\plotpoint}}
\multiput(410,191)(20.595,2.574){0}{\usebox{\plotpoint}}
\put(419.40,192.00){\usebox{\plotpoint}}
\multiput(426,192)(20.595,2.574){0}{\usebox{\plotpoint}}
\put(440.10,193.00){\usebox{\plotpoint}}
\multiput(441,193)(20.595,2.574){0}{\usebox{\plotpoint}}
\multiput(449,194)(20.756,0.000){0}{\usebox{\plotpoint}}
\put(460.76,194.47){\usebox{\plotpoint}}
\multiput(465,195)(20.756,0.000){0}{\usebox{\plotpoint}}
\multiput(472,195)(20.595,2.574){0}{\usebox{\plotpoint}}
\put(481.42,196.00){\usebox{\plotpoint}}
\multiput(488,196)(20.595,2.574){0}{\usebox{\plotpoint}}
\put(502.11,197.00){\usebox{\plotpoint}}
\multiput(504,197)(20.547,2.935){0}{\usebox{\plotpoint}}
\multiput(511,198)(20.595,2.574){0}{\usebox{\plotpoint}}
\put(522.74,199.00){\usebox{\plotpoint}}
\multiput(527,199)(20.595,2.574){0}{\usebox{\plotpoint}}
\multiput(535,200)(20.756,0.000){0}{\usebox{\plotpoint}}
\put(543.43,200.06){\usebox{\plotpoint}}
\multiput(550,201)(20.595,2.574){0}{\usebox{\plotpoint}}
\put(564.05,202.00){\usebox{\plotpoint}}
\multiput(566,202)(20.595,2.574){0}{\usebox{\plotpoint}}
\multiput(574,203)(20.595,2.574){0}{\usebox{\plotpoint}}
\put(584.68,204.00){\usebox{\plotpoint}}
\multiput(589,204)(20.595,2.574){0}{\usebox{\plotpoint}}
\multiput(597,205)(20.595,2.574){0}{\usebox{\plotpoint}}
\put(605.31,206.00){\usebox{\plotpoint}}
\multiput(613,206)(20.547,2.935){0}{\usebox{\plotpoint}}
\put(625.95,207.74){\usebox{\plotpoint}}
\multiput(628,208)(20.756,0.000){0}{\usebox{\plotpoint}}
\multiput(636,208)(20.595,2.574){0}{\usebox{\plotpoint}}
\put(646.60,209.37){\usebox{\plotpoint}}
\multiput(651,210)(20.595,2.574){0}{\usebox{\plotpoint}}
\multiput(659,211)(20.756,0.000){0}{\usebox{\plotpoint}}
\put(667.25,211.03){\usebox{\plotpoint}}
\multiput(675,212)(20.547,2.935){0}{\usebox{\plotpoint}}
\put(687.83,213.73){\usebox{\plotpoint}}
\multiput(690,214)(20.595,2.574){0}{\usebox{\plotpoint}}
\multiput(698,215)(20.756,0.000){0}{\usebox{\plotpoint}}
\put(708.49,215.31){\usebox{\plotpoint}}
\multiput(714,216)(20.547,2.935){0}{\usebox{\plotpoint}}
\multiput(721,217)(20.595,2.574){0}{\usebox{\plotpoint}}
\put(729.06,218.01){\usebox{\plotpoint}}
\multiput(737,219)(20.595,2.574){0}{\usebox{\plotpoint}}
\put(749.65,220.66){\usebox{\plotpoint}}
\multiput(752,221)(20.595,2.574){0}{\usebox{\plotpoint}}
\multiput(760,222)(20.756,0.000){0}{\usebox{\plotpoint}}
\put(770.30,222.29){\usebox{\plotpoint}}
\multiput(776,223)(20.547,2.935){0}{\usebox{\plotpoint}}
\put(790.88,224.98){\usebox{\plotpoint}}
\multiput(791,225)(20.595,2.574){0}{\usebox{\plotpoint}}
\multiput(799,226)(20.595,2.574){0}{\usebox{\plotpoint}}
\put(811.46,227.64){\usebox{\plotpoint}}
\multiput(814,228)(20.595,2.574){0}{\usebox{\plotpoint}}
\multiput(822,229)(20.595,2.574){0}{\usebox{\plotpoint}}
\put(832.05,230.26){\usebox{\plotpoint}}
\multiput(838,231)(20.547,2.935){0}{\usebox{\plotpoint}}
\put(852.63,232.95){\usebox{\plotpoint}}
\multiput(853,233)(20.136,5.034){0}{\usebox{\plotpoint}}
\multiput(861,235)(20.595,2.574){0}{\usebox{\plotpoint}}
\put(873.03,236.58){\usebox{\plotpoint}}
\multiput(876,237)(20.595,2.574){0}{\usebox{\plotpoint}}
\multiput(884,238)(20.595,2.574){0}{\usebox{\plotpoint}}
\put(893.62,239.20){\usebox{\plotpoint}}
\multiput(900,240)(20.547,2.935){0}{\usebox{\plotpoint}}
\put(914.04,242.76){\usebox{\plotpoint}}
\multiput(915,243)(20.595,2.574){0}{\usebox{\plotpoint}}
\multiput(923,244)(20.595,2.574){0}{\usebox{\plotpoint}}
\put(934.61,245.52){\usebox{\plotpoint}}
\multiput(938,246)(20.595,2.574){0}{\usebox{\plotpoint}}
\put(955.08,248.82){\usebox{\plotpoint}}
\multiput(956,249)(20.629,2.292){0}{\usebox{\plotpoint}}
\multiput(965,250)(20.352,4.070){0}{\usebox{\plotpoint}}
\put(975.56,252.08){\usebox{\plotpoint}}
\multiput(982,253)(20.595,2.574){0}{\usebox{\plotpoint}}
\put(996.14,254.77){\usebox{\plotpoint}}
\multiput(998,255)(20.595,2.574){0}{\usebox{\plotpoint}}
\multiput(1006,256)(20.595,2.574){0}{\usebox{\plotpoint}}
\put(1016.74,257.34){\usebox{\plotpoint}}
\multiput(1022,258)(20.547,2.935){0}{\usebox{\plotpoint}}
\put(1029,259){\usebox{\plotpoint}}
\end{picture}
 \vspace{1ex}
% GNUPLOT: LaTeX picture
\setlength{\unitlength}{0.240900pt}
\ifx\plotpoint\undefined\newsavebox{\plotpoint}\fi
\sbox{\plotpoint}{\rule[-0.200pt]{0.400pt}{0.400pt}}%
\begin{picture}(1200,900)(0,0)
\font\gnuplot=cmr10 at 10pt
\gnuplot
\sbox{\plotpoint}{\rule[-0.200pt]{0.400pt}{0.400pt}}%
\put(220.0,113.0){\rule[-0.200pt]{220.664pt}{0.400pt}}
\put(220.0,401.0){\rule[-0.200pt]{4.818pt}{0.400pt}}
\put(198,401){\makebox(0,0)[r]{0.2}}
\put(1116.0,401.0){\rule[-0.200pt]{4.818pt}{0.400pt}}
\put(220.0,688.0){\rule[-0.200pt]{4.818pt}{0.400pt}}
\put(198,688){\makebox(0,0)[r]{0.4}}
\put(1116.0,688.0){\rule[-0.200pt]{4.818pt}{0.400pt}}
\put(449.0,113.0){\rule[-0.200pt]{0.400pt}{4.818pt}}
\put(449,68){\makebox(0,0){0.001}}
\put(449.0,812.0){\rule[-0.200pt]{0.400pt}{4.818pt}}
\put(678.0,113.0){\rule[-0.200pt]{0.400pt}{4.818pt}}
\put(678,68){\makebox(0,0){0.01}}
\put(678.0,812.0){\rule[-0.200pt]{0.400pt}{4.818pt}}
\put(907.0,113.0){\rule[-0.200pt]{0.400pt}{4.818pt}}
\put(907,68){\makebox(0,0){0.1}}
\put(907.0,812.0){\rule[-0.200pt]{0.400pt}{4.818pt}}
\put(220.0,113.0){\rule[-0.200pt]{220.664pt}{0.400pt}}
\put(1136.0,113.0){\rule[-0.200pt]{0.400pt}{173.207pt}}
\put(220.0,832.0){\rule[-0.200pt]{220.664pt}{0.400pt}}
\put(45,472){\makebox(0,0){$\chi_{22}$}}
\put(678,-1){\makebox(0,0){loop size (fm)}}
\put(678,877){\makebox(0,0){$\chi_{22}$ Loop Ratio\,---\,$2^{\rm nd}$ Order}}
\put(220.0,113.0){\rule[-0.200pt]{0.400pt}{173.207pt}}
\put(518,688){\makebox(0,0)[r]{exact}}
\put(562,688){\circle{24}}
\put(1029,650){\circle{24}}
\put(975,495){\circle{24}}
\put(946,448){\circle{24}}
\put(605,259){\circle{24}}
\put(254,208){\circle{24}}
\put(518,643){\makebox(0,0)[r]{new P.Th.}}
\put(540.0,643.0){\rule[-0.200pt]{15.899pt}{0.400pt}}
\put(254,207){\usebox{\plotpoint}}
\put(254,207.17){\rule{1.700pt}{0.400pt}}
\multiput(254.00,206.17)(4.472,2.000){2}{\rule{0.850pt}{0.400pt}}
\put(262,208.67){\rule{1.927pt}{0.400pt}}
\multiput(262.00,208.17)(4.000,1.000){2}{\rule{0.964pt}{0.400pt}}
\put(270,209.67){\rule{1.927pt}{0.400pt}}
\multiput(270.00,209.17)(4.000,1.000){2}{\rule{0.964pt}{0.400pt}}
\put(278,210.67){\rule{1.686pt}{0.400pt}}
\multiput(278.00,210.17)(3.500,1.000){2}{\rule{0.843pt}{0.400pt}}
\put(285,211.67){\rule{1.927pt}{0.400pt}}
\multiput(285.00,211.17)(4.000,1.000){2}{\rule{0.964pt}{0.400pt}}
\put(293,212.67){\rule{1.927pt}{0.400pt}}
\multiput(293.00,212.17)(4.000,1.000){2}{\rule{0.964pt}{0.400pt}}
\put(301,213.67){\rule{1.927pt}{0.400pt}}
\multiput(301.00,213.17)(4.000,1.000){2}{\rule{0.964pt}{0.400pt}}
\put(309,214.67){\rule{1.927pt}{0.400pt}}
\multiput(309.00,214.17)(4.000,1.000){2}{\rule{0.964pt}{0.400pt}}
\put(317,215.67){\rule{1.686pt}{0.400pt}}
\multiput(317.00,215.17)(3.500,1.000){2}{\rule{0.843pt}{0.400pt}}
\put(324,217.17){\rule{1.700pt}{0.400pt}}
\multiput(324.00,216.17)(4.472,2.000){2}{\rule{0.850pt}{0.400pt}}
\put(332,218.67){\rule{1.927pt}{0.400pt}}
\multiput(332.00,218.17)(4.000,1.000){2}{\rule{0.964pt}{0.400pt}}
\put(340,219.67){\rule{1.927pt}{0.400pt}}
\multiput(340.00,219.17)(4.000,1.000){2}{\rule{0.964pt}{0.400pt}}
\put(348,220.67){\rule{1.927pt}{0.400pt}}
\multiput(348.00,220.17)(4.000,1.000){2}{\rule{0.964pt}{0.400pt}}
\put(356,221.67){\rule{1.686pt}{0.400pt}}
\multiput(356.00,221.17)(3.500,1.000){2}{\rule{0.843pt}{0.400pt}}
\put(363,222.67){\rule{1.927pt}{0.400pt}}
\multiput(363.00,222.17)(4.000,1.000){2}{\rule{0.964pt}{0.400pt}}
\put(371,223.67){\rule{1.927pt}{0.400pt}}
\multiput(371.00,223.17)(4.000,1.000){2}{\rule{0.964pt}{0.400pt}}
\put(379,224.67){\rule{1.927pt}{0.400pt}}
\multiput(379.00,224.17)(4.000,1.000){2}{\rule{0.964pt}{0.400pt}}
\put(387,225.67){\rule{1.927pt}{0.400pt}}
\multiput(387.00,225.17)(4.000,1.000){2}{\rule{0.964pt}{0.400pt}}
\put(395,227.17){\rule{1.500pt}{0.400pt}}
\multiput(395.00,226.17)(3.887,2.000){2}{\rule{0.750pt}{0.400pt}}
\put(402,228.67){\rule{1.927pt}{0.400pt}}
\multiput(402.00,228.17)(4.000,1.000){2}{\rule{0.964pt}{0.400pt}}
\put(410,229.67){\rule{1.927pt}{0.400pt}}
\multiput(410.00,229.17)(4.000,1.000){2}{\rule{0.964pt}{0.400pt}}
\put(418,230.67){\rule{1.927pt}{0.400pt}}
\multiput(418.00,230.17)(4.000,1.000){2}{\rule{0.964pt}{0.400pt}}
\put(426,231.67){\rule{1.927pt}{0.400pt}}
\multiput(426.00,231.17)(4.000,1.000){2}{\rule{0.964pt}{0.400pt}}
\put(434,232.67){\rule{1.686pt}{0.400pt}}
\multiput(434.00,232.17)(3.500,1.000){2}{\rule{0.843pt}{0.400pt}}
\put(441,233.67){\rule{1.927pt}{0.400pt}}
\multiput(441.00,233.17)(4.000,1.000){2}{\rule{0.964pt}{0.400pt}}
\put(449,235.17){\rule{1.700pt}{0.400pt}}
\multiput(449.00,234.17)(4.472,2.000){2}{\rule{0.850pt}{0.400pt}}
\put(457,236.67){\rule{1.927pt}{0.400pt}}
\multiput(457.00,236.17)(4.000,1.000){2}{\rule{0.964pt}{0.400pt}}
\put(465,237.67){\rule{1.686pt}{0.400pt}}
\multiput(465.00,237.17)(3.500,1.000){2}{\rule{0.843pt}{0.400pt}}
\put(472,238.67){\rule{1.927pt}{0.400pt}}
\multiput(472.00,238.17)(4.000,1.000){2}{\rule{0.964pt}{0.400pt}}
\put(480,239.67){\rule{1.927pt}{0.400pt}}
\multiput(480.00,239.17)(4.000,1.000){2}{\rule{0.964pt}{0.400pt}}
\put(488,240.67){\rule{1.927pt}{0.400pt}}
\multiput(488.00,240.17)(4.000,1.000){2}{\rule{0.964pt}{0.400pt}}
\put(496,242.17){\rule{1.700pt}{0.400pt}}
\multiput(496.00,241.17)(4.472,2.000){2}{\rule{0.850pt}{0.400pt}}
\put(504,243.67){\rule{1.686pt}{0.400pt}}
\multiput(504.00,243.17)(3.500,1.000){2}{\rule{0.843pt}{0.400pt}}
\put(511,244.67){\rule{1.927pt}{0.400pt}}
\multiput(511.00,244.17)(4.000,1.000){2}{\rule{0.964pt}{0.400pt}}
\put(519,245.67){\rule{1.927pt}{0.400pt}}
\multiput(519.00,245.17)(4.000,1.000){2}{\rule{0.964pt}{0.400pt}}
\put(527,246.67){\rule{1.927pt}{0.400pt}}
\multiput(527.00,246.17)(4.000,1.000){2}{\rule{0.964pt}{0.400pt}}
\put(535,247.67){\rule{1.927pt}{0.400pt}}
\multiput(535.00,247.17)(4.000,1.000){2}{\rule{0.964pt}{0.400pt}}
\put(543,249.17){\rule{1.500pt}{0.400pt}}
\multiput(543.00,248.17)(3.887,2.000){2}{\rule{0.750pt}{0.400pt}}
\put(550,250.67){\rule{1.927pt}{0.400pt}}
\multiput(550.00,250.17)(4.000,1.000){2}{\rule{0.964pt}{0.400pt}}
\put(558,251.67){\rule{1.927pt}{0.400pt}}
\multiput(558.00,251.17)(4.000,1.000){2}{\rule{0.964pt}{0.400pt}}
\put(566,252.67){\rule{1.927pt}{0.400pt}}
\multiput(566.00,252.17)(4.000,1.000){2}{\rule{0.964pt}{0.400pt}}
\put(574,254.17){\rule{1.700pt}{0.400pt}}
\multiput(574.00,253.17)(4.472,2.000){2}{\rule{0.850pt}{0.400pt}}
\put(582,255.67){\rule{1.686pt}{0.400pt}}
\multiput(582.00,255.17)(3.500,1.000){2}{\rule{0.843pt}{0.400pt}}
\put(589,256.67){\rule{1.927pt}{0.400pt}}
\multiput(589.00,256.17)(4.000,1.000){2}{\rule{0.964pt}{0.400pt}}
\put(597,257.67){\rule{1.927pt}{0.400pt}}
\multiput(597.00,257.17)(4.000,1.000){2}{\rule{0.964pt}{0.400pt}}
\put(605,258.67){\rule{1.927pt}{0.400pt}}
\multiput(605.00,258.17)(4.000,1.000){2}{\rule{0.964pt}{0.400pt}}
\put(613,260.17){\rule{1.500pt}{0.400pt}}
\multiput(613.00,259.17)(3.887,2.000){2}{\rule{0.750pt}{0.400pt}}
\put(620,261.67){\rule{1.927pt}{0.400pt}}
\multiput(620.00,261.17)(4.000,1.000){2}{\rule{0.964pt}{0.400pt}}
\put(628,262.67){\rule{1.927pt}{0.400pt}}
\multiput(628.00,262.17)(4.000,1.000){2}{\rule{0.964pt}{0.400pt}}
\put(636,264.17){\rule{1.700pt}{0.400pt}}
\multiput(636.00,263.17)(4.472,2.000){2}{\rule{0.850pt}{0.400pt}}
\put(644,265.67){\rule{1.686pt}{0.400pt}}
\multiput(644.00,265.17)(3.500,1.000){2}{\rule{0.843pt}{0.400pt}}
\put(651,267.17){\rule{1.700pt}{0.400pt}}
\multiput(651.00,266.17)(4.472,2.000){2}{\rule{0.850pt}{0.400pt}}
\put(659,268.67){\rule{1.927pt}{0.400pt}}
\multiput(659.00,268.17)(4.000,1.000){2}{\rule{0.964pt}{0.400pt}}
\put(667,270.17){\rule{1.700pt}{0.400pt}}
\multiput(667.00,269.17)(4.472,2.000){2}{\rule{0.850pt}{0.400pt}}
\put(675,271.67){\rule{1.686pt}{0.400pt}}
\multiput(675.00,271.17)(3.500,1.000){2}{\rule{0.843pt}{0.400pt}}
\put(682,273.17){\rule{1.700pt}{0.400pt}}
\multiput(682.00,272.17)(4.472,2.000){2}{\rule{0.850pt}{0.400pt}}
\put(690,275.17){\rule{1.700pt}{0.400pt}}
\multiput(690.00,274.17)(4.472,2.000){2}{\rule{0.850pt}{0.400pt}}
\put(698,277.17){\rule{1.700pt}{0.400pt}}
\multiput(698.00,276.17)(4.472,2.000){2}{\rule{0.850pt}{0.400pt}}
\put(706,279.17){\rule{1.700pt}{0.400pt}}
\multiput(706.00,278.17)(4.472,2.000){2}{\rule{0.850pt}{0.400pt}}
\put(714,281.17){\rule{1.500pt}{0.400pt}}
\multiput(714.00,280.17)(3.887,2.000){2}{\rule{0.750pt}{0.400pt}}
\multiput(721.00,283.61)(1.579,0.447){3}{\rule{1.167pt}{0.108pt}}
\multiput(721.00,282.17)(5.579,3.000){2}{\rule{0.583pt}{0.400pt}}
\put(729,286.17){\rule{1.700pt}{0.400pt}}
\multiput(729.00,285.17)(4.472,2.000){2}{\rule{0.850pt}{0.400pt}}
\multiput(737.00,288.61)(1.579,0.447){3}{\rule{1.167pt}{0.108pt}}
\multiput(737.00,287.17)(5.579,3.000){2}{\rule{0.583pt}{0.400pt}}
\multiput(745.00,291.61)(1.355,0.447){3}{\rule{1.033pt}{0.108pt}}
\multiput(745.00,290.17)(4.855,3.000){2}{\rule{0.517pt}{0.400pt}}
\multiput(752.00,294.61)(1.579,0.447){3}{\rule{1.167pt}{0.108pt}}
\multiput(752.00,293.17)(5.579,3.000){2}{\rule{0.583pt}{0.400pt}}
\multiput(760.00,297.61)(1.579,0.447){3}{\rule{1.167pt}{0.108pt}}
\multiput(760.00,296.17)(5.579,3.000){2}{\rule{0.583pt}{0.400pt}}
\multiput(768.00,300.60)(1.066,0.468){5}{\rule{0.900pt}{0.113pt}}
\multiput(768.00,299.17)(6.132,4.000){2}{\rule{0.450pt}{0.400pt}}
\multiput(776.00,304.61)(1.355,0.447){3}{\rule{1.033pt}{0.108pt}}
\multiput(776.00,303.17)(4.855,3.000){2}{\rule{0.517pt}{0.400pt}}
\multiput(783.00,307.60)(1.066,0.468){5}{\rule{0.900pt}{0.113pt}}
\multiput(783.00,306.17)(6.132,4.000){2}{\rule{0.450pt}{0.400pt}}
\multiput(791.00,311.60)(1.066,0.468){5}{\rule{0.900pt}{0.113pt}}
\multiput(791.00,310.17)(6.132,4.000){2}{\rule{0.450pt}{0.400pt}}
\multiput(799.00,315.60)(1.066,0.468){5}{\rule{0.900pt}{0.113pt}}
\multiput(799.00,314.17)(6.132,4.000){2}{\rule{0.450pt}{0.400pt}}
\multiput(807.00,319.59)(0.710,0.477){7}{\rule{0.660pt}{0.115pt}}
\multiput(807.00,318.17)(5.630,5.000){2}{\rule{0.330pt}{0.400pt}}
\multiput(814.00,324.59)(0.821,0.477){7}{\rule{0.740pt}{0.115pt}}
\multiput(814.00,323.17)(6.464,5.000){2}{\rule{0.370pt}{0.400pt}}
\multiput(822.00,329.59)(0.821,0.477){7}{\rule{0.740pt}{0.115pt}}
\multiput(822.00,328.17)(6.464,5.000){2}{\rule{0.370pt}{0.400pt}}
\multiput(830.00,334.59)(0.821,0.477){7}{\rule{0.740pt}{0.115pt}}
\multiput(830.00,333.17)(6.464,5.000){2}{\rule{0.370pt}{0.400pt}}
\multiput(838.00,339.59)(0.581,0.482){9}{\rule{0.567pt}{0.116pt}}
\multiput(838.00,338.17)(5.824,6.000){2}{\rule{0.283pt}{0.400pt}}
\multiput(845.00,345.59)(0.821,0.477){7}{\rule{0.740pt}{0.115pt}}
\multiput(845.00,344.17)(6.464,5.000){2}{\rule{0.370pt}{0.400pt}}
\multiput(853.00,350.59)(0.569,0.485){11}{\rule{0.557pt}{0.117pt}}
\multiput(853.00,349.17)(6.844,7.000){2}{\rule{0.279pt}{0.400pt}}
\multiput(861.00,357.59)(0.671,0.482){9}{\rule{0.633pt}{0.116pt}}
\multiput(861.00,356.17)(6.685,6.000){2}{\rule{0.317pt}{0.400pt}}
\multiput(869.00,363.59)(0.492,0.485){11}{\rule{0.500pt}{0.117pt}}
\multiput(869.00,362.17)(5.962,7.000){2}{\rule{0.250pt}{0.400pt}}
\multiput(876.00,370.59)(0.569,0.485){11}{\rule{0.557pt}{0.117pt}}
\multiput(876.00,369.17)(6.844,7.000){2}{\rule{0.279pt}{0.400pt}}
\multiput(884.00,377.59)(0.569,0.485){11}{\rule{0.557pt}{0.117pt}}
\multiput(884.00,376.17)(6.844,7.000){2}{\rule{0.279pt}{0.400pt}}
\multiput(892.00,384.59)(0.494,0.488){13}{\rule{0.500pt}{0.117pt}}
\multiput(892.00,383.17)(6.962,8.000){2}{\rule{0.250pt}{0.400pt}}
\multiput(900.59,392.00)(0.485,0.569){11}{\rule{0.117pt}{0.557pt}}
\multiput(899.17,392.00)(7.000,6.844){2}{\rule{0.400pt}{0.279pt}}
\multiput(907.00,400.59)(0.494,0.488){13}{\rule{0.500pt}{0.117pt}}
\multiput(907.00,399.17)(6.962,8.000){2}{\rule{0.250pt}{0.400pt}}
\multiput(915.59,408.00)(0.488,0.560){13}{\rule{0.117pt}{0.550pt}}
\multiput(914.17,408.00)(8.000,7.858){2}{\rule{0.400pt}{0.275pt}}
\multiput(923.59,417.00)(0.488,0.560){13}{\rule{0.117pt}{0.550pt}}
\multiput(922.17,417.00)(8.000,7.858){2}{\rule{0.400pt}{0.275pt}}
\multiput(931.59,426.00)(0.485,0.721){11}{\rule{0.117pt}{0.671pt}}
\multiput(930.17,426.00)(7.000,8.606){2}{\rule{0.400pt}{0.336pt}}
\multiput(938.59,436.00)(0.488,0.626){13}{\rule{0.117pt}{0.600pt}}
\multiput(937.17,436.00)(8.000,8.755){2}{\rule{0.400pt}{0.300pt}}
\multiput(946.58,446.00)(0.491,0.652){17}{\rule{0.118pt}{0.620pt}}
\multiput(945.17,446.00)(10.000,11.713){2}{\rule{0.400pt}{0.310pt}}
\multiput(956.59,459.00)(0.489,0.786){15}{\rule{0.118pt}{0.722pt}}
\multiput(955.17,459.00)(9.000,12.501){2}{\rule{0.400pt}{0.361pt}}
\multiput(965.58,473.00)(0.491,0.808){17}{\rule{0.118pt}{0.740pt}}
\multiput(964.17,473.00)(10.000,14.464){2}{\rule{0.400pt}{0.370pt}}
\multiput(975.59,489.00)(0.485,1.103){11}{\rule{0.117pt}{0.957pt}}
\multiput(974.17,489.00)(7.000,13.013){2}{\rule{0.400pt}{0.479pt}}
\multiput(982.59,504.00)(0.488,1.088){13}{\rule{0.117pt}{0.950pt}}
\multiput(981.17,504.00)(8.000,15.028){2}{\rule{0.400pt}{0.475pt}}
\multiput(990.59,521.00)(0.488,1.220){13}{\rule{0.117pt}{1.050pt}}
\multiput(989.17,521.00)(8.000,16.821){2}{\rule{0.400pt}{0.525pt}}
\multiput(998.59,540.00)(0.488,1.220){13}{\rule{0.117pt}{1.050pt}}
\multiput(997.17,540.00)(8.000,16.821){2}{\rule{0.400pt}{0.525pt}}
\multiput(1006.59,559.00)(0.488,1.286){13}{\rule{0.117pt}{1.100pt}}
\multiput(1005.17,559.00)(8.000,17.717){2}{\rule{0.400pt}{0.550pt}}
\multiput(1014.59,579.00)(0.488,1.352){13}{\rule{0.117pt}{1.150pt}}
\multiput(1013.17,579.00)(8.000,18.613){2}{\rule{0.400pt}{0.575pt}}
\multiput(1022.59,600.00)(0.485,1.560){11}{\rule{0.117pt}{1.300pt}}
\multiput(1021.17,600.00)(7.000,18.302){2}{\rule{0.400pt}{0.650pt}}
\sbox{\plotpoint}{\rule[-0.500pt]{1.000pt}{1.000pt}}%
\put(518,598){\makebox(0,0)[r]{old P.Th.}}
\multiput(540,598)(20.756,0.000){4}{\usebox{\plotpoint}}
\put(606,598){\usebox{\plotpoint}}
\put(254,199){\usebox{\plotpoint}}
\put(254.00,199.00){\usebox{\plotpoint}}
\multiput(262,200)(20.595,2.574){0}{\usebox{\plotpoint}}
\put(274.63,201.00){\usebox{\plotpoint}}
\multiput(278,201)(20.547,2.935){0}{\usebox{\plotpoint}}
\multiput(285,202)(20.595,2.574){0}{\usebox{\plotpoint}}
\put(295.25,203.00){\usebox{\plotpoint}}
\multiput(301,203)(20.595,2.574){0}{\usebox{\plotpoint}}
\put(315.89,204.86){\usebox{\plotpoint}}
\multiput(317,205)(20.756,0.000){0}{\usebox{\plotpoint}}
\multiput(324,205)(20.595,2.574){0}{\usebox{\plotpoint}}
\put(336.54,206.57){\usebox{\plotpoint}}
\multiput(340,207)(20.756,0.000){0}{\usebox{\plotpoint}}
\multiput(348,207)(20.595,2.574){0}{\usebox{\plotpoint}}
\put(357.20,208.17){\usebox{\plotpoint}}
\multiput(363,209)(20.756,0.000){0}{\usebox{\plotpoint}}
\put(377.84,209.85){\usebox{\plotpoint}}
\multiput(379,210)(20.595,2.574){0}{\usebox{\plotpoint}}
\multiput(387,211)(20.595,2.574){0}{\usebox{\plotpoint}}
\put(398.46,212.00){\usebox{\plotpoint}}
\multiput(402,212)(20.595,2.574){0}{\usebox{\plotpoint}}
\multiput(410,213)(20.595,2.574){0}{\usebox{\plotpoint}}
\put(419.08,214.14){\usebox{\plotpoint}}
\multiput(426,215)(20.756,0.000){0}{\usebox{\plotpoint}}
\put(439.73,215.82){\usebox{\plotpoint}}
\multiput(441,216)(20.595,2.574){0}{\usebox{\plotpoint}}
\multiput(449,217)(20.595,2.574){0}{\usebox{\plotpoint}}
\put(460.32,218.41){\usebox{\plotpoint}}
\multiput(465,219)(20.756,0.000){0}{\usebox{\plotpoint}}
\multiput(472,219)(20.595,2.574){0}{\usebox{\plotpoint}}
\put(480.97,220.12){\usebox{\plotpoint}}
\multiput(488,221)(20.595,2.574){0}{\usebox{\plotpoint}}
\put(501.56,222.70){\usebox{\plotpoint}}
\multiput(504,223)(20.547,2.935){0}{\usebox{\plotpoint}}
\multiput(511,224)(20.595,2.574){0}{\usebox{\plotpoint}}
\put(522.14,225.39){\usebox{\plotpoint}}
\multiput(527,226)(20.595,2.574){0}{\usebox{\plotpoint}}
\put(542.80,227.00){\usebox{\plotpoint}}
\multiput(543,227)(20.547,2.935){0}{\usebox{\plotpoint}}
\multiput(550,228)(20.595,2.574){0}{\usebox{\plotpoint}}
\put(563.26,230.31){\usebox{\plotpoint}}
\multiput(566,231)(20.595,2.574){0}{\usebox{\plotpoint}}
\multiput(574,232)(20.595,2.574){0}{\usebox{\plotpoint}}
\put(583.79,233.26){\usebox{\plotpoint}}
\multiput(589,234)(20.595,2.574){0}{\usebox{\plotpoint}}
\put(604.37,235.92){\usebox{\plotpoint}}
\multiput(605,236)(20.595,2.574){0}{\usebox{\plotpoint}}
\multiput(613,237)(20.547,2.935){0}{\usebox{\plotpoint}}
\put(624.95,238.62){\usebox{\plotpoint}}
\multiput(628,239)(20.136,5.034){0}{\usebox{\plotpoint}}
\multiput(636,241)(20.595,2.574){0}{\usebox{\plotpoint}}
\put(645.36,242.19){\usebox{\plotpoint}}
\multiput(651,243)(20.595,2.574){0}{\usebox{\plotpoint}}
\put(665.79,245.70){\usebox{\plotpoint}}
\multiput(667,246)(20.595,2.574){0}{\usebox{\plotpoint}}
\multiput(675,247)(20.547,2.935){0}{\usebox{\plotpoint}}
\put(686.24,249.06){\usebox{\plotpoint}}
\multiput(690,250)(20.595,2.574){0}{\usebox{\plotpoint}}
\multiput(698,251)(20.136,5.034){0}{\usebox{\plotpoint}}
\put(706.57,253.07){\usebox{\plotpoint}}
\multiput(714,254)(20.547,2.935){0}{\usebox{\plotpoint}}
\put(727.01,256.50){\usebox{\plotpoint}}
\multiput(729,257)(20.136,5.034){0}{\usebox{\plotpoint}}
\multiput(737,259)(20.595,2.574){0}{\usebox{\plotpoint}}
\put(747.30,260.66){\usebox{\plotpoint}}
\multiput(752,262)(20.595,2.574){0}{\usebox{\plotpoint}}
\put(767.57,264.89){\usebox{\plotpoint}}
\multiput(768,265)(20.136,5.034){0}{\usebox{\plotpoint}}
\multiput(776,267)(20.547,2.935){0}{\usebox{\plotpoint}}
\put(787.85,269.21){\usebox{\plotpoint}}
\multiput(791,270)(20.136,5.034){0}{\usebox{\plotpoint}}
\multiput(799,272)(20.136,5.034){0}{\usebox{\plotpoint}}
\put(808.01,274.14){\usebox{\plotpoint}}
\multiput(814,275)(20.136,5.034){0}{\usebox{\plotpoint}}
\put(828.26,278.57){\usebox{\plotpoint}}
\multiput(830,279)(20.136,5.034){0}{\usebox{\plotpoint}}
\multiput(838,281)(19.957,5.702){0}{\usebox{\plotpoint}}
\put(848.33,283.83){\usebox{\plotpoint}}
\multiput(853,285)(20.136,5.034){0}{\usebox{\plotpoint}}
\put(868.47,288.87){\usebox{\plotpoint}}
\multiput(869,289)(19.957,5.702){0}{\usebox{\plotpoint}}
\multiput(876,291)(20.136,5.034){0}{\usebox{\plotpoint}}
\put(888.54,294.14){\usebox{\plotpoint}}
\multiput(892,295)(19.434,7.288){0}{\usebox{\plotpoint}}
\multiput(900,298)(19.957,5.702){0}{\usebox{\plotpoint}}
\put(908.33,300.33){\usebox{\plotpoint}}
\multiput(915,302)(20.136,5.034){0}{\usebox{\plotpoint}}
\put(928.27,305.98){\usebox{\plotpoint}}
\multiput(931,307)(19.957,5.702){0}{\usebox{\plotpoint}}
\multiput(938,309)(20.136,5.034){0}{\usebox{\plotpoint}}
\put(948.22,311.67){\usebox{\plotpoint}}
\multiput(956,314)(19.690,6.563){0}{\usebox{\plotpoint}}
\put(968.01,317.90){\usebox{\plotpoint}}
\multiput(975,320)(19.957,5.702){0}{\usebox{\plotpoint}}
\put(987.79,324.17){\usebox{\plotpoint}}
\multiput(990,325)(20.136,5.034){0}{\usebox{\plotpoint}}
\multiput(998,327)(20.136,5.034){0}{\usebox{\plotpoint}}
\put(1007.84,329.46){\usebox{\plotpoint}}
\multiput(1014,331)(20.136,5.034){0}{\usebox{\plotpoint}}
\put(1027.93,334.69){\usebox{\plotpoint}}
\put(1029,335){\usebox{\plotpoint}}
\end{picture}
\end{center}
\caption{The $\chi_{22}$ Creutz ratio of Wilson loops versus loop
size. Results from Monte Carlo simulations (exact), and from
tadpole-improved (new) and traditional (old) lattice perturbation theory are
shown.}
\label{creutz-fig}
\end{figure}
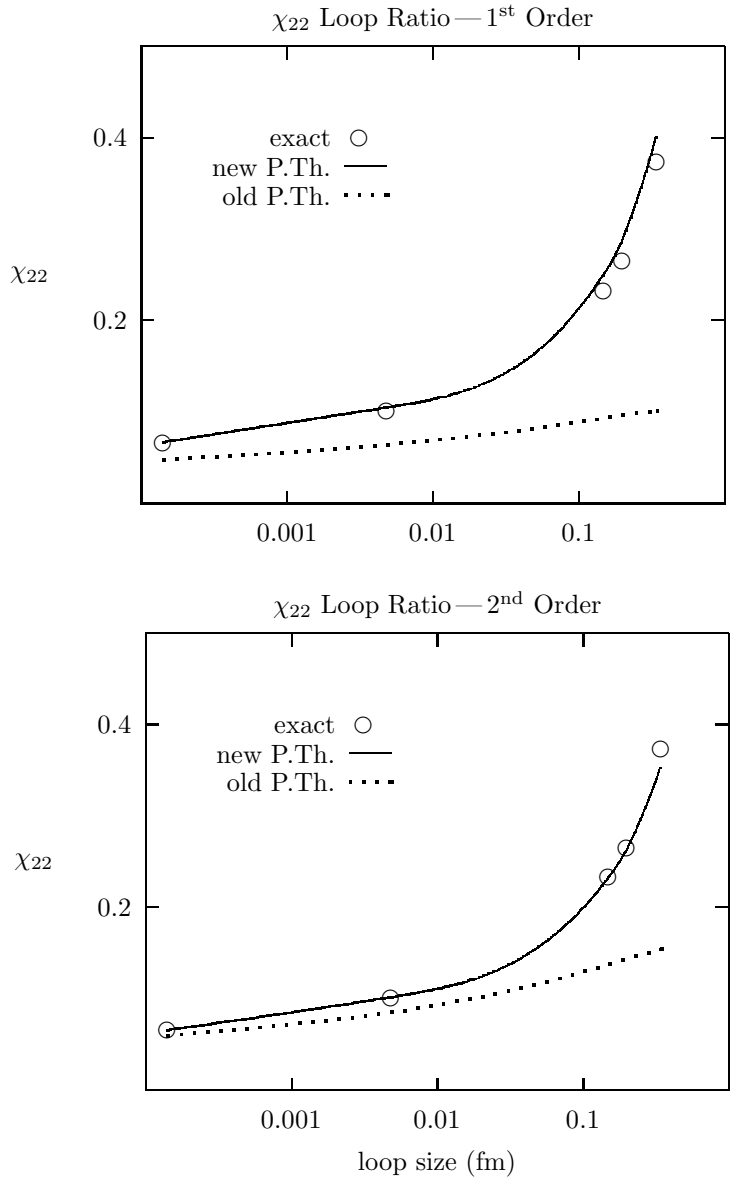

Fig.~\ref{creutz-fig} illustrates the highly unsatisfactory state of
traditional lattice-QCD perturbation theory.
It shows the ``Creutz ratio'' of $2a\times2a$, $2a\times a$ and
$a\times a$ Wilson loops,
\be \label{creutz-ratio}
\chi_{2,2} \equiv -\ln\left(\frac{W(2a\times2a)\,W(a\times a)}{W^2(2a\times
a)}\right),
\ee
plotted versus the size~$2a$ of the largest loop. Traditional perturbation
theory (dotted lines) 
underestimates the exact result by factors of three or four for
loops of order~$1/2$\,fm; only when the loops are smaller than $1/20$\,fm
does perturbation theory begin to give accurate results.

The problem with traditional lattice-QCD perturbation theory is that the
coupling it uses is much too small. The standard practice was to express
perturbative expansions of short-distance lattice quantities in terms of
the bare coupling~$\alpha_{\rm lat}$ used in the lattice lagrangian. This
practice followed from the notion that the bare coupling in a cutoff theory
is approximately equal to the running coupling evaluated at the cutoff
scale, here~$\alpha_s(\pi/a)$, and therefore that it is the appropriate
coupling for quantities dominated by momenta near the cutoff. In fact the
bare coupling in traditional lattice QCD is much smaller than true
effective coupling at large lattice spacings: for example,
\bearray
\alpha_{\rm lat} &=& \alpha_V(\pi/a) - 4.7\,\alpha_V^2 + \cdots \\
&\le& \half\alpha_V(\pi/a)\qquad\mbox{for $a>.1$\,fm}
\eearray
where $\alpha_V(q)$ is a continuum coupling defined by the static-quark
potential,
\be
V_{\rm Q\overline{Q}}(q) \equiv -4\,\pi\,C_{\rm F} \frac{\alpha_V(q)}{q^2}.
\ee
Consequently $\alpha_{\rm lat}$~expansions, though formally correct, badly
underestimate perturbative effects, and converge poorly.

The anomalously small bare coupling in the traditional lattice theory is a
symptom of the ``tadpole problem''. 
As we discuss later, all gluonic operators in lattice QCD
are built from the link operator
\be
U_\mu(x) \equiv \Porder \e^{-\I \int_x^{x+a\hat\mu}gA\cdot \dd x}
\approx \e^{-\I agA_\mu}
\ee
rather than from the vector potential~$A_\mu$. Thus, for example, the
leading term in the lagrangian that couples quarks and gluons
is~$\psib U_\mu\gamma_\mu\psi/a$.  
Such a term contains the usual $\psib gA\cdot\gamma\psi$~vertex, but, in
addition, it contains vertices with any number of additional powers of
$agA_\mu$. These extra vertices are irrelevant for classical fields since
they are suppressed by powers of the lattice spacing. For quantum fields,
however, the situation is quite different since pairs of~$A_\mu$'s, if
contracted with each other, generate ultraviolet divergent factors
of~$1/a^2$ that precisely cancel the extra~$a$'s. Consequently the 
contributions generated by the extra vertices are suppressed by powers
of~$g^2$ (not $a$), and turn out to be uncomfortably large. These are the
tadpole contributions.

The tadpoles  result in
large renormalizations\,---\,often as large as a factor of two or
three\,---\,that spoil naive perturbation theory, and with it our
intuition about the connection between lattice operators and the continuum.
However tadpole contributions are generically process independent and so it
is possible to measure their contribution in one quantity and then correct
for them in all other quantities.  

The simplest way to do this is to cancel
them out. The mean value~$u_0$ of~$\third\Re\Tr U_\mu$ consists of only
tadpoles and so we can largely cancel the tadpole contributions by dividing
every link operator by~$u_0$. That is, in every lattice operator we replace
\be
U_\mu(x) \to \frac{U_\mu(x)}{u_0}
\ee
where~$u_0$ is computed numerically in a simulation.  

The $u_0$'s cancel tadpole
contributions, making lattice operators and perturbation theory far more
continuum-like in their behavior. Thus, for example, the only change
in the standard gluon action when it is tadpole-improved is that the new
bare coupling~$\alpha_{\rm TI}$ is enhanced by a factor of $1/u_0^4$
relative to the coupling~$\alpha_{\rm lat}$ in the unimproved theory:
\be
\alpha_{\rm TI} = \frac{\alpha_{\rm lat}}{u_0^4}.
\ee
Since $u_0^4\!<\!.6$ when $a\!>\!.1$\,fm, the
tadpole-improved coupling is typically more than twice as large for coarse
lattices. Expressing~$\alpha_{\rm TI}$ in terms of the continuum
coupling~$\alpha_V$, we find that now our intuition is satisfied:
\bearray
{\alpha}_{\rm TI} &=& \alpha_V(\pi/a) - .5\,\alpha_V^2 +\cdots \\
&\approx& \alpha_V(\pi/a). 
\eearray

Perturbation theory for the Creutz ratio \eq{creutz-ratio}
converges rapidly to the correct answer when it is reexpressed in terms
of~$\alpha_{\rm TI}$. An even better result is obtained if the expansion is
reexpressed as a series in a coupling constant defined in terms of a
physical quantity, like the static-quark potential, where that coupling
constant is measured in a simulation.  By measuring the coupling we
automatically include any large renormalizations of the coupling due to
tadpoles. It is important that the scale~$q^*$ at which the running
coupling constant is evaluated be chosen appropriately for the quantity
being studied\,\cite{gpl93,gpl94}. 
When these refinements are added, perturbation theory is
dramatically improved, and, as illustrated in Fig.~\ref{creutz-fig}, is
still quite accurate for loops as large as $1/2$\,fm.

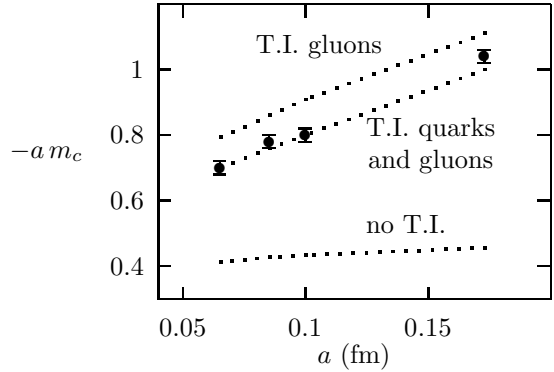
\begin{figure}
\begin{center} 
% GNUPLOT: LaTeX picture
\setlength{\unitlength}{0.240900pt}
\ifx\plotpoint\undefined\newsavebox{\plotpoint}\fi
\sbox{\plotpoint}{\rule[-0.200pt]{0.400pt}{0.400pt}}%
\begin{picture}(900,600)(0,0)
\font\gnuplot=cmr10 at 10pt
\gnuplot
\sbox{\plotpoint}{\rule[-0.200pt]{0.400pt}{0.400pt}}%
\put(220.0,165.0){\rule[-0.200pt]{4.818pt}{0.400pt}}
\put(198,165){\makebox(0,0)[r]{$0.4$}}
\put(816.0,165.0){\rule[-0.200pt]{4.818pt}{0.400pt}}
\put(220.0,268.0){\rule[-0.200pt]{4.818pt}{0.400pt}}
\put(198,268){\makebox(0,0)[r]{$0.6$}}
\put(816.0,268.0){\rule[-0.200pt]{4.818pt}{0.400pt}}
\put(220.0,371.0){\rule[-0.200pt]{4.818pt}{0.400pt}}
\put(198,371){\makebox(0,0)[r]{$0.8$}}
\put(816.0,371.0){\rule[-0.200pt]{4.818pt}{0.400pt}}
\put(220.0,474.0){\rule[-0.200pt]{4.818pt}{0.400pt}}
\put(198,474){\makebox(0,0)[r]{$1$}}
\put(816.0,474.0){\rule[-0.200pt]{4.818pt}{0.400pt}}
\put(259.0,113.0){\rule[-0.200pt]{0.400pt}{4.818pt}}
\put(259,68){\makebox(0,0){$0.05$}}
\put(259.0,557.0){\rule[-0.200pt]{0.400pt}{4.818pt}}
\put(451.0,113.0){\rule[-0.200pt]{0.400pt}{4.818pt}}
\put(451,68){\makebox(0,0){$0.1$}}
\put(451.0,557.0){\rule[-0.200pt]{0.400pt}{4.818pt}}
\put(644.0,113.0){\rule[-0.200pt]{0.400pt}{4.818pt}}
\put(644,68){\makebox(0,0){$0.15$}}
\put(644.0,557.0){\rule[-0.200pt]{0.400pt}{4.818pt}}
\put(220.0,113.0){\rule[-0.200pt]{148.394pt}{0.400pt}}
\put(836.0,113.0){\rule[-0.200pt]{0.400pt}{111.778pt}}
\put(220.0,577.0){\rule[-0.200pt]{148.394pt}{0.400pt}}
\put(45,345){\makebox(0,0){$-a\,m_c$}}
\put(528,23){\makebox(0,0){$a$ (fm)}}
\put(547,232){\makebox(0,0)[l]{no T.I.}}
\put(374,510){\makebox(0,0)[l]{T.I. gluons}}
\put(547,355){\makebox(0,0)[l]{\shortstack{T.I. quarks\\and gluons}}}
\put(220.0,113.0){\rule[-0.200pt]{0.400pt}{111.778pt}}
\put(732,495){\circle*{18}}
\put(451,371){\circle*{18}}
\put(394,360){\circle*{18}}
\put(317,319){\circle*{18}}
\put(732.0,484.0){\rule[-0.200pt]{0.400pt}{5.059pt}}
\put(722.0,484.0){\rule[-0.200pt]{4.818pt}{0.400pt}}
\put(722.0,505.0){\rule[-0.200pt]{4.818pt}{0.400pt}}
\put(451.0,360.0){\rule[-0.200pt]{0.400pt}{5.059pt}}
\put(441.0,360.0){\rule[-0.200pt]{4.818pt}{0.400pt}}
\put(441.0,381.0){\rule[-0.200pt]{4.818pt}{0.400pt}}
\put(394.0,350.0){\rule[-0.200pt]{0.400pt}{5.059pt}}
\put(384.0,350.0){\rule[-0.200pt]{4.818pt}{0.400pt}}
\put(384.0,371.0){\rule[-0.200pt]{4.818pt}{0.400pt}}
\put(317.0,309.0){\rule[-0.200pt]{0.400pt}{5.059pt}}
\put(307.0,309.0){\rule[-0.200pt]{4.818pt}{0.400pt}}
\put(307.0,330.0){\rule[-0.200pt]{4.818pt}{0.400pt}}
\sbox{\plotpoint}{\rule[-0.500pt]{1.000pt}{1.000pt}}%
\put(732,194){\usebox{\plotpoint}}
\multiput(732,194)(-20.737,-0.886){14}{\usebox{\plotpoint}}
\multiput(451,182)(-20.727,-1.091){3}{\usebox{\plotpoint}}
\multiput(394,179)(-20.647,-2.118){2}{\usebox{\plotpoint}}
\multiput(355,175)(-20.691,-1.634){2}{\usebox{\plotpoint}}
\put(317,172){\usebox{\plotpoint}}
\put(732,531){\usebox{\plotpoint}}
\multiput(732,531)(-19.465,-7.204){15}{\usebox{\plotpoint}}
\multiput(451,427)(-19.008,-8.337){3}{\usebox{\plotpoint}}
\multiput(394,402)(-18.845,-8.698){2}{\usebox{\plotpoint}}
\multiput(355,384)(-19.129,-8.054){2}{\usebox{\plotpoint}}
\put(317,368){\usebox{\plotpoint}}
\put(732,474){\usebox{\plotpoint}}
\multiput(732,474)(-19.488,-7.143){15}{\usebox{\plotpoint}}
\multiput(451,371)(-19.476,-7.175){3}{\usebox{\plotpoint}}
\multiput(394,350)(-19.254,-7.751){4}{\usebox{\plotpoint}}
\put(317,319){\usebox{\plotpoint}}
\end{picture}
\end{center}
\caption{The critical bare quark mass for Wilson's lattice quark action
versus lattice spacing. Monte Carlo data points are compared with
perturbation theories in a theory with no tadpole improvement (T.I),
tadpole-improved gluon dynamics, and tadpole-improved quark and gluon
dynamics.}
\label{mc-fig}
\end{figure}

This same conclusion follows from Fig.~\ref{mc-fig} which shows
the value of the bare quark mass needed to obtain zero-mass pions using
Wilson's lattice action for quarks. This quantity diverges linearly
as the lattice spacing vanishes, and so should be quite perturbative. Here
we see dramatic improvements as the tadpoles are removed first from the
gluon action, through use of an improved coupling, and then also from the
quark action. 

The Creutz ratio and the critical quark mass are both very
similar to the couplings we need to compute for improved lagrangians.
Tadpole improvement has been very successful in a wide range of
applications.

Asymptotic freedom implies that short-distance QCD is simple
(perturbative) while long-distance QCD is difficult (nonperturbative).
The lattice separates short from long distances, allowing us to exploit
this dichotomy to create highly efficient algorithms for solving the
entire theory:
$p\!>\!\pi/a$~QCD is included via corrections~$\delta\Lag$ to the lattice
lagrangian that are computed using perturbation theory; $p\!<\!\pi/a$~QCD
is handled nonperturbatively using Monte Carlo integration. Thus, while we
wish to make the lattice spacing~$a$ as large as possible, we are
constrained by two requirements. First $a$~must be sufficiently small that
our finite-difference approximations for derivatives in the lagrangian and
field equations are sufficiently accurate. Second $a$~must be sufficiently
small that $\pi/a$~is a perturbative momentum. Numerical experiments
indicate that both constraints can be satisfied when $a\approx1/2$\,fm or
smaller, provided all lattice operators are tadpole improved.

\section{QCD on a Lattice}

\subsection{Classical Gluons\,\protect\cite{schlad}}

The continuum action for QCD is
\be
S = \int d^4x\,\half \sum_{\mu,\nu} \Tr \F\mu\nu^2(x)
\ee
where 
\be
\F\mu\nu \equiv \partial_\mu A_\nu - \partial_\nu A_\mu + ig [A_\mu,A_\nu] 
\ee
is the field tensor, a traceless $3\times3$ hermitian matrix. The defining
characteristic of the theory is its invariance with respect to gauge
transformations where
\be
\F\mu\nu \to \Omega(x)\,\F\mu\nu\,\Omega(x)^\dagger
\ee
and $\Omega(x)$ is an arbitrary $x$-dependent \su\ matrix.

The standard discretization of this theory seems perverse at first sight.
Rather than specifying the gauge field by the values of~$A_\mu(x)$ at the
sites of the lattice, the field is specified by variables on the links
joining the sites. In the classical theory, the ``link variable'' on
the link joining a site at~$x$ to one at $x+a\hat\mu$ is determined by the
line integral of $A_\mu$ along the link: 
\be
\U\mu(x) \equiv \P \exp\left(
-i\int_x^{x+a\hat\mu} gA\cdot \dd y \right)
\ee
where the $\P$-operator path-orders the $A_\mu$'s along the integration path.
We use~$U_\mu$'s in place of~$A_\mu$'s on
the lattice, because it is impossible to formulate a lattice version
of QCD directly in terms of~$A_\mu$'s that has exact gauge invariance. The
$U_\mu$'s, on the other hand, are \su\ matrices that transform very
simply under a gauge  transformation:
\be\label{gauge-tnfm}
\U\mu(x)\to\Omega(x)\,\U\mu(x)\,\Omega(x+a\hat\mu)^\dagger.
\ee
This makes it easy to build a discrete theory with exact gauge invariance.

A link variable~$\U\mu(x)$ is represented pictorially by a directed line
from~$x$ to~$x+\hat\mu$, where this line is the integration path for the
line integral in the exponent of~$\U\mu(x)$:
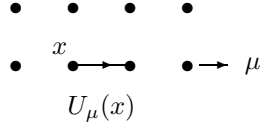
\begin{center}
\setlength{\unitlength}{.03in}
\begin{picture}(75,25)(0,15)
\multiput(25,25)(10,0){4}{\circle*{2}}
\multiput(25,35)(10,0){4}{\circle*{2}}

\put(35,25){\vector(1,0){7}}\put(35,25){\line(1,0){10}}
\put(40,20){\makebox(0,0)[t]{$\U\mu(x)$}}
\put(34,28){\makebox(0,0)[r]{$x$}}
\put(57,25){\vector(1,0){5}}
\put(65,25){\makebox(0,0)[l]{$\mu$}}
\end{picture}
\end{center}
In the conjugate matrix~$\Udag\mu(x)$ the direction of the line integral
is flipped and so we represent~$\Udag\mu(x)$ by a line going backwards
from~$x+\hat\mu$ to~$x$:
\begin{center}
\setlength{\unitlength}{.03in}
\begin{picture}(75,25)(0,15)
\multiput(25,25)(10,0){4}{\circle*{2}}
\multiput(25,35)(10,0){4}{\circle*{2}}
\put(45,25){\vector(-1,0){7}}\put(35,25){\line(1,0){10}}
\put(40,20){\makebox(0,0)[t]{$\Udag\mu(x)$}}
\put(34,28){\makebox(0,0)[r]{$x$}}
\put(57,25){\vector(1,0){5}}
\put(65,25){\makebox(0,0)[l]{$\mu$}}
\end{picture}
\end{center}
A Wilson loop function,
\be
W(\C) \equiv \third\Tr\P\e^{-\I \oint_\C gA\cdot \dd x},
\ee
for any closed path~$\C$ built of links on the lattice can be computed from the
path-ordered product of the~$\U\mu$'s  and~$\Udag\mu$'s associated with each
link. For example, if~$\C$ is the loop 
\begin{center}
\setlength{\unitlength}{.03in}
\begin{picture}(40,40)(25,20)
\multiput(25,25)(10,0){4}{\circle*{2}}
\multiput(25,35)(10,0){4}{\circle*{2}}
\multiput(25,45)(10,0){4}{\circle*{2}}
\multiput(25,55)(10,0){4}{\circle*{2}}
\put(35,25){\vector(1,0){7}}\put(35,25){\line(1,0){10}}
\put(45,25){\vector(0,1){7}}\put(45,25){\line(0,1){10}}
\put(45,35){\vector(1,0){7}}\put(45,35){\line(1,0){10}}
\put(55,35){\vector(0,1){7}}\put(55,35){\line(0,1){10}}
\put(55,45){\vector(-1,0){7}}\put(55,45){\line(-1,0){10}}
\put(45,45){\vector(-1,0){7}}\put(45,45){\line(-1,0){10}}
\put(35,45){\vector(0,-1){7}}\put(35,45){\line(0,-1){10}}
\put(35,35){\vector(0,-1){7}}\put(35,35){\line(0,-1){10}}
\put(34,22){\makebox(0,0)[r]{$x$}}
\put(60,25){\vector(1,0){5}}\put(67,25){\makebox(0,0)[l]{$\mu$}}
\put(60,27){\vector(0,1){5}}\put(60,34){\makebox(0,0)[b]{$\nu$}}
\end{picture}
\end{center}
then
\be
W(\C) = \third\Tr\!\left(\U\mu(x)\U\nu(x+a\hat\mu)\ldots\Udag\nu(x)\right).
\ee
Such quantities are obviously invariant under arbitrary
gauge transformations~\eq{gauge-tnfm}.

One might wonder why we go to so much trouble to preserve gauge invariance
when we quite willing give up Lorentz invariance, rotation
invariance, etc. The reason is quite practical. With gauge invariance, the
quark-gluon, three-gluon, and four-gluon couplings in QCD are all equal,
and the bare gluon mass is zero. Without gauge invariance, each of these
couplings must be tuned independently and a gluon mass introduced if one is
to recover QCD. Tuning this many parameters in a numerical simulation is
very expensive. This is not much of a problem in the classical theory,
where approximate gauge invariance keeps the couplings approximately equal;
but it is serious in the quantum theory because quantum fluctuations
(loop-effects) renormalize the various couplings differently in the absence
of exact gauge invariance. So while it is quite possible to formulate
lattice QCD directly in terms of~$\A\mu$'s, the resulting theory would have
only approximate gauge invariance, and thus would be prohibitively expensive
to simulate. Symmetries like Lorentz invariance can be given up with little
cost because the symmetries of the lattice, though far less restrictive, are
still sufficient to prevent the introduction of new interactions with new
couplings (at least to lowest order in~$a$).

We must now build a lattice lagrangian from the link operators. We require
that the lagrangian be gauge invariant, local, and symmetric with respect
to axis interchanges (which is all that is left of Lorentz invariance). The
most local nontrivial gauge invariant object one can build from the link
operators is the ``plaquette operator,'' which involves the product of link
variables around the smallest square at site~$x$ in the $\mu\nu$~plane:
\be
\Pl\mu\nu(x) \equiv \third\Re\Tr\!\left(\U\mu(x)\U\nu(x+a\hat\mu)
\Udag\mu(x+a\hat\mu+a\hat\nu)\Udag\nu(x)\right).
\ee
To see what this object is, consider evaluating the plaquette centered
about a point~$x_0$ for a very smooth weak classical $\A\mu$~field.
In this limit, 
\be \label{uplaqa}
\Pl\mu\nu \approx 1
\ee
since
\be
\U\mu \approx \e^{-\I ga\A\mu} \approx 1 .
\ee
Given that $\A\mu$~is slowly varying, its value anywhere on the plaquette
should be accurately specified by its value and derivatives at~$x_0$. Thus
the corrections to \eq{uplaqa} should be a polynomial in~$a$ with
coefficients formed from gauge-invariant combinations of~$\A\mu(x_0)$ and
its derivatives: that is,
\begin{eqnarray}
\Pl\mu\nu \,= \, 1 
&-&c_1\,a^4\,\Tr\!\left(g\F\mu\nu(x_0)\right)^2 \nl 
&-&c_2\,a^6\,\Tr\!\left(g\F\mu\nu(x_0)(\D_\mu^2+\D_\nu^2)g\F\mu\nu(x_0)
 \right)\nl
&+&\order(a^8) 
\label{ope}
\end{eqnarray}
where $c_1$ and $c_2$ are constants, and $\D_\mu$ is the
gauge-covariant derivative.  The leading correction is
order~$a^4$ because $\F\mu\nu^2$ is the lowest-dimension gauge-invariant
combination of derivatives of $\A\mu$, and it has dimension~4. 
(There are no $F^3$~terms because
$\Pl\mu\nu$~is invariant under $\U\mu\to\Udag\mu$ or,
equivalently, $F\to -F$.)

It is straightforward to find the coefficients~$c_1$ and~$c_2$. We need
only examine terms in the expansion of~$\Pl\mu\nu$ that are quadratic
in~$A_\mu$; the cubic and quartic parts of $\F\mu\mu^2$ then follow
automatically, by gauge invariance. Because of the trace, the path
ordering is irrelevant to this order. Thus
\bearray
\Pl\mu\nu &=&  \third\Re\Tr\P\e^{-\I \oint_\Box gA\cdot \dd x} \nl
&=&\third\Re\Tr\left[1 - i\oint_\Box gA\cdot \dd x -\half\left(\oint_\Box
gA\cdot \dd x\right)^2 + \order(A^3)\right]
\eearray
where, by Stoke's Theorem,
\bearray
\oint_\Box  A\cdot \dd x &=& \int_{-a/2}^{a/2}\dd x_\mu \dd x_\nu\left[\partial_\mu
A_\nu(x_0+x) - \partial_\nu A_\mu(x_0+x)\right] \nl
&=& \int_{-a/2}^{a/2}\dd x_\mu \dd x_\nu \left[\F\mu\nu(x_0) +(x_\mu\D_\mu 
+x_\nu\D_\mu)\F\mu\nu(x_0) +\cdots\right] \nl
&=& a^2\,\F\mu\nu(x_0)  + \frac{a^4}{24}\,(\D_\mu^2+\D_\nu^2)\F\mu\nu(x_0)
+\order(a^6,A^2).
\eearray
Thus~$c_1 = 1/6$ and~$c_2=1/72$ in~\eq{ope}.

The expansion in \eq{ope} is the classical analogue of an operator product
expansion. Using this expansion, we find that the traditional ``Wilson
action'' for gluons on a lattice,
\be\label{Wil-S}
S_{\rm Wil} = \beta\,\sum_{x,\mu>\nu} \left(1-\Pl\mu\nu(x)\right)
\ee
where $\beta\!=\!6/g^2$, has the correct limit for small lattice spacing
up to corrections of order~$a^2$:
\be \label{Wil-cont}
S_{\rm Wil} = \int d^4x\,\sum_{\mu,\nu} \left\{\half\Tr \F\mu\nu^2
+ \frac{a^2}{24}\Tr\F\mu\nu(\D_\mu^2+\D_\nu^2)\F\mu\nu +\cdots\right\}.
\ee

We can cancel the $a^2$~error in the Wilson action by adding other Wilson
loops. For example, the~$2a\times a$``rectangle operator''
\be
\Rt\mu\nu = \third\Re\Tr\!
\setlength{\unitlength}{.015in}
\begin{picture}(75,13)(0,17)
  \put(10,10){\vector(0,1){12.5}}
  \put(10,10){\line(0,1){20}}
  \put(10,30){\vector(1,0){12.5}}
  \put(10,30){\vector(1,0){32.5}}
  \put(10,30){\line(1,0){40}}
  \put(50,30){\vector(0,-1){12.5}}
  \put(50,30){\line(0,-1){20}}
  \put(50,10){\vector(-1,0){12.5}}
  \put(50,10){\vector(-1,0){32.5}}
  \put(50,10){\line(-1,0){40}}
  \put(61,10){\vector(1,0){10}}\put(75,10){\makebox(0,0){$\mu$}}
  \put(60,11){\vector(0,1){10}}\put(60,25){\makebox(0,0){$\nu$}}
\end{picture}
\ee
has expansion
\be
\Rt\mu\nu = 1 -\frac{4}{6}\,a^4\Tr(g\F\mu\nu)^2
-\frac{4}{72}\,a^6\Tr\left(g\F\mu\nu(4\,\D_\mu^2+\D_\nu^2)g\F\mu\nu\right)
-\cdots .
\ee
The mix of $a^4$~terms and $a^6$~terms in the rectangle is different from
that in the plaquette. Therefore we can combine the two operators to
obtain an improved classical lattice action that is accurate up
to~$\order(a^4)$\,\cite{curci83,luscher85a}:
\bearray
S_{\rm classical} &\equiv& -\beta \sum_{x,\mu>\nu}\left\{
\frac{5\Pl\mu\nu}{3} - \frac{\Rt\mu\nu+\Rt\nu\mu}{12}\right\}  + {\rm const}
\\
&=& \int d^4x\,\sum_{\mu,\nu}\half\Tr\F\mu\nu^2 + \order(a^4).
\eearray
This process is the analogue of improving the derivatives in
discretizations of 
non-gauge theories.\footnote{An important step that I have not
discussed is to show that the gluon action is positive for any
configuration of link variables\cite{luscher85a}. This guarantees that
the classical 
ground state of the lattice action corresponds to
$\F\mu\nu\!=\!0$.}

\begin{exercise} Defining the ``twisted-rectangle operator''
\be
T_{\mu\nu} = \third\Re\Tr\!
\setlength{\unitlength}{.015in}
\begin{picture}(60,13)(0,17)
% Left staple:
% Go left:
  \put(27.5,10){\vector(-1,0){10}}
  \put(27.5,10){\line(-1,0){17.5}}
% up:
  \put(10,10){\vector(0,1){12.5}}
  \put(10,10){\line(0,1){20}}
% right:
  \put(10,30){\vector(1,0){12.5}}
  \put(10,30){\line(1,0){17.5}}

% Right staple:
% Go right:
  \put(32.5,10){\vector(1,0){10}}
  \put(32.5,10){\line(1,0){17.5}}
% up: 
  \put(50,10){\vector(0,1){12.5}}
  \put(50,10){\line(0,1){20}}
% left:
  \put(50,30){\vector(-1,0){12.5}}
  \put(50,30){\line(-1,0){17.5}}

% Diagonal struts:
% bottom right to top left strut:
  \put(32.5,10){\line(-1,4){5.0}}
% positive slope strut:
  \put(27.5,10){\line( 1,4){5.0}}
\end{picture}.
\ee
show that
\bearray \label{trt-S}
S_{\rm classical}  &\equiv& -\beta \sum_{x,\mu>\nu}
\left\{\Pl\mu\nu + \frac{T_{\mu\nu}+T_{\nu\mu}}{12}\right\}
+{\rm const}
\\
&=& \int d^4x\,\sum_{\mu,\nu}\half\Tr\F\mu\nu^2 + \order(a^4).
\eearray
This is an alternative to the improved gluon action derived in the previous
exercise.
\end{exercise}

\subsection{Quantum Gluons\protect\footnotemark}
\footnotetext{This section is based on work with M.\,Alford,
W.\,Dimm, G.\,Hockney and P.\,Mackenzie that is described
in~\cite{alford95}.} 
In the previous section we derived improved classical actions for gluons
that are accurate through order~$a^4$. We now turn these into quantum
actions. The most important step is to tadpole improve the action by
dividing each link operator~$\U\mu$ by the mean link~$u_0$: for
example, the action built of plaquette and rectangle operators becomes
\be \label{imp-S}
S = -\beta\sum_{x,\mu>\nu}\left\{\frac{5}{3}\frac{\Pl\mu\nu}{u_0^4}
-\frac{\Rt\mu\nu+\Rt\nu\mu}{12\,u_0^6}\right\}.
\ee
The $u_0$'s cancel lattice tadpole contributions that
otherwise would spoil weak-coupling perturbation theory in the lattice theory
and undermine our procedure for improving the lattice discretization.
Note that $u_0\!\approx\!3/4$ when $a\!=\!.4$\,fm, and therefore the
relative importance of the~$\Rt\mu\nu$'s is larger by a
factor of~$1/u_0^2\!\approx\!2$ than without tadpole improvement. 	
Without tadpole improvement, we cancel only half of the $a^2$ error.

The mean link~$u_0$ is computed numerically by guessing a value for use in
the action,  measuring the mean link in a simulation, and then readjusting
the value used in the action accordingly. This tuning cycle converges
rapidly to selfconsistent values, and can be done very quickly using small
lattice volumes. The $u_0$'s depend only on lattice spacing, and
become equal to one as the lattice spacing vanishes.

The expectation value of the link operator is gauge dependent. Thus to
minimize gauge artifacts, $u_0$ is commonly defined  as the
Landau-gauge expectation value, $\langle0|\third\Tr\U\mu|0\rangle_{\rm LG}$.
Landau gauge is the axis-symmetric gauge that maximizes $u_0$, thereby
minimizing the tadpole contribution; any tadpole contribution that is left
in Landau gauge cannot be a gauge artifact. An alternative procedure is to
define
$u_0$ as the fourth root of the plaquette expectation value,
\be
u_0 = \langle 0|\Pl\mu\nu |0\rangle^{1/4}.
\ee
This definition gives almost identical results and is more convenient for
numerical work since gauge fixing is unnecessary.

Tadpole improvement is the first step in a systematic procedure for
improving the action. The next step is to add in renormalizations due to
contributions from $k\!>\!\pi/a$~physics not already included in the
tadpole improvement. These renormalizations induce
$a^2\,\alpha_s(\pi/a)$~corrections,
\bearray
\delta\Lag &=& 
\alpha_s\,r_1\,a^2\sum_{\mu,\nu}\Tr\!(\F\mu\nu\D_\mu^2\F\mu\nu)
\nl
&+&\alpha_s\,r_2\,a^2\sum_{\mu,\nu}\Tr\!(\D_\mu\F\nu\sigma
\D_\mu\F\nu\sigma)
\nl
&+&\alpha_s\,r_3\,a^2\sum_{\mu,\nu}\Tr\!(\D_\mu\F\mu\sigma
\D_\nu\F\nu\sigma)
\nl
&+&\cdots,
\eearray
that must be removed. The last term is harmless; its coefficient can be set
to zero by a change of field variable (in the path integral) of the form
\be \label{Atransform}
A_\mu \to A_\mu + a^2\,\alpha_s\,f(\alpha_s)\,\sum_\nu \D_\nu F_{\nu\mu}.
\ee
Since changing integration variables does not change the value of an integral,
such field transformations must leave the physics unchanged.\footnote{One
must, of course, include the jacobian for the transformation in the transformed
path integral. This contributes only in one-loop order and higher; it has no
effect on tree-level calculations.} Operators that can be removed by a field
transformation are called ``redundant.'' The other corrections are removed by
renormalizing the coefficient of the rectangle operator~$\Rt\mu\nu$ in the
action, and by adding an additional operator. One choice for the extra operator
is 
\be
C_{\mu\nu\sigma}  \equiv \third\Re\Tr\!\!
\setlength{\unitlength}{.015in}
\begin{picture}(60,20)(0,17)
  \put(10,10){\vector(0,1){12.5}}
  \put(10,10){\line(0,1){20}}
  \put(10,30){\vector(2,1){10}}
  \put(10,30){\line(2,1){15}}
  \put(25.2,37.6){\vector(1,0){12.5}}
  \put(25.2,37.6){\line(1,0){20}}
  \put(45.2,37.6){\vector(0,-1){12.5}}
  \put(45.2,37.6){\line(0,-1){20}}
  \put(45.2,17.6){\vector(-2,-1){10}}
  \put(45.2,17.6){\line(-2,-1){15}}
  \put(30,10){\vector(-1,0){12.5}}
  \put(30,10){\line(-1,0){20}}
\end{picture}.
\ee
Then the action, correct up to~$\order(a^2\alpha_s^2,a^4)$,
is\,\cite{luscher85b}  
\be \label{LW-S}
S = -\beta\sum_{x,\mu>\nu}\left\{\frac{5}{3}\frac{\Pl\mu\nu}{u_0^4}
-r_{\rm g}\,\frac{\Rt\mu\nu+\Rt\nu\mu}{12\,u_0^6}\right\}
+c_{\rm g}\,\beta\sum_{x,\mu>\nu>\sigma} \frac{C_{\mu\nu\sigma}}{u_0^6},
\ee
where
\bearray
r_{\rm g} &=& 1 + .48\,\alpha_s(\pi/a) \\
c_{\rm g} &=& .055\,\alpha_s(\pi/a).
\eearray

The coefficients~$r_{\rm g}$ and~$c_{\rm g}$ are computed by ``matching'' physical
quantities, like low-energy scattering amplitudes, computed using
perturbation theory in the lattice theory with the analogous quantity in
the continuum theory. The lattice result depends upon~$r_{\rm g}$ and~$c_{\rm g}$;
these parameters are tuned until the lattice amplitude agrees with the
continuum amplitude to the order in~$a$ and~$\alpha_s$ required:
\be
T_{\rm lat}(r_{\rm g},c_{\rm g}) = T_{\rm contin}.
\ee
Note that tadpole improvement has a big effect on these coefficients.
Without tadpole improvement, $r_{\rm g}=1+2\alpha_s$; that is, the coefficient of
the radiative correction is four times larger. 
Tadpole improvement automatically supplies 75\% of the one-loop
contribution needed without improvement.
Since $\alpha_s\!\approx\!0.3$, the unimproved expansion for~$r_{\rm g}$ is not
particularly convergent. However, with tadpole improvement, the one-loop
correction is only about 10--20\% of~$r_{\rm g}$. Indeed, for most
current applications, one-loop corrections to tadpole-improved actions
are negligible.

\begin{exercise} Show that the gauge that maximizes 
$\langle0|\third\Tr\U\mu|0\rangle$ becomes Landau gauge ($\partial\cdot
A\!=\!0$) in the $a\!\to\!0$~limit.
\end{exercise}

\subsection{Monte Carlo Evaluation of Gluonic Path Integrals}
A computer code for the Monte Carlo evaluation of gluonic path
integrals can be designed in close analogy with our code for
one-dimensional quantum mechanics. The Metropolis algorithm for
generating random configurations must be adapted to work with \su\
matrices. In our quantum mechanics example, the coordinate was updated
by adding a random number. In QCD the gluon field is specified by 
link variables $\U\mu(x)$, which are exponentials of the
fundamental field. Thus to update a link variable we must multiply by the
exponential of a random field; that is we must multiply by a random
\su\ matrix~$M$:
\be
\U\mu \to M\,\U\mu
\ee
Typically the matrix $M$ is chosen randomly from a set of 50 or~100
random \su\ matrices that is generated once, at the start of the
simulation; the only restrictions on this set are that it be large
enough so that products of the various $M$'s cover the entire space of
\su\ matrices, and that the inverse, $M^\dagger$, for each matrix~$M$
in the set also be included in the set. The $M$'s can be generated by
first creating a set of hermitian matrices~$H$ whose matrix elements
are random numbers between $-1$ and~$1$. These are converted to \su\
matrices by forming $1+i\epsilon H$ and unitarizing it.\footnote{To
convert an arbitrary matrix 
$M=(m_1\,m_2\,m_3)$ into an \su\ matrix: first normalize the
first column~$m_1$ to unity, then make the second column~$m_2$
orthogonal to the (new) first column and normalize it, and replace the
third column by the cross product of the (new) first two columns.}
Parameter~$\epsilon$ determines the size of the update; as before, it
is adjusted so that roughly half of all trial updates are accepted.

A second modification of the Metropolis algorithm that is
useful for QCD is to update each link variable
several times (rather than just once) before moving on to the next
variable in a single sweep through the lattice. This allows the link
variable to come into statistical equilibrium with its immediate
neighbors on the lattice. The additional cost for these extra updates
is relatively small because standard gluon actions are linear in the
link variables. Thus the part of the action that must be computed when
updating a particular~$\U\mu(x)$ can be written 
\be
\Delta S(x,\mu) = \Re\Tr(\U\mu(x)\Gamma_\mu(x)) 
\ee
where $\Gamma_\mu(x)$, which is a sum of products of the link
variables, is independent of~$\U\mu(x)$. Computing~$\Gamma_\mu(x)$ is
the most expensive part of the Metropolis update, but it need be
computed only once for each set of successive updates of
$\U\mu(x)$. Typically one does about 10 ``hits'' of the Metropolis
algorithm before moving on the the next link variable.

\begin{exercise} Design a computer code for evaluating gluonic path
integrals using the Metropolis algorithm. Do this first for the
simplest lattice action, the Wilson action
(\eq{Wil-S}):
\be
S_{\rm Wil} = -\tilde{\beta}\,\sum_{x,\mu>\nu} \frac{\Pl\mu\nu(x)}{u_0^4}
\ee
Run simulations at $\beta\equiv\tilde{\beta}/u_0^4 = 5.5$, which
corresponds to a lattice spacing of around 0.25\,fm. The lattice
volume should be of order 2\,fm on a side for typical QCD simulations;
use $L/a=8$ points on a side in your simulation. Set the Metropolis
step size $\epsilon=0.24$ and omit $\Ncorr=50$ sweeps between
Monte Carlo measurements. Compute Monte Carlo averages of $a\times a$
and $a\times 2a$ Wilson loops; you should obtain about 0.50 and 0.26,
respectively.

Also try the improved action, \eq{imp-S}. Use $\beta=1.719$ and
$u_0=0.797$ to again obtain $a\approx0.25$\,fm\cite{alford}. 
The $a\times a$ and
$a\times 2a$ Wilson loops have values of 0.54 and 0.28,
respectively. (Wilson loops are unrenormalized and so these
values need not agree with those from the Wilson action.)
\end{exercise}

\subsection{A First Simulation}

Perhaps the simplest physics calculation that involves just gluons is
to compute the potential energy between a static quark and a static
antiquark separated by a distance~$r$. This ``static potential'' should be
approximately Coulombic at short distances, but grow linearly at large
distances, demonstrating quark confinement. It can be used in a
Schr\"odinger equation to predict energy levels for the $\psi/J$ and
$\Upsilon$~families of mesons.

The euclidean Green's function or propagator~$G$ for a heavy nonrelativistic
quark in a background gauge field~$A_\mu$ satisfies the equation
\be
\left(\D_t-\frac{\Dv^2}{2M}\right)G(x) = \delta^4(x)
\ee
where $\D_\mu=\partial_\mu - igA_\mu(x)$ is the gauge-covariant
derivative. This equation is easily solved in the static-quark limit,
where $M\to\infty$, to obtain
\be
G_\infty(\xv,t) = \left[\P\e^{-\I \int^t_0
gA_0(\xv,t)\dd t}\right]^\dagger\,\delta^3(\xv),
\ee
which on the lattice becomes
\be
G_\infty(\xv,t) = \Udag t(\xv,t\!-\!a)\,\Udag t(\xv,t\!-\! 2a)\,\ldots\,
\Udag t(\xv,0).
\ee
Propagation of a static antiquark is described by~$G_\infty^\dagger$.
Therefore we obtain the static potential~$V(r)$, which is the
energy of a static 
quark and antiquark a distance~$r$ apart, from expectation values
of $r\times t$~Wilson loops:
\be
W(r,t) \equiv \langle 0| \third\Tr
\setlength{\unitlength}{.02in}
\begin{picture}(50,12)(-5,18)
%\multiput(0,0)(10,0){5}{\circle*{2}}
\multiput(0,10)(10,0){5}{\circle*{2}}
\multiput(0,20)(10,0){5}{\circle*{2}}
\multiput(0,30)(10,0){5}{\circle*{2}}

\multiput(0,10)(10,0){4}{\vector(1,0){7}}
\multiput(0,10)(10,0){4}{\line(1,0){10}}

\multiput(40,10)(0,10){2}{\vector(0,1){7}}
\multiput(40,10)(0,10){2}{\line(0,1){10}}

\multiput(40,30)(-10,0){4}{\vector(-1,0){7}}
\multiput(40,30)(-10,0){4}{\line(-1,0){10}}

\multiput(0,30)(0,-10){2}{\vector(0,-1){7}}
\multiput(0,30)(0,-10){2}{\line(0,-1){10}}

\put(20,0){\makebox(0,0){$t$}}\put(17,0){\vector(-1,0){17}}
\put(23,0){\vector(1,0){17}}

\put(60,20){\makebox(0,0){$r$}}\put(60,23){\vector(0,1){7}}
\put(60,17){\vector(0,-1){7}}

\end{picture}
| 0 \rangle\rule[-.5in]{0pt}{1in}
\ee
where for large~$t$
\be \label{static-V}
W(r,t)\to {\rm const}\,\,\e^{-V(r)\,t} .
\ee
Thus, to calculate the static potential~$V(r)$, we
compute $W(r,t)$ for a variety of $t$'s and then take the large-$t$ limit 
\be
W(r,t)/W(r,t+a) \to  a\,V(r).
\ee

There is one modification of this procedure that greatly improves the
results. This is to replace the spatial link matrices, in the $r$~direction,
by ``smeared'' link variables~$\tilde{U}_\mu(x)$:
\be
\tilde{U}_\mu(x) \equiv
\left(1+\epsilon a^2\Delta^{(2)}\right)^n\,\U\mu(x)
\ee
where 
\be
\Delta^{(2)}\U\mu(x) \equiv \sum_\rho \Delta^{(2)}_\rho \U\mu(x)
\ee
and $\Delta^{(2)}_\rho$ is a discretized, gauge-covariant derivative:
\bearray
\Delta^{(2)}_\rho\,U_\mu(x) &\equiv&
\frac{1}{u_0^2\,a^2} \left(
U_\rho(x)\,U_\mu(x+a\hat\rho)\,U_\rho^\dagger(x+a\hat\mu) 
- 2\,u_0^2\,U_\mu(x) \right. \nl
&&\quad\quad + \left.U_\rho^\dagger(x-a\hat\rho)\,U_\mu(x-a\hat\rho)
\,U_\rho(x-a\hat\rho+a\hat\mu)\right).
\eearray
The low-momentum components of the smeared link variable are unchanged
by the smearing, up to corrections of order $a^2p^2$; but at high
momentum the smearing acts as an ultraviolet cutoff:
\be
\left(1+\epsilon a^2\Delta^{(2)}\right)^n
\to \left(1 -\epsilon a^2 p^2\right)^n
\approx \e^{-\epsilon a^2 p^2 n}.
\ee
By suppressing high-momentum gluons in the initial and final states of
our matrix element, we suppress contributions from gluonic excitations
and thereby hasten the convergence of $W(r,t)/W(r,t+a)$ to
its asymptotic value. This significantly reduces the Monte Carlo
statistical errors since these are 
generally much smaller at smaller~$t$'s.

\begin{exercise} Add measurement code for $W(r,t)$ to your gluon Monte
Carlo program. Run this code for the parameter sets given above and
extract the static potential. Try this with unsmeared spatial
$\U\mu$'s on the ends of the Wilson loop, and again with smeared links
(try $n=4$ smearings with $\epsilon=1/12$). In the smeared case 
asymptotic results appear at the first or second time step

My results for the Wilson and improved actions are shown in
Fig.~\ref{v}. Each run took about 3~hours on a 300\,MHz personal
computer (with 64\,MB of RAM). The potential was computed for quarks
separated along the lattice axes, as well as for separations along
diagonal directions on the lattice (using nonplanar Wilson loops).
The plots clearly show the linear rise
of the potential at large~$r$, which results in quark confinement; 
and careful examination shows the
onset of Coulombic behavior at small~$r$, which is due to asymptotic
freedom. 
The Monte Carlo results are compared in the figure with global fits
to the form 
\be
V(r) = \sigma r - b/r +c
\ee
where $\sigma$ is the string tension. This parameterization works well
for the range of $r$'s we are considering (0.25\,fm--1\,fm). Note that
the Monte Carlo data is not as smooth in the Wilson case as it is for
the improved action. This is caused the $\order(a^2)$~errors in
the Wilson action, coming from the second term in \eq{Wil-cont}. This
term breaks rotational invariance; it is the leading manifestation in
the gluon dynamics of the cubic structure of our lattice. While the
true potential is a function only of the magnitude of the
quark separation, 
we expect differences on the lattice between potentials for
separations that are along the grid axes and those for separations
along diagonals. This is particularly apparent 
at $r=3a$ where there are two Monte Carlo points, one for separation
$(3a,0,0)$ and the other for $(2a,2a,a)$. The two points are clearly
separated on the plot for the Wilson action. 
The $a^2$~errors are removed from the improved action; the two $r=3a$
points merge on the plot for that action. The difference
$a(V(2a,2a,a)-V(3a,0,0))$ 
is reduced from $0.065\pm0.007$ with the Wilson action
to $0.003\pm0.006$ with the improved action in these
particular simulations. The $a^2$ errors are larger at
smaller~$r$'s, but more difficult to quantify there.
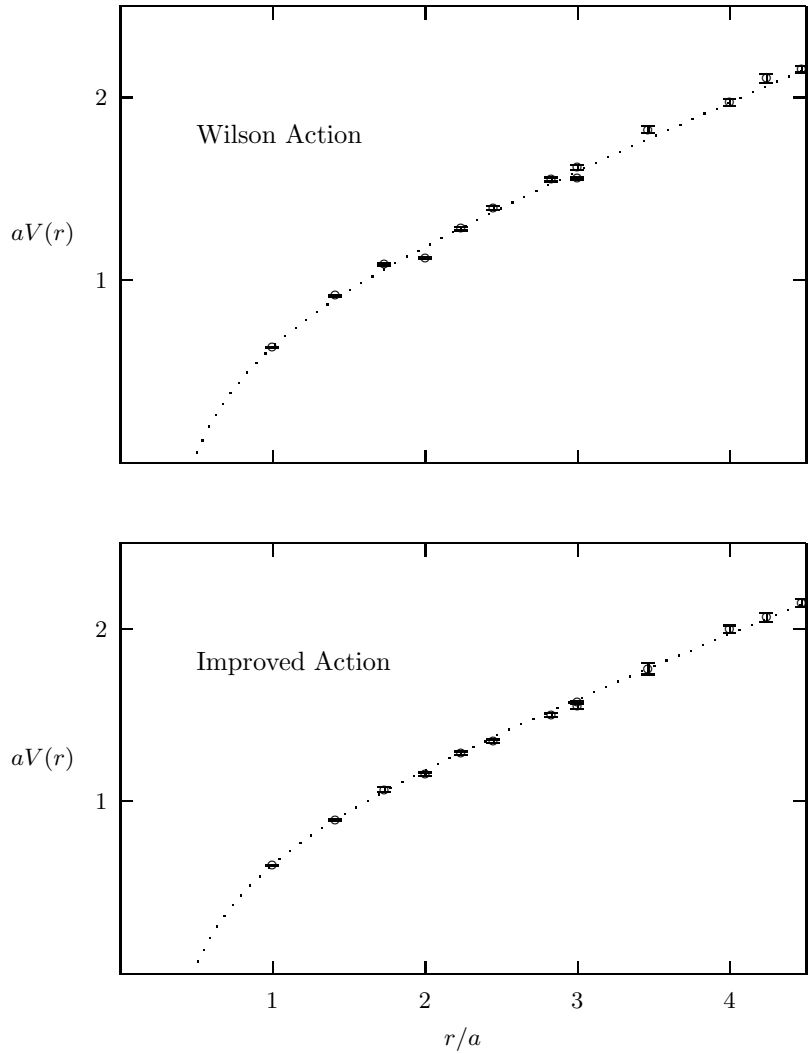
\begin{figure}
\begin{center}
% GNUPLOT: LaTeX picture
\setlength{\unitlength}{0.240900pt}
\ifx\plotpoint\undefined\newsavebox{\plotpoint}\fi
\sbox{\plotpoint}{\rule[-0.200pt]{0.400pt}{0.400pt}}%
\begin{picture}(1259,840)(0,0)
\font\gnuplot=cmr10 at 10pt
\gnuplot
\sbox{\plotpoint}{\rule[-0.200pt]{0.400pt}{0.400pt}}%
\put(161.0,369.0){\rule[-0.200pt]{4.818pt}{0.400pt}}
\put(141,369){\makebox(0,0)[r]{$1$}}
\put(1218.0,369.0){\rule[-0.200pt]{4.818pt}{0.400pt}}
\put(161.0,656.0){\rule[-0.200pt]{4.818pt}{0.400pt}}
\put(141,656){\makebox(0,0)[r]{$2$}}
\put(1218.0,656.0){\rule[-0.200pt]{4.818pt}{0.400pt}}
\put(400.0,82.0){\rule[-0.200pt]{0.400pt}{4.818pt}}
\put(400.0,780.0){\rule[-0.200pt]{0.400pt}{4.818pt}}
\put(640.0,82.0){\rule[-0.200pt]{0.400pt}{4.818pt}}
\put(640.0,780.0){\rule[-0.200pt]{0.400pt}{4.818pt}}
\put(879.0,82.0){\rule[-0.200pt]{0.400pt}{4.818pt}}
\put(879.0,780.0){\rule[-0.200pt]{0.400pt}{4.818pt}}
\put(1118.0,82.0){\rule[-0.200pt]{0.400pt}{4.818pt}}
\put(1118.0,780.0){\rule[-0.200pt]{0.400pt}{4.818pt}}
\put(161.0,82.0){\rule[-0.200pt]{259.449pt}{0.400pt}}
\put(1238.0,82.0){\rule[-0.200pt]{0.400pt}{172.966pt}}
\put(161.0,800.0){\rule[-0.200pt]{259.449pt}{0.400pt}}
\put(40,441){\makebox(0,0){$aV(r)$}}
\put(281,599){\makebox(0,0)[l]{Wilson Action}}
\put(161.0,82.0){\rule[-0.200pt]{0.400pt}{172.966pt}}
\put(400.0,263.0){\usebox{\plotpoint}}
\put(390.0,263.0){\rule[-0.200pt]{4.818pt}{0.400pt}}
\put(390.0,264.0){\rule[-0.200pt]{4.818pt}{0.400pt}}
\put(499.0,343.0){\rule[-0.200pt]{0.400pt}{0.723pt}}
\put(489.0,343.0){\rule[-0.200pt]{4.818pt}{0.400pt}}
\put(489.0,346.0){\rule[-0.200pt]{4.818pt}{0.400pt}}
\put(576.0,392.0){\rule[-0.200pt]{0.400pt}{0.964pt}}
\put(566.0,392.0){\rule[-0.200pt]{4.818pt}{0.400pt}}
\put(566.0,396.0){\rule[-0.200pt]{4.818pt}{0.400pt}}
\put(640.0,402.0){\rule[-0.200pt]{0.400pt}{0.723pt}}
\put(630.0,402.0){\rule[-0.200pt]{4.818pt}{0.400pt}}
\put(630.0,405.0){\rule[-0.200pt]{4.818pt}{0.400pt}}
\put(696.0,447.0){\rule[-0.200pt]{0.400pt}{1.204pt}}
\put(686.0,447.0){\rule[-0.200pt]{4.818pt}{0.400pt}}
\put(686.0,452.0){\rule[-0.200pt]{4.818pt}{0.400pt}}
\put(747.0,479.0){\rule[-0.200pt]{0.400pt}{1.445pt}}
\put(737.0,479.0){\rule[-0.200pt]{4.818pt}{0.400pt}}
\put(737.0,485.0){\rule[-0.200pt]{4.818pt}{0.400pt}}
\put(838.0,524.0){\rule[-0.200pt]{0.400pt}{1.445pt}}
\put(828.0,524.0){\rule[-0.200pt]{4.818pt}{0.400pt}}
\put(828.0,530.0){\rule[-0.200pt]{4.818pt}{0.400pt}}
\put(879.0,542.0){\rule[-0.200pt]{0.400pt}{1.927pt}}
\put(869.0,542.0){\rule[-0.200pt]{4.818pt}{0.400pt}}
\put(869.0,550.0){\rule[-0.200pt]{4.818pt}{0.400pt}}
\put(879.0,527.0){\rule[-0.200pt]{0.400pt}{0.964pt}}
\put(869.0,527.0){\rule[-0.200pt]{4.818pt}{0.400pt}}
\put(869.0,531.0){\rule[-0.200pt]{4.818pt}{0.400pt}}
\put(990.0,600.0){\rule[-0.200pt]{0.400pt}{2.650pt}}
\put(980.0,600.0){\rule[-0.200pt]{4.818pt}{0.400pt}}
\put(980.0,611.0){\rule[-0.200pt]{4.818pt}{0.400pt}}
\put(1118.0,642.0){\rule[-0.200pt]{0.400pt}{2.891pt}}
\put(1108.0,642.0){\rule[-0.200pt]{4.818pt}{0.400pt}}
\put(1108.0,654.0){\rule[-0.200pt]{4.818pt}{0.400pt}}
\put(1176.0,679.0){\rule[-0.200pt]{0.400pt}{3.373pt}}
\put(1166.0,679.0){\rule[-0.200pt]{4.818pt}{0.400pt}}
\put(1166.0,693.0){\rule[-0.200pt]{4.818pt}{0.400pt}}
\put(1231.0,695.0){\rule[-0.200pt]{0.400pt}{2.409pt}}
\put(1221.0,695.0){\rule[-0.200pt]{4.818pt}{0.400pt}}
\put(400,264){\circle{12}}
\put(499,345){\circle{12}}
\put(576,394){\circle{12}}
\put(640,403){\circle{12}}
\put(696,450){\circle{12}}
\put(747,482){\circle{12}}
\put(838,527){\circle{12}}
\put(879,546){\circle{12}}
\put(879,529){\circle{12}}
\put(990,605){\circle{12}}
\put(1118,648){\circle{12}}
\put(1176,686){\circle{12}}
\put(1231,700){\circle{12}}
\put(1221.0,705.0){\rule[-0.200pt]{4.818pt}{0.400pt}}
\put(275.00,82.00){\usebox{\plotpoint}}
\multiput(281,98)(8.648,18.868){2}{\usebox{\plotpoint}}
\put(300.07,138.95){\usebox{\plotpoint}}
\put(310.49,156.89){\usebox{\plotpoint}}
\put(321.66,174.38){\usebox{\plotpoint}}
\put(333.72,191.26){\usebox{\plotpoint}}
\put(346.51,207.60){\usebox{\plotpoint}}
\put(360.05,223.33){\usebox{\plotpoint}}
\put(374.36,238.36){\usebox{\plotpoint}}
\put(388.55,253.50){\usebox{\plotpoint}}
\put(403.86,267.51){\usebox{\plotpoint}}
\put(419.60,281.03){\usebox{\plotpoint}}
\put(435.78,294.02){\usebox{\plotpoint}}
\put(452.56,306.23){\usebox{\plotpoint}}
\put(469.23,318.59){\usebox{\plotpoint}}
\put(486.20,330.49){\usebox{\plotpoint}}
\put(503.48,341.98){\usebox{\plotpoint}}
\put(521.18,352.75){\usebox{\plotpoint}}
\put(538.69,363.89){\usebox{\plotpoint}}
\put(556.63,374.31){\usebox{\plotpoint}}
\put(574.31,385.17){\usebox{\plotpoint}}
\put(592.24,395.61){\usebox{\plotpoint}}
\put(610.31,405.80){\usebox{\plotpoint}}
\put(628.92,414.96){\usebox{\plotpoint}}
\put(647.14,424.90){\usebox{\plotpoint}}
\put(665.12,435.25){\usebox{\plotpoint}}
\put(683.74,444.40){\usebox{\plotpoint}}
\put(702.25,453.75){\usebox{\plotpoint}}
\put(720.74,463.16){\usebox{\plotpoint}}
\put(739.21,472.60){\usebox{\plotpoint}}
\put(757.95,481.52){\usebox{\plotpoint}}
\put(776.44,490.93){\usebox{\plotpoint}}
\put(795.21,499.75){\usebox{\plotpoint}}
\put(813.82,508.92){\usebox{\plotpoint}}
\put(832.58,517.79){\usebox{\plotpoint}}
\put(851.43,526.47){\usebox{\plotpoint}}
\put(869.92,535.87){\usebox{\plotpoint}}
\put(888.81,544.46){\usebox{\plotpoint}}
\put(907.71,553.05){\usebox{\plotpoint}}
\put(926.54,561.77){\usebox{\plotpoint}}
\put(945.32,570.60){\usebox{\plotpoint}}
\put(964.22,579.19){\usebox{\plotpoint}}
\put(983.11,587.78){\usebox{\plotpoint}}
\put(1002.10,596.13){\usebox{\plotpoint}}
\put(1021.07,604.49){\usebox{\plotpoint}}
\put(1039.96,613.07){\usebox{\plotpoint}}
\put(1058.86,621.66){\usebox{\plotpoint}}
\put(1077.75,630.25){\usebox{\plotpoint}}
\put(1096.99,638.00){\usebox{\plotpoint}}
\put(1115.71,646.96){\usebox{\plotpoint}}
\put(1134.61,655.55){\usebox{\plotpoint}}
\put(1153.58,663.94){\usebox{\plotpoint}}
\put(1172.74,671.88){\usebox{\plotpoint}}
\put(1191.50,680.75){\usebox{\plotpoint}}
\put(1210.53,689.01){\usebox{\plotpoint}}
\put(1229.59,697.18){\usebox{\plotpoint}}
\put(1238,701){\usebox{\plotpoint}}
\end{picture}
% GNUPLOT: LaTeX picture
\setlength{\unitlength}{0.240900pt}
\ifx\plotpoint\undefined\newsavebox{\plotpoint}\fi
\sbox{\plotpoint}{\rule[-0.200pt]{0.400pt}{0.400pt}}%
\begin{picture}(1259,840)(0,0)
\font\gnuplot=cmr10 at 10pt
\gnuplot
\sbox{\plotpoint}{\rule[-0.200pt]{0.400pt}{0.400pt}}%
\put(161.0,394.0){\rule[-0.200pt]{4.818pt}{0.400pt}}
\put(141,394){\makebox(0,0)[r]{$1$}}
\put(1218.0,394.0){\rule[-0.200pt]{4.818pt}{0.400pt}}
\put(161.0,665.0){\rule[-0.200pt]{4.818pt}{0.400pt}}
\put(141,665){\makebox(0,0)[r]{$2$}}
\put(1218.0,665.0){\rule[-0.200pt]{4.818pt}{0.400pt}}
\put(400.0,123.0){\rule[-0.200pt]{0.400pt}{4.818pt}}
\put(400,82){\makebox(0,0){$1$}}
\put(400.0,780.0){\rule[-0.200pt]{0.400pt}{4.818pt}}
\put(640.0,123.0){\rule[-0.200pt]{0.400pt}{4.818pt}}
\put(640,82){\makebox(0,0){$2$}}
\put(640.0,780.0){\rule[-0.200pt]{0.400pt}{4.818pt}}
\put(879.0,123.0){\rule[-0.200pt]{0.400pt}{4.818pt}}
\put(879,82){\makebox(0,0){$3$}}
\put(879.0,780.0){\rule[-0.200pt]{0.400pt}{4.818pt}}
\put(1118.0,123.0){\rule[-0.200pt]{0.400pt}{4.818pt}}
\put(1118,82){\makebox(0,0){$4$}}
\put(1118.0,780.0){\rule[-0.200pt]{0.400pt}{4.818pt}}
\put(161.0,123.0){\rule[-0.200pt]{259.449pt}{0.400pt}}
\put(1238.0,123.0){\rule[-0.200pt]{0.400pt}{163.089pt}}
\put(161.0,800.0){\rule[-0.200pt]{259.449pt}{0.400pt}}
\put(40,461){\makebox(0,0){$aV(r)$}}
\put(699,21){\makebox(0,0){$r/a$}}
\put(281,610){\makebox(0,0)[l]{Improved Action}}
\put(161.0,123.0){\rule[-0.200pt]{0.400pt}{163.089pt}}
\put(400.0,293.0){\usebox{\plotpoint}}
\put(390.0,293.0){\rule[-0.200pt]{4.818pt}{0.400pt}}
\put(390.0,294.0){\rule[-0.200pt]{4.818pt}{0.400pt}}
\put(499.0,363.0){\rule[-0.200pt]{0.400pt}{0.723pt}}
\put(489.0,363.0){\rule[-0.200pt]{4.818pt}{0.400pt}}
\put(489.0,366.0){\rule[-0.200pt]{4.818pt}{0.400pt}}
\put(576.0,408.0){\rule[-0.200pt]{0.400pt}{1.927pt}}
\put(566.0,408.0){\rule[-0.200pt]{4.818pt}{0.400pt}}
\put(566.0,416.0){\rule[-0.200pt]{4.818pt}{0.400pt}}
\put(640.0,434.0){\rule[-0.200pt]{0.400pt}{1.204pt}}
\put(630.0,434.0){\rule[-0.200pt]{4.818pt}{0.400pt}}
\put(630.0,439.0){\rule[-0.200pt]{4.818pt}{0.400pt}}
\put(696.0,467.0){\rule[-0.200pt]{0.400pt}{1.204pt}}
\put(686.0,467.0){\rule[-0.200pt]{4.818pt}{0.400pt}}
\put(686.0,472.0){\rule[-0.200pt]{4.818pt}{0.400pt}}
\put(747.0,485.0){\rule[-0.200pt]{0.400pt}{1.445pt}}
\put(737.0,485.0){\rule[-0.200pt]{4.818pt}{0.400pt}}
\put(737.0,491.0){\rule[-0.200pt]{4.818pt}{0.400pt}}
\put(838.0,526.0){\rule[-0.200pt]{0.400pt}{1.686pt}}
\put(828.0,526.0){\rule[-0.200pt]{4.818pt}{0.400pt}}
\put(828.0,533.0){\rule[-0.200pt]{4.818pt}{0.400pt}}
\put(879.0,539.0){\rule[-0.200pt]{0.400pt}{2.168pt}}
\put(869.0,539.0){\rule[-0.200pt]{4.818pt}{0.400pt}}
\put(869.0,548.0){\rule[-0.200pt]{4.818pt}{0.400pt}}
\put(879.0,547.0){\rule[-0.200pt]{0.400pt}{1.204pt}}
\put(869.0,547.0){\rule[-0.200pt]{4.818pt}{0.400pt}}
\put(869.0,552.0){\rule[-0.200pt]{4.818pt}{0.400pt}}
\put(990.0,593.0){\rule[-0.200pt]{0.400pt}{4.336pt}}
\put(980.0,593.0){\rule[-0.200pt]{4.818pt}{0.400pt}}
\put(980.0,611.0){\rule[-0.200pt]{4.818pt}{0.400pt}}
\put(1118.0,658.0){\rule[-0.200pt]{0.400pt}{2.891pt}}
\put(1108.0,658.0){\rule[-0.200pt]{4.818pt}{0.400pt}}
\put(1108.0,670.0){\rule[-0.200pt]{4.818pt}{0.400pt}}
\put(1176.0,676.0){\rule[-0.200pt]{0.400pt}{3.373pt}}
\put(1166.0,676.0){\rule[-0.200pt]{4.818pt}{0.400pt}}
\put(1166.0,690.0){\rule[-0.200pt]{4.818pt}{0.400pt}}
\put(1231.0,699.0){\rule[-0.200pt]{0.400pt}{3.132pt}}
\put(1221.0,699.0){\rule[-0.200pt]{4.818pt}{0.400pt}}
\put(400,294){\circle{12}}
\put(499,365){\circle{12}}
\put(576,412){\circle{12}}
\put(640,437){\circle{12}}
\put(696,470){\circle{12}}
\put(747,488){\circle{12}}
\put(838,530){\circle{12}}
\put(879,544){\circle{12}}
\put(879,550){\circle{12}}
\put(990,602){\circle{12}}
\put(1118,664){\circle{12}}
\put(1176,683){\circle{12}}
\put(1231,705){\circle{12}}
\put(1221.0,712.0){\rule[-0.200pt]{4.818pt}{0.400pt}}
\put(275.00,123.00){\usebox{\plotpoint}}
\put(283.06,142.11){\usebox{\plotpoint}}
\multiput(292,160)(9.282,18.564){2}{\usebox{\plotpoint}}
\put(312.38,196.99){\usebox{\plotpoint}}
\put(324.09,214.12){\usebox{\plotpoint}}
\put(337.01,230.37){\usebox{\plotpoint}}
\put(350.62,246.03){\usebox{\plotpoint}}
\put(364.64,261.33){\usebox{\plotpoint}}
\put(379.65,275.65){\usebox{\plotpoint}}
\put(394.57,290.07){\usebox{\plotpoint}}
\put(410.39,303.50){\usebox{\plotpoint}}
\put(426.92,316.03){\usebox{\plotpoint}}
\put(443.44,328.59){\usebox{\plotpoint}}
\put(460.68,340.13){\usebox{\plotpoint}}
\put(477.66,352.06){\usebox{\plotpoint}}
\put(495.17,363.20){\usebox{\plotpoint}}
\put(513.11,373.62){\usebox{\plotpoint}}
\put(531.05,384.03){\usebox{\plotpoint}}
\put(548.83,394.72){\usebox{\plotpoint}}
\put(566.98,404.79){\usebox{\plotpoint}}
\put(585.03,415.02){\usebox{\plotpoint}}
\put(603.52,424.42){\usebox{\plotpoint}}
\put(622.01,433.82){\usebox{\plotpoint}}
\put(640.50,443.23){\usebox{\plotpoint}}
\put(658.90,452.74){\usebox{\plotpoint}}
\put(677.67,461.58){\usebox{\plotpoint}}
\put(696.16,470.98){\usebox{\plotpoint}}
\put(715.05,479.57){\usebox{\plotpoint}}
\put(733.95,488.16){\usebox{\plotpoint}}
\put(752.50,497.45){\usebox{\plotpoint}}
\put(771.15,506.52){\usebox{\plotpoint}}
\put(790.05,515.11){\usebox{\plotpoint}}
\put(809.29,522.86){\usebox{\plotpoint}}
\put(828.13,531.56){\usebox{\plotpoint}}
\put(846.90,540.41){\usebox{\plotpoint}}
\put(865.80,549.00){\usebox{\plotpoint}}
\put(885.04,556.74){\usebox{\plotpoint}}
\put(903.93,565.33){\usebox{\plotpoint}}
\put(922.83,573.92){\usebox{\plotpoint}}
\put(941.92,582.05){\usebox{\plotpoint}}
\put(961.00,590.18){\usebox{\plotpoint}}
\put(980.05,598.39){\usebox{\plotpoint}}
\put(999.29,606.13){\usebox{\plotpoint}}
\put(1018.04,615.02){\usebox{\plotpoint}}
\put(1037.25,622.84){\usebox{\plotpoint}}
\put(1056.49,630.58){\usebox{\plotpoint}}
\put(1075.39,639.14){\usebox{\plotpoint}}
\put(1094.62,646.92){\usebox{\plotpoint}}
\put(1113.71,655.05){\usebox{\plotpoint}}
\put(1132.95,662.80){\usebox{\plotpoint}}
\put(1152.19,670.54){\usebox{\plotpoint}}
\put(1171.38,678.41){\usebox{\plotpoint}}
\put(1190.45,686.58){\usebox{\plotpoint}}
\put(1209.56,694.66){\usebox{\plotpoint}}
\put(1228.71,702.62){\usebox{\plotpoint}}
\put(1238,706){\usebox{\plotpoint}}
\end{picture}
\end{center}
\caption{Static-quark potential computed using the Wilson gluon action
and the $\order(a^2)$ improved action. The dotted line in each case is
a fit of the Monte Carlo results to $\sigma r - b/r + c$. The lattice
spacing $a\approx0.25$\,fm in each case. Each plot required about
3~hours of computer time using a 300\,MHz personal computer. The
improved action gives a smoother curve.}
\label{v}
\end{figure}
\end{exercise}

\section{Conclusions}
The QCD simulation discussed in the previous section is, of course, just
a beginning. A 
reasonable next step would be to add heavy quarks to the simulation,
using nonrelativistic QCD, and to compute spectra and wavefunctions
for $\psi/J$ and $\Upsilon$~mesons\cite{schlad,nrqcd1,nrqcd2}. 
Far more costly to simulate, although no
more subtle theoretically, are light quarks\cite{schlad,alford}. 
These are needed to
analyze protons, neutrons and other standard hadrons. Finally, and
most costly,  one might include light-quark vacuum polarization, which
is an essential step for high-precision calculations (better than 10--15\%).

\section*{Acknowledgements} This work was supported by a grant from
the National Science Foundation.

\section*{Appendix}
The sample simulation code in Section~2 is written in Python, a simple
but powerful computer language that is freely available for just about
any type of computer from the web site www.python.org. These
simulations are most efficient when the Numeric package is used with
Python; this is included with some Python distributions but must
downloaded separately (from www.python.org) for others. To use the
code given in the text, collect it together in a single file as
follows:
\begin{verbatim}

    import Numeric
    from whrandom import uniform
    from math import *

    # ... code from text goes here

    # set parameters:
    N = 20                      
    N_cor = 20
    N_cf = 100
    a = 0.5
    eps = 1.4

    # create arrays:
    x = Numeric.zeros((N,),Numeric.Float)       
    G = Numeric.zeros((N_cf,N),Numeric.Float)

    # do the simulation:
    MCaverage(x,G)

\end{verbatim}
If the file is called \verb|simulation.py|, it is run with the command
\verb|python simulation.py|. 

To test the binning and bootstrap codes add the following to the the
file:
\begin{verbatim}

    def avg(G):                       # MC avg of G
        return Numeric.sum(G)/len(G)

    def sdev(G):                      # std dev of G
        g = Numeric.asarray(G)
        return Numeric.absolute(avg(g**2)-avg(g)**2)**0.5

    print 'avg G\n',avg(G)
    print 'avg G (binned)\n',avg(bin(G,4))
    print 'avg G (bootstrap)\n',avg(bootstrap(G))          

\end{verbatim}
The average of the binned copy of \verb|G| should be the same as the
average of \verb|G| itself; the average of the bootstrap copy should
be different by an amount of order the Monte Carlo error. Compute
averages for several bootstrap copies to get a good feel for the
errors.

Finally one wants to extract energies. This is done by adding code to
compute $\Delta E(t)$:
\begin{verbatim}

    def deltaE(G):                   # Delta E(t)
        avgG = avg(G)
        adE = Numeric.log(Numeric.absolute(avgG[:-1]/avgG[1:]))
        return adE/a

    print 'Delta E\n',deltaE(G)
    print 'Delta E (bootstrap)\n',deltaE(bootstrap(G))

\end{verbatim}
Again repeating the evaluation for 50 or 100 bootstrap copies of
\verb|G| gives an estimate of the statistical errors in the
energies. Additional 
code can be added to evaluate standard deviations from these
copies:
\begin{verbatim}

    def bootstrap_deltaE(G,nbstrap=100): # Delta E + errors
        avgE = deltaE(G)            # avg deltaE
        bsE = []
        for i in range(nbstrap):    # bs copies of deltaE
            g = bootstrap(G)
            bsE.append(deltaE(g))
        bsE = Numeric.array(bsE)
        sdevE = sdev(bsE)           # spread of deltaE's
        print "\n%2s  %10s  %10s" % ("t","Delta E(t)","error")
        print 26*"-"
        for i in range(len(avgE)/2):
            print "%2d  %10g  %10g" % (i,avgE[i],sdevE[i])

    bootstrap_deltaE(G)

\end{verbatim}
The entire program should take only a few seconds to run on a modern
personal computer.

This example almost completely ignores the powerful object-oriented features
that distinguish Python from most other scripting languages.


\section*{References}
\begin{thebibliography}{99}
\bibitem{baym} See for example Chapter~3 in G.~Baym's {\em Lectures on
Quantum Mechanics} (Benjamin/Cummings, Menlo Park, 1973).

\bibitem{vegas} G.P. Lepage, J. Comp. Phys. {\bf 27} (1978) 192.

\bibitem{tasi-alg} An extensive discussion of the systematics of Monte
Carlo errors can be found in: 
G.P.\,Lepage, {\em The Analysis of Algorithms for
Lattice Field 
Theory}, in {\em From Actions to Answers}, edited by T.\,DeGrand and D.
Toussaint (World Scientific, Singapore, 1989).

\bibitem{metropolis} See for example M. Creutz, {\em Quarks, gluons,
and lattices} (Cambridge University Press, Cambridge, 1985).

\bibitem{python} Download python from www.python.org. See also M. Lutz and
D. Ascher, {\em Learning Python} (O'Reilly and Associates, 1999).

\bibitem{gpl93} G.P.\,Lepage and P.B.\,Mackenzie, {\em Phys.\ Rev.\/} {\bf
D48} (1993) 2250.

\bibitem{wilson83} For an 
overview see K.G.\,Wilson, {\em Rev.\ Mod.\ Phys.\/} {\bf 55} 
(1983) 583.

\bibitem{gpl94} G.P.\,Lepage, {\em Lattice QCD for Small Computers},
in {\em The Building Blocks of Creation,} edited by 
S.\,Raby and T.\,Walker (World Scientific Press, Singapore, 1994).

\bibitem{tsukuba}
G.P.~Lepage,
%``Perturbative improvement for lattice QCD: An update,''
Nucl.\ Phys.\ Proc.\ Suppl.\ {\bf 60A} (1998) 267

\bibitem{schlad} Parts of Section~3 and~4 are based on the 
following article; it is more detailed and also discusses
simulations of light and heavy quarks:
G.P.~Lepage, {\em Redesigning Lattice QCD}, in 
{\em Perturbative and Nonperturbative Aspects of Quantum Field
Theory} (Lecture Notes in Physics, 479),
edited by H.~Latal and W.~Schweiger (Springer-Verlag, Berlin, 1997).

\bibitem{kawai81} See, for example, 
H.\,Kawai, R.\,Nakayama and K.\,Seo, {\em Nucl.\
Phys.\/} {\bf B189} (1981) 40.

\bibitem{curci83} G.\,Curci, P.\,Menotti, and G.\,Paffuti,
    {\em Phys.\ Lett.} {\bf 130B} (1983) 205;
     Erratum:  {\em ibid.} {\bf 135B} (1984) 516.

\bibitem{luscher85a} M.~L\"uscher and P.~Weisz, {\em
Comm.\ Math.\ Phys.\/} {\bf 97} 
(1985) 59.

\bibitem{alford95} M.\,Alford, W.\,Dimm, G.P.\,Lepage, G.\,Hockney and
P.\,Mackenzie, {\em Phys.\ Lett.\/} {\bf B361} (1995) 87.

\bibitem{luscher85b} M.~L\"uscher and P.~Weisz, {\em
Phys.\ Lett.\/} {\bf 158B},
(1985) 250, and references therein.

\bibitem{alford}
See for example: M.~Alford, T.R.~Klassen and G.P.~Lepage,
%``A quark action for very coarse lattices,''
Phys.\ Rev.\ {\bf D58} (1998) 034503.


\bibitem{nrqcd1} G.P.\,Lepage and B.\,Thacker, {\em Nucl.\ Phys.\/}
{\bf B}(Proc.\ Suppl.){\bf 4} (1988) 199; G.P.\,Lepage, K.\,Hornbostel,
L.\,Magnea, U.\,Magnea and C.\,Nakhleh,
{\em Phys.\ Rev.\/} {\bf D46} (1992) 4052.

\bibitem{nrqcd2} C.\,Davies  et al., {\em Phys.\ Rev.} 
{\bf D50} (1994) 6963; P.\,McCallum and J.\,Shigemitsu,
{\em Nucl.\ Phys.\/} {\bf B47} (Proc.\ Suppl.) (1996) 409;
C.\,Davies et al., {\em Phys.\ Rev.\ Lett.\/}
{\bf 73} (1994) 2654; C.\,Davies et al., {\em Phys.\ Lett.\/}
{\bf B345} (1995) 42; J.\,Shigemitsu's talk at {\em Lattice '96} (June
1996, St.\ Louis).

\end{thebibliography}
\end{document}